\documentclass[letterpaper]{article} 
\usepackage{amsmath}
\usepackage{amssymb}
\usepackage{aaai2027}  
\usepackage[hyphens]{url}  
\usepackage{graphicx} 
\usepackage{natbib}  
\usepackage{caption} 
\usepackage{placeins}
\usepackage{algorithm}
\usepackage{algorithmic}
\usepackage{xcolor}
\usepackage{tikz}
\usetikzlibrary{backgrounds}
\usetikzlibrary{positioning,fit,quotes,arrows,calc,backgrounds}

\definecolor{color1}{HTML}{53446B}
\definecolor{color2}{HTML}{807DBA}
\definecolor{color3}{HTML}{C0AECE}
\definecolor{color4}{HTML}{16045C}
\definecolor{color5}{HTML}{A40407}
\definecolor{color6}{HTML}{DAB6B6}
\definecolor{color7}{HTML}{CE9292}
\definecolor{color8}{HTML}{5A0404}
\definecolor{color9}{HTML}{9EC109}
\definecolor{color10}{HTML}{D27C5A}
\definecolor{color11}{HTML}{6E9C91}

\usepackage{newfloat}
\usepackage{listings}
\DeclareCaptionStyle{ruled}{labelfont=normalfont,labelsep=colon,strut=off} 
\floatstyle{ruled}
\newfloat{listing}{tb}{lst}{}
\floatname{listing}{Listing}

\usepackage{booktabs}

\title{Causal DAG-Guided Reinforcement Learning for Long-Term Physical Activity Maintenance: Using Prior Data to Inform Future Deployment}
\author{
    Xueqing Liu\textsuperscript{\rm 1}, 
    Young Hyun Cho\textsuperscript{\rm 1},
    Bhanu Teja Gullapalli\textsuperscript{\rm 1},\\
    Predrag Klasnja\textsuperscript{\rm 2},
    Susan Murphy\textsuperscript{\rm 1}
}
\affiliations{
    \textsuperscript{\rm 1} Harvard University\\
    \textsuperscript{\rm 2} University of Michigan

}

\usepackage{mathtools}
\usepackage{tabularx}
\usepackage{amsmath,amssymb,amsthm,graphicx,bm}
\usepackage{float}
\usepackage{subcaption}
\usepackage{xcolor}
\usepackage{url}
\usepackage{mathtools}
\usepackage{float}
\usepackage{booktabs,multirow,bigstrut,makecell}
\usepackage{algorithm}
\usepackage{algorithmic}
\usepackage{listings}
\usepackage{tikz}
\usetikzlibrary{backgrounds}
\usetikzlibrary{positioning,fit,quotes,arrows, calc, backgrounds}

\definecolor{color1}{HTML}{53446B}
\definecolor{color2}{HTML}{807DBA}
\definecolor{color3}{HTML}{C0AECE}
\definecolor{color4}{HTML}{16045C}
\definecolor{color5}{HTML}{A40407}
\definecolor{color6}{HTML}{DAB6B6}
\definecolor{color7}{HTML}{CE9292}
\definecolor{color8}{HTML}{5A0404}
\definecolor{color9}{HTML}{9EC109}
\definecolor{color10}{HTML}{D27C5A}
\definecolor{color11}{HTML}{6E9C91}

\definecolor{dkgreen}{rgb}{0,0.6,0}
\definecolor{gray}{rgb}{0.5,0.5,0.5}
\definecolor{mauve}{rgb}{0.58,0,0.82}

\theoremstyle{definition}

\theoremstyle{remark}

\providecommand{\by}{\bm{y}}

\providecommand{\bX}{\bm{X}}

\providecommand{\bbeta}{\bm{\beta}}

\providecommand{\btheta}{\bm{\theta}}

\begin{document}

\maketitle

\begin{abstract}
Digital health interventions can provide scalable, personalized support for physical activity (PA), yet sustaining PA over the long-term remains difficult. Because affective experiences with PA are important for activity sustainment, we developed an online reinforcement learning (RL) system for the forthcoming ADAPT factorial trial that personalizes twice-daily activity suggestions aimed at improving affective associations with PA.  This setting poses several challenges: participant-specific data are scarce relative to the candidate state space, important states and rewards are partially observed, interventions are frequent while the primary outcome is measured weekly, and activity suggestions can both improve PA-related experiences and increase intervention burden. We address these challenges using an expert-specified causal directed acyclic graph as an integrated RL design tool, guiding state construction, estimation of partially observed states, delayed credit assignment, and representation of beneficial and adverse intervention pathways. We further construct a longitudinal simulator calibrated from a prior micro-randomized trial to compare candidate RL designs under realistic state evolution, missingness, and participant heterogeneity before deployment. Together, this work illustrates how causal domain knowledge and prior trial data can make long-horizon, participant-specific RL more practical for digital health interventions targeting sustained behavior change.
\end{abstract}


\begin{links}
\link{Code}{https://github.com/xueqingliuu/ADAPT_Experiment}
\end{links}
\section{Introduction}
\label{sec:intro}

Digital health interventions (DHIs) offer a scalable way to deliver personalized support for physical activity (PA) \citep{bi2024effectiveness,klasnja2019efficacy}. However, maintaining long-term PA behavior remains difficult: interventions may increase activity in the short term without necessarily supporting long-term maintenance \citep{amagai2022challenges}. One potential determinant of maintenance is how individuals feel about physical activity. 
Affective associations with PA---automatic affective reactions to the thought of being active and cognitive attitudes about liking or disliking activity---can shape future behavioral choices \citep{brand2018affective,stevens2020affective}, and more positive affective associations have been associated with greater subsequent PA participation \citep{williams2008acute,rhodes2015can}. These findings motivate intervention strategies that target not only the amount of PA performed, but also the affective experience associated with being active \citep{simonton2021testing,choi2025improving}.

The forthcoming 9-month ADAPT factorial trial, scheduled to begin in fall 2026, aims to improve participants' affective associations with PA through a multicomponent DHI. Among participants randomized to the ADAPT-on conditions\footnote{The ADAPT study includes multiple intervention components. RL is used only for the activity suggestions component.}, 
an online reinforcement learning (RL) algorithm \citep{sutton1998reinforcement} will personalize the delivery of activity suggestions at two decision times per day. The RL algorithm uses data from wristband activity tracker, the HeartSteps physical activity mHealth app, and self-reported surveys for learning and decision-making. The primary outcome optimized by the RL algorithm is the core affective experience (CAE) subscale of the Affective Experiences (AFFEXX) scale \citep{ekkekakis2021you}, a weekly measure of how participants feel about being physically active. Thus, the algorithm must learn when to deliver activity suggestions at frequent decision times in order to improve an outcome that is observed and evolves on a much slower, weekly timescale.

Designing and deploying a participant-specific online RL algorithm in this setting raises several interconnected challenges. First, participant-specific data are scarce relative to the potentially large state space. Each participant provides only a limited number of interactions with the RL algorithm, while wristband activity tracker, app, and survey data provide many candidate state variables. Learning a high-dimensional decision rule from scratch can therefore lead to high estimation variance and poor performance early in deployment \citep{liao2020personalized,gazi2026statistical}. Second, important state information may be partially observed. Sensor measurements can be delayed or unavailable, and participants may not complete all surveys, resulting in missing or delayed state variables and outcomes \citep{seewald2019practical}. Third, more intervention is not necessarily better. Although activity suggestions may improve affective associations with PA, frequent or poorly timed suggestions may increase intervention burden or reduce perceived utility, producing adverse effects that may only become apparent later \citep{nahum2016just}. An effective policy must therefore account for both the potential benefit of an activity suggestion and its possible downstream costs. Finally, the reward is sparse and delayed. Activity suggestions can be delivered twice daily from Monday through Saturday, resulting in 12 decision times associated with only one weekly CAE reward. This creates a difficult credit-assignment problem: the RL algorithm must learn which earlier actions contributed to the eventual weekly outcome \citep{arjona2019rudder}.

We use an expert-specified causal directed acyclic graph (DAG) to incorporate domain knowledge into the RL system and address these challenges in a principled manner. The DAG guides four linked design choices. First, it informs state construction by identifying which aspects of the observed history are relevant for future states and outcomes, allowing us to build lower-dimensional state summaries rather than include every available sensor, app, and survey variable \citep{gao2025harnessing,xu2026integrating}. Second, it specifies how latent weekly CAE and perceived utility relate to observed survey, sensor, and engagement measures, motivating the models used to estimate these states when direct measurements are missing or incomplete. Third, the DAG represents both a beneficial pathway to CAE-assessed affective associations through anticipated affect and a potentially adverse pathway through perceived utility. This allows the RL design to account for both the intended effects of activity suggestions and their possible delayed effects on future engagement and outcomes. Finally, the DAG helps address sparse and delayed rewards by identifying intermediate variables along the pathways from activity suggestions to future CAE, which provide structured information for delayed credit assignment and reward redistribution across within-week decisions \citep{arjona2019rudder,zou2025causal}. 

Our main contributions are summarized as follows:
\begin{enumerate}
    \item We develop a participant-specific online RL framework that uses an expert-specified causal DAG to guide state construction, latent-state estimation, representation of benefit and burden pathways, and delayed credit assignment in a sparse-data digital health setting.
    \item We construct an ADAPT simulation testbed calibrated using data from a prior micro-randomized trial (MRT). It reproduces the key decision schedule, longitudinal feedback pathways, and observation processes of the forthcoming ADAPT trial, enabling evaluation of candidate RL variants before deployment.
\end{enumerate}

\section{Related Work}
\label{sec:related_work}

Our work is related to three streams of research:
\paragraph{Online RL in digital health.} Online RL has increasingly been deployed to personalize DHIs, including interventions for physical activity \citep{yom2017encouraging,zhou2018personalizing,liao2020personalized,aguilera2024effectiveness}, medication adherence \citep{lauffenburger2024impact}, chronic pain \citep{piette2022patient}, cannabis use \citep{ghosh2024rebandit}, oral health \citep{trella2025deployed}, and dietary behavior \citep{gonzalez2026practical}. Most deployed systems use contextual bandits targeting relatively proximal outcomes, although some account for longer-term effects through engineered state features, reward adjustments, or nonstationary dynamics \citep{liao2020personalized,mintz2020nonstationary,baek2025policy}. More closely related, \citet{pulick2026optimizing} explicitly model a dynamic engagement state and plan over its future consequences. 

ADAPT combines several additional long-horizon challenges: multiple within-week decisions share a weekly outcome, key weekly states are partially observed, and effects may propagate across weeks through distinct affective-associations and perceived-utility pathways.

\paragraph{Causal RL.}
Causal structure has been used to improve sequential decision-making by reducing exploration complexity \citep{lattimore2016causal,lu2020regret,lu2022efficient}, restricting policy search \citep{zhang2020designing}, constructing state representations \citep{huang2022action,xu2026integrating}, and estimating long-term treatment effects from short-term data \citep{tran2024inferring}. Closest to our setting, \citet{gao2025harnessing} use an expert-specified DAG to construct Markovian states for multiple decisions sharing a delayed reward, while \citet{gao2026active} use causal structure and belief-state modeling for partially observed states. Causal approaches have also addressed delayed credit assignment through latent reward redistribution \citep{zhang2023interpretable} and counterfactual Shapley attribution \citep{li2026counterfactual}.

Our work uses a single expert-specified DAG to jointly guide state construction, partially observed-state estimation, representation of beneficial and adverse pathways, and mediator-based credit assignment in a participant-specific online RL system.

\paragraph{Pre-deployment evaluation.} 
Simulation testbeds informed by historical data are commonly used to evaluate digital health RL algorithms before deployment\citep{liao2020personalized,trella2022designing,mintz2020nonstationary,ghosh2024rebandit}. Related model-based approaches estimate participant dynamics from behavioral-intervention data and use the resulting models for closed-loop simulation and robustness analysis \citep{el2023system,el2025data}. More recently, \citet{pulick2026optimizing} used an MRT-calibrated hierarchical dynamical model to generate synthetic patients for evaluating long-horizon RL algorithms, while \citet{xu2025reinforcement} constructed a simulator from clinical data to evaluate multi-agent RL before an upcoming medication-adherence trial.

The ADAPT simulator extends this line of work by modeling multi-timescale states, mediators, outcomes, and observation processes, enabling pre-deployment evaluation under delayed rewards, participant heterogeneity, and incomplete observations.


\begin{figure}[t]
    \centering
    \begin{tikzpicture}[->, thick, main/.style={font=\sffamily}]
    \tikzset{
        fixedwidth/.style={draw=none, text width=0.75cm, align=center},
        circled/.style={fixedwidth, draw=blue, thick, circle, inner sep=1.5pt}
    }

\matrix [column sep=0.3cm, row sep=0.45cm] {
    \node[circled] (R0) {$Y_w$};
    & & &
    \node[circled] (RK) {$Y_{w+1}$};
    & & &
    \node[circled] (RK1) {$Y_{w+2}$}; \\

    &
    \node[fixedwidth] (MR1) {$M^Y_{w,1}$};
    &
    \node[fixedwidth] (MR2) {$M^Y_{w,2}$};
    & &
    \node[fixedwidth] (MR11) {$M^Y_{w+1,1}$};
    &
    \node[fixedwidth] (MR21) {$M^Y_{w+1,2}$};
    & \\

    \node[fixedwidth] (I0) {$J_w$};
    &
    \node[fixedwidth] (A1) {$A_{w,1}$};
    &
    \node[fixedwidth] (A2) {$A_{w,2}$};
    &
    \node[fixedwidth] (IK) {$J_{w+1}$};
    &
    \node[fixedwidth] (A11) {$A_{w+1,1}$};
    &
    \node[fixedwidth] (A21) {$A_{w+1,2}$};
    & \\

    &
    \node[fixedwidth] (ME1) {$M^E_{w,1}$};
    &
    \node[fixedwidth] (ME2) {$M^E_{w,2}$};
    & &
    \node[fixedwidth] (ME11) {$M^E_{w+1,1}$};
    &
    \node[fixedwidth] (ME21) {$M^E_{w+1,2}$};
    & \\

    \node[circled] (E0) {$E_w$};
    & & &
    \node[circled] (EK) {$E_{w+1}$};
    & & &
    \node[circled] (EK1) {$E_{w+2}$}; \\
};

    \path[every node/.style={font=\sffamily}]
        (R0) edge [color4] node [right] {} (RK)
        (RK) edge [color4] node [right] {} (RK1)

        (I0) edge [color11] node [right] {} (ME1)
        (I0) edge [color11] node [right] {} (ME2)
        (IK) edge [color11] node [right] {} (ME11)
        (IK) edge [color11] node [right] {} (ME21)

        (E0) edge [color4] node [right] {} (EK)
        (EK) edge [color4] node [right] {} (EK1)

        (E0) edge [color4] node [right] {} (I0)
        (EK) edge [color4] node [right] {} (IK)

        (R0) edge [color2] node [right] {} (MR1)
        (R0) edge [color2] node [right] {} (MR2)
        (MR1) edge [color2] node [right] {} (RK)
        (MR2) edge [color2] node [right] {} (RK)
        (MR1) edge [color2] node [right] {} (MR2)

        (ME1) edge [color2] node [right] {} (ME2)
        (ME1) edge [color2] node [right] {} (EK)
        (ME2) edge [color2] node [right] {} (EK)

        (RK) edge [color2] node [right] {} (MR11)
        (RK) edge [color2] node [right] {} (MR21)
        (MR11) edge [color2] node [right] {} (RK1)
        (MR21) edge [color2] node [right] {} (RK1)
        (MR11) edge [color2] node [right] {} (MR21)

        (ME11) edge [color2] node [right] {} (ME21)
        (ME11) edge [color2] node [right] {} (EK1)
        (ME21) edge [color2] node [right] {} (EK1)

        (E0) edge [bend left=18, color2] node [right] {} (MR1)
        (E0) edge [bend left=18, color2] node [right] {} (MR2)
        (EK) edge [bend left=18, color2] node [right] {} (MR11)
        (EK) edge [bend left=18, color2] node [right] {} (MR21)

        (E0) edge [color2] node [right] {} (ME1)
        (E0) edge [color2] node [right] {} (ME2)
        (EK) edge [color2] node [right] {} (ME11)
        (EK) edge [color2] node [right] {} (ME21)

        (A1) edge node [right] {} (MR1)
        (A2) edge node [right] {} (MR2)
        (A1) edge node [right] {} (ME1)
        (A2) edge node [right] {} (ME2)

        (A11) edge node [right] {} (MR11)
        (A21) edge node [right] {} (MR21)
        (A11) edge node [right] {} (ME11)
        (A21) edge node [right] {} (ME21)
    ;
    \end{tikzpicture}
    \caption{
Simplified working causal DAG for the ADAPT RL system. 
$A$ denotes activity-suggestion actions, $Y$ weekly CAE, $E$ weekly perceived utility, $M^Y$ CAE-related mediators, $M^E$ perceived-utility-related mediators, and $J$ weekly survey completion.
To keep the diagram readable, the 12 within-week decision times and their associated mediators are represented by two illustrative decision points, indexed by $1$ and $2$; some daily mediators may depend on both within-day actions.
Context variables $C$, CAE observation variables $O^Y$, and perceived-utility emissions $U$ are omitted.}
    \label{fig:dag1sam}
\end{figure}
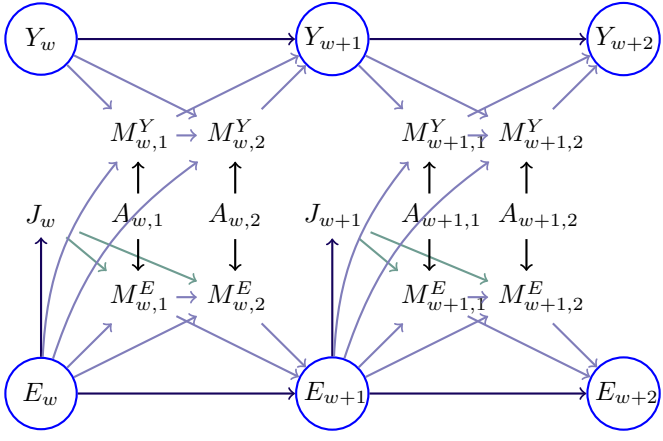


\section{The ADAPT RL System}
\label{sec:RL}
The ADAPT RL system will personalize mobile app-delivered activity suggestions for each participant. In the planned trial, 75 participants will use the system for 36 weeks. Learning draws on three sources: causal domain knowledge encoded in a DAG, historical ADAPT MRT data, and accumulating participant-specific data. Details of the MRT and the planned factorial trial are provided in Appendix~A.

At the decision time $t$ on day $d$ of week $w$, the algorithm observes the current state of the user $S_{w,d,t}$ and chooses $A_{w,d,t}\in\{0,1\}$, indicating whether to provide an activity suggestion. Decisions occur twice daily from Monday through Saturday when the participant is available\footnote{Participants may temporarily turn off activity-suggestion delivery, in which case only $A_{w,d,t}=0$ is available.}. The RL system is updated once per week on Monday night\footnote{Although the update could be performed once the Monday-morning state becomes available, we defer it until Monday night to accommodate potential delay in data collection. Monday's decisions therefore use the most recently available policy.}, after the preceding weekly survey-response window closes. 

The objective is to maximize cumulative weekly CAE over the study horizon. Let $Y_w$ denote the full 12-item weekly CAE outcome. At a weekly assessment, the trial may instead request a short 3-item CAE measure, denoted $\widetilde Y_w$, which provides partial information about $Y_w$ but is not itself the optimization target. The choice of CAE instrument is randomized separately from the activity-suggestion policy.

Thus, decisions made during week $w$ target $Y_{w+1}$, but the weekly outcome may be observed only partially or not at all. Combined with 12 decisions per week, this mismatch between frequent actions and sparse, partially observed outcomes motivates the causal-DAG-guided RL design described next.

\subsection{Causal-DAG-Guided RL Design}
Figure~\ref{fig:dag1sam} encodes the domain assumptions used to structure the RL problem. We use the same expert-specified DAG to guide four linked design choices: state construction,  latent state estimation, representation of beneficial and adverse intervention pathways, and mediator-based delayed credit assignment.

\paragraph{State construction.}  
The DAG provides a principled way to compress the observed history rather than including every available sensor, survey, and app measurement, with the goal of preserving the Markov structure relevant for decision-making~\citep{gao2025harnessing}. In the working DAG, the weekly states $Y_w$ and $E_w$ serve as bottlenecks through which information from prior weeks propagate forward, while within-week actions can affect future states and outcomes through the $Y$- and $E$-related mediators. We therefore retain past information that, under the working DAG, can still influence future mediators, weekly states, or outcomes. The resulting state is intended to render the earlier history approximately conditionally irrelevant for predicting the future, given subsequent actions:
\begin{align} S_{w,d,t} = \big[ &\widehat b_w^Y,\, \widetilde b_w^Y,\, \widehat e_w,\, \overline{M}^{Y}_{w,d,t},\, 
\overline{M}^{E}_{w,d,t},\,
C_{w,d,t}^{\top} \big]^{\top}, 
\end{align} 
Here, $\widehat b_w^Y$ and $\widetilde b_w^Y$ are the posterior mean and standard deviation of CAE, and $\widehat e_w$ is the estimated perceived utility. 
As shown in Figure~1, $\overline M^Y_{w,d,t}\in \mathbb R^2$ contains recency-weighted summaries of the two $Y$-related mediator streams: daily anticipated affect and next four hour step count, whereas $\overline M^E_{w,d,t}\in\mathbb R^3$ summarizes the three $E$-related streams: next four hour app page views, daily Fitbit wear, and daily survey completion. Only mediator measurements observed before the current decision are included. The vector $C_{w,d,t}$ contains currently available contextual variables that may affect subsequent mediators or outcomes or modify the effect of the current action. Full definitions are provided in Appendix~B.

\paragraph{Partial observability.} 
Both weekly CAE $Y_w$ and perceived utility $E_w$ are incompletely observed, but they are estimated differently.

For CAE, variables from week $w-1$ with directed paths into $Y_w$ are used in the transition model to predict the week-$w$ CAE state. This prediction is then updated using week-$w$ measurements that are descendants of $Y_w$. Completion of the full 12-item CAE survey reveals $Y_w$ directly, whereas the short CAE measure $\widetilde Y_w$ and available week-$w$ $Y$-related mediators provide indirect information about it. Thus, the same mediator stream can play different roles across time: week-$(w-1)$ measurements help predict $Y_w$, while week-$w$ measurements help infer its current value.

The algorithm maintains the posterior distribution
\[
b_w(y)
=
p\!\left(Y_w=y\mid\mathcal H_w^{\mathrm{obs}}\right),
\]
where $\mathcal H_w^{\mathrm{obs}}$ denotes the observed history available at the weekly update. A particle filter first propagates the previous CAE state through the transition model using relevant week-$(w-1)$ information and then updates this distribution using current observations: the full CAE survey when observed, or otherwise the short-survey emission and available week-$w$ mediator emissions. The posterior mean $\widehat b_w^Y$ serves as the weekly CAE learning signal, while $\widehat b_w^Y$ and its posterior standard deviation $\widetilde b_w^Y$ enter the RL state.


Perceived utility is estimated separately from CAE. Guided by the incoming arrows to $E_w$ in Figure~\ref{fig:dag1sam}, we estimate the current state from the previous perceived-utility estimate and the $E$-related mediators observed during week $w-1$, while weekly survey completion $J_w$ and usefulness/helpfulness responses $U_{w,1},U_{w,2}$ provide contemporaneous information:
\[
\widehat e_w
=
g_E\!\left(
\widehat e_{w-1},
\overline M^{E}_{w-1,d,t},
J_w,
U_{w,1},
U_{w,2};
\widehat\alpha^{E}
\right).
\]
Unlike CAE, we use a point estimate for perceived utility. The burden-related signal is weaker and more sparsely observed, so maintaining and updating a participant-specific latent-state distribution would add estimation variability early in deployment. We therefore estimate $\widehat\alpha^E$ from the  MRT and keep it fixed during the factorial trial, while updating $\widehat e_w$ as participant-specific observations accumulate. 
Survey-completion indicators are modeled as emissions of $E_w$ rather than $Y_w$, reflecting the working assumption that nonresponse is driven by engagement-related burden and provides information about $E_w$ but not additional information about $Y_w$ conditional on $E_w$ and the observed history. See Supplement Section 2 for details.


\paragraph{Benefit--burden pathways}
Under the working DAG, activity suggestions may affect future CAE through two distinct pathways:
\begin{align*}
\text{Benefit pathway:}\quad
&A_{w,d,t}
\longrightarrow
M^Y_{w,\cdot}
\longrightarrow
Y_{w+1}
\longrightarrow \cdots,\\
\text{Burden pathway:}\quad
&A_{w,d,t}
\longrightarrow
M^E_{w,\cdot}
\longrightarrow
E_{w+1}
\longrightarrow
M^Y_{w+1,\cdot}\\
&\longrightarrow
Y_{w+2}
\longrightarrow \cdots.
\end{align*}
The benefit pathway represents the hypothesized effects of an activity suggestion on anticipated affect and physical activity, which may improve CAE at the end of the week. The burden pathway represents the possibility that repeated or poorly timed suggestions adversely affect engagement-related mediators $M^E_{w,\cdot}$ and perceived utility $E_{w+1}$, with downstream consequences for later CAE-related mediators $M^Y_{w+1,\cdot}$ and CAE $Y_{w+2}$. 
Because the weekly states evolve recursively, these effects may propagate further into the future.
Incorporating summaries of both pathways into the state and value-learning design allows the RL system to account for delayed benefits and more delayed intervention costs rather than optimizing only proximal responses.

\paragraph{Temporal structure and delayed credit assignment.}
Figure~1 highlights two sources of delayed credit. First, 12 within-week $w$ decisions share a single weekly CAE outcome $Y_{w+1}$. Second, effects operating through perceived utility $E_{w+1}$ require an additional week before affecting CAE $Y_{w+2}$. The DAG also identifies two forms of temporal structure that can be exploited for learning. First, $Y_w$ and $E_w$ form a low-dimensional weekly \textit{bottleneck state} between consecutive weeks: effects of earlier actions and mediators on later weeks propagate through these weekly states. Second, the within-week mediators provide intermediate information along the pathways from activity suggestions to the weekly states and outcomes.


These structures motivate three modifications to reward design. First, the \emph{bottleneck-value} variant introduces an auxiliary week-level value function
$V(\widehat b_w, \widetilde b_w. \widehat e_w)$, providing a direct value bridge between within-week decisions and future weeks. Second, \emph{potential-based reward shaping} uses perceived utility as a potential function,
\begin{align*}
R_{w+1}^{\mathrm{PB}}
=
\widehat b_{w+1}^{Y}
+
\gamma\widehat e_{w+1}
-
\widehat e_w.
\end{align*}
The additional term has the standard potential-based form $\gamma\Phi_{w+1}-\Phi_w$, with $\Phi_w=\widehat e_w$. In the infinite-horizon discounted problem this leaves the optimal policy for ($\sum \gamma^w\widehat b_{w+1}^{Y}$) unchanged \citep{ng1999policy}.
Third, \emph{reward redistribution} uses post-action mediator information from the pathways in Figure~1 to construct retrospective decision-level learning signals\citep{arjona2019rudder}. Specifically, the weekly reward is decomposed across the 12 within-week decisions using the corresponding states and subsequently observed mediators.
Full formulations are provided in Supplement Section 2.

\subsection{Online Personalization with RLSVI}
We use randomized least-squares value iteration (RLSVI) to learn a participant-specific activity-suggestion policy online \citep{osband2016generalization}. Its Bayesian formulation allows historical MRT data to initialize learning while retaining randomized exploration as participant-specific data accumulate. We approximate the action-value function using
\[
Q(S,a;\beta)
=
\phi(S,a)^\top\beta,
\qquad
\beta\sim\mathcal N(\mu_0,\Sigma_0),
\]
where $\phi(S,a)$ combines the DAG-guided information state with a parsimonious set of action-interaction features. The prior $(\mu_0,\Sigma_0)$ is constructed from fitted Q-regression on historical ADAPT MRT data, while the linear representation limits model complexity in the data-scarce participant-specific setting.

Following ensemble-based randomized value-function methods \citep{osband2016deep}, we maintain $B$ randomized Q-functions. This is to ensure explicit randomization probabality calculation for after-study analyses. At each available decision time, the fraction of ensemble members that favor an activity suggestion approximates the posterior probability that action $1$ has greater value than action $0$:
\[
\widehat{\pi}_{w,d,t}
=
\frac{1}{B}
\sum_{b=1}^{B}
\mathbf{1}
\left\{
Q(S_{w,d,t},1;\beta^{(b)})
>
Q(S_{w,d,t},0;\beta^{(b)})
\right\}.
\]
We then clip $\widehat{\pi}_{w,d,t}$ to
$[\epsilon_0,1-\epsilon_0]$ with $\epsilon=0.1$ and sample
$A_{w,d,t}\sim\operatorname{Bernoulli}(\pi_{w,d,t})$. The clipping preserves a minimum probability of both actions, supporting continued exploration and subsequent statistical analyses.
Activity suggestions are randomized with probability $0.5$ during the initial week; the participant-specific ensemble is subsequently updated weekly.

In the baseline algorithm, the reward for decisions made during week $w$ is the posterior mean of the subsequent weekly CAE outcome,
$
R_{w+1}^{\mathrm{base}}
=
\widehat b_{w+1}^{Y}
$.
Within-week transitions are undiscounted, and this reward enters directly only at the final decision of the week. The week-boundary target is
\[
R_{w+1}
+
\gamma
\max_{a\in\{0,1\}}
Q(S_{w+1,1,1},a),
\]
where $\gamma$
is the cross-week discount factor. Earlier within-week decisions receive information about the weekly outcome through successive value backups.

All candidate algorithms retain this participant-specific RLSVI backbone and action-selection rule, while varying the cross-week discount factor, bottleneck-value specification, and reward construction. Full feature definitions, posterior updates, and implementation details are provided in Appendix~C; the complete set of RL variants is summarized in Appendix~D.


\section{Pre-Deployment Evaluation}
\label{sec:evaluation}
We use data from the prior ADAPT micro-randomized trial to construct an interactive simulation environment for comparing candidate RL algorithms before deployment in the forthcoming factorial trial. 

\subsection{The ADAPT MRT Data}
The ADAPT MRT closely aligns with the forthcoming factorial trial in intervention structure and data collection: activity suggestions are delivered twice daily, weekly surveys measure CAE and perceived utility, and additional information is collected through daily surveys, wristband activity tracker, and app use. Some variables differ across studies; for example, step count is used as a surrogate for active minutes when the corresponding factorial trial variable is unavailable in the MRT. A complete mapping between MRT and factorial trial variables is provided in Appendix~A.

We use MRT data available through July 26, 2026. Among 43 participants who completed the 12-week active phase, eight were excluded based on prespecified data-availability criteria involving wearable and survey/CAE measurements. An additional seven were excluded based on prespecified CAE ceiling criteria, because their persistently high CAE made them less representative of the population for whom the RL intervention is intended. This yielded 28 participants for simulator construction.
Details are provided in Supplement Section 1. 

The MRT serves two roles: estimating the data-generating models for the simulator and constructing informative priors for participant-specific RL.  During pre-deployment evaluation, these priors are constructed in a leave-one-participant-out manner so that the simulated participant's own MRT data are not used for initialization.

\subsection{Simulation Environment Design}
The simulator generates policy-dependent participant trajectories using three interacting components: a CAE-related process, a perceived-utility process, and additional time-varying state variables. We summarize the resulting deployment choices and remaining limitations.

\paragraph{CAE-related process.} 
This process includes weekly CAE $Y_w$, weekly short CAE $\widetilde{Y}_w$, daily anticipated affect $M_{w,d}^{Y,AA}$, and post-decision next-four-hour step count $M_{w,d,t}^{Y,SC}$. We model these variables with Bayesian hierarchical linear models that partially pool across participants to stabilize estimation and retain participant heterogeneity.


\paragraph{Perceived-utility process.} 
This process includes latent weekly perceived utility $E_w$, weekly survey completion $J_w$, weekly app-usefulness and helpfulness measurements $U_{w,1}$ and $U_{w,2}$, wristband activity tracker wear $M_{w,d}^{E,FW}$, daily survey completion $M_{w,d}^{E,PJ}$, and next-four-hour app page views $M_{w,d,t}^{E,PV}$. We jointly model these variables with a participant-specific state-space model, treating $E_w$ as a latent state inferred from its observed mediators and emissions.


\paragraph{Additional time-varying state variables.} 
The simulator also models recent step count $C_{w,d,t}^{SC}$, recent interaction with activity suggestions $C_{w,d,t}^{AS}$, and daily activity status $C_{w,d}^{AD}$. These processes are estimated separately for each participant using ridge-penalized regression.

To carry temporal dependence remaining in the fitted-model residuals into the simulator, we generate participant-specific residual sequences with an AR(1)-style bootstrap based on the fitted residuals. 


For alternative testbed variants, we retune the fitted environment so that mean STE equals Cohen’s conventional small (0.2) and large (0.8) values \citep{cohen2013statistical}, by strengthening the burden and beneficial pathways, respectively. Full model specifications and simulation procedures are provided in Appendix~E.

\subsection{Experiments and Results}
\label{subsec:results}
We conduct two sets of experiments. We first validate the simulator against observed MRT trajectories and then use it to compare candidate RL variants under the planned 36-week factorial trial setting. RL performance is evaluated by cumulative CAE over the study horizon.


\subsubsection{Simulator Validation}
\label{subsubsec:simu_validation}

We assess simulator fidelity by comparing simulated and observed MRT trajectories. For each participant and process, we calculate the proportion of observed MRT measurements that fall within the pointwise 95\% simulation interval across 100 simulated trajectories. We then average these proportions equally across the 28 participants. For perceived-utility-related mediators, the reported value is the average across app page views, Fitbit wear, and daily survey completion.

\begin{table}[htbp]
\centering
\small
\caption{Participant-averaged coverage of observed MRT measurements by pointwise 95\% simulation intervals.}
\label{tab:simulator_validation}
\begin{tabular}{@{}lc@{}}
\toprule
Process & Coverage \\
\midrule
Weekly CAE & 92.7\% \\
Next-four-hour step count & 89.9\% \\
Anticipated affect & 93.8\% \\
Perceived-utility-related mediators & 97.7\% \\
\bottomrule
\end{tabular}
\end{table}

Table~\ref{tab:simulator_validation} shows that, for the average participant, 89.9--97.7\% of observed MRT measurements across these processes fall within the pointwise 95\% simulation intervals. Thus, the simulator broadly reproduces the observed participant-level trajectories across the main outcome and mediator processes. This agreement is encouraging given the limited longitudinal data available for fitting participant-specific dynamics. The DAG-guided model structure provides a parsimonious representation of how weekly states, mediators, and outcomes evolve across time scales while retaining participant heterogeneity. Appendix~F provides additional model-fit, temporal-dependence, and numerical-stability diagnostics over the planned 36-week horizon.

\subsubsection{RL Variants Comparison}
\label{subsubsec:simu_comparison}
We conduct 200 simulation replications. In each replication, 75 simulated participants are sampled with replacement from the 28 MRT participant models. When the same participant model is sampled more than once, each replicate follows the same fitted participant-specific dynamics but receives an independently generated stochastic trajectory through residual resampling. 

\begin{figure}[htbp]
    \centering
    \includegraphics[width=1\linewidth]{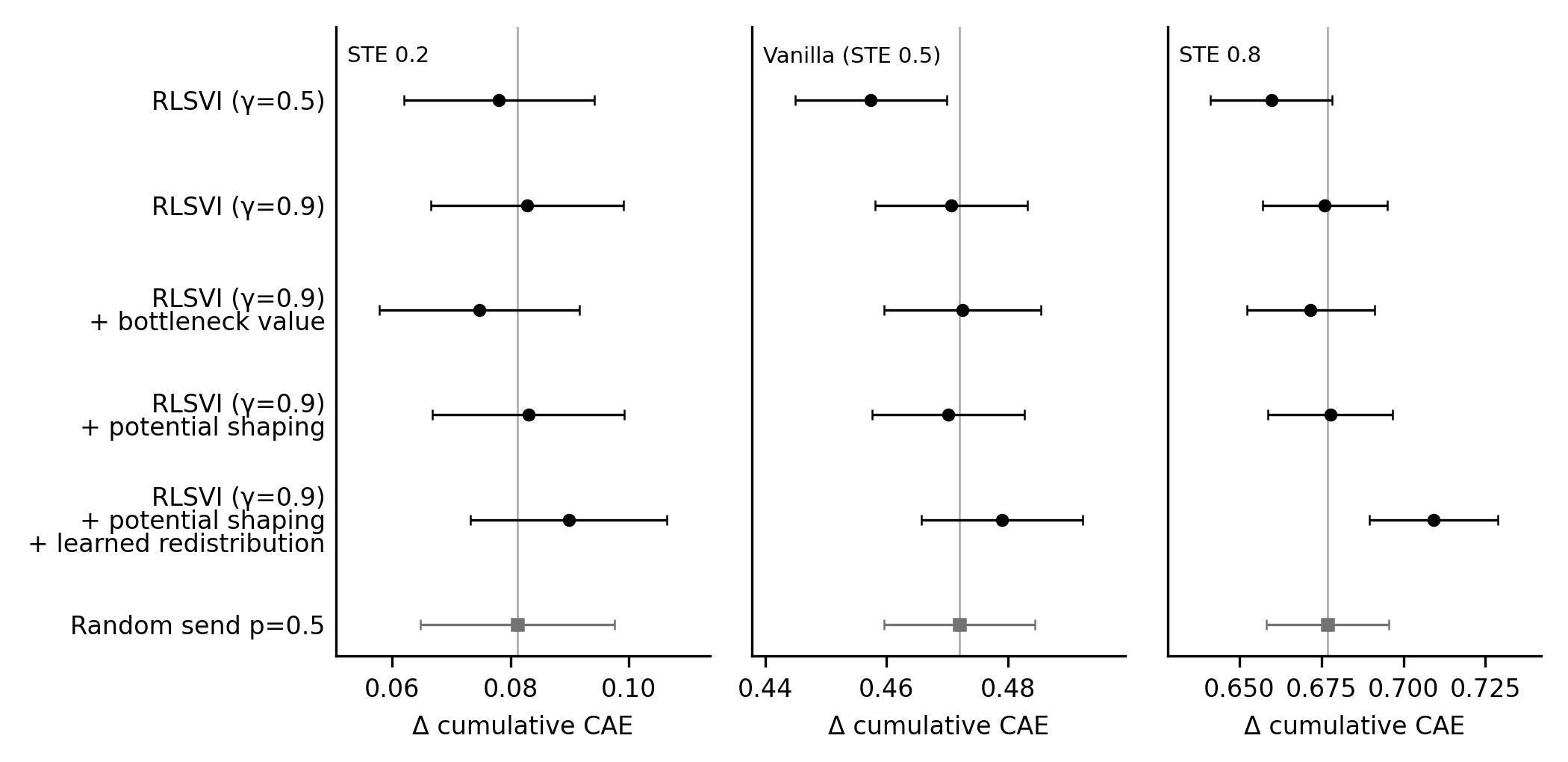}
    \caption{Cumulative 36-week CAE for each policy relative to never send, paired by experiment and participant draw. Panels are the three testbed environments. Dots are means over 200 simulated trials of 75 participants; error bars are 95\% confidence intervals.}
    \label{fig:forest}
\end{figure}

Figure~\ref{fig:forest} reports paired 36-week cumulative CAE relative to never-send (point estimate and 95\% CI) for five RLSVI variants and a non-learning random-send baseline ($p=0.5$) in three
environments of increasing intervention signal: STE $\approx0.2$, vanilla
(STE $\approx0.5$), and STE $\approx0.8$. 

Among RLSVI variants, two comparisons are stable across panels. First, RLSVI with weekly discount $\gamma=0.9$ outperforms $\gamma=0.5$: $\gamma=0.5$ places almost no weight on outcomes more than two weeks ahead, so it optimizes a short-horizon criterion that is misaligned with 36-week cumulative CAE. Second, a bottleneck-value parameterization does not improve on standard value propagation and is slightly worse when the signal is weak, consistent with additional parameters that must be estimated from sparse participant-specific data.

Standard RLSVI at $\gamma=0.9$ and potential-based shaping alone are
indistinguishable from random send in all three environments. Shaping adds a week-boundary potential $F=\gamma\,\hat e_{w+1}-\hat e_w$ so that delayed perceived utility enters the TD target, but the weekly outcome remains a single scalar.
Stacking learned reward redistribution on shaping is the only variant that improves on random send. The gain is small at STE $\approx 0.2$ and in the vanilla environment, and becomes clear at STE $\approx 0.8$, where the proximal $A\to\mathrm{four ~hour ~step~ count}\to\mathrm{CAE}$ pathway is amplified and the slot-level reward model therefore receives a stronger immediate learning signal. Appendix~G reports morning and afternoon send probabilities and shows how the variants learn: redistribution retimes sends toward morning, $\gamma=0.5$ cuts afternoon probability without raising morning probability, and standard RLSVI at $\gamma=0.9$ barely moves in either slot.

\section{Discussion}
\label{sec:discussion}
This work develops a causal-DAG-guided RL system for the forthcoming ADAPT factorial trial and evaluates candidate designs in an MRT-calibrated simulator.

\subsection{Path to Deployment}

The pre-deployment experiments inform the final RL configuration.  We select a cross-week discount factor of $\gamma=0.9$ and use potential-based shaping combined with learned reward redistribution to provide denser within-week learning signals. Because participant-specific data are limited early in the study, the deployed algorithm will be initialized using informative priors estimated from all eligible ADAPT MRT participants.

Deployment requires safeguards for missing or delayed data and failed weekly updates. Each decision uses only information available at that time; later-arriving measurements are incorporated only into subsequent state estimation and model updates. Monday-night updates include CAE and perceived-utility estimation, empirical-Bayes variance updates, RLSVI, and, when applicable, reward redistribution. Prespecified validity checks and fallback rules retain the most recent validated component when an update fails.

Deployment will be supported by systematic monitoring of data availability and latency, state construction, weekly model updates, action probabilities, and fallback events, enabling both real-time detection of operational failures and post-trial assessment of how data quality and algorithm reliability affected policy behavior \citep{trella2026effective,ghosh2025reproducible}.

\subsection{Limitations and Future Work}
Several limitations remain. First, although the simulator explicitly models missing survey responses, it does not fully reproduce delayed availability of wristband activity tracker, survey, and app data; incorporating realistic data-latency processes would enable a more stringent robustness assessment. 
Furthermore, the system relies on the working causal DAG and associated models, so sensitivity analyses and post-deployment model checking will be important for assessing robustness to structural misspecification.

\bibliography{aaai2027}

\clearpage
\onecolumn
\appendix
\setcounter{secnumdepth}{3}
\begin{center}
{\Large\bfseries Supplementary Material}
\end{center}

\section{ADAPT Study and Data Details}
\label{app:simulation}

\subsection{ADAPT Factorial Trial Design}

The ADAPT factorial trial will enroll $n=150$ cardiac rehabilitation patients, each followed for approximately 36 weeks. The study uses a $2\times2$ factorial design defined by two baseline-randomized intervention factors: ADAPT and Intrinsic Motivation (IM), with participants remaining in their assigned conditions throughout the study. Because RL is used only in the ADAPT-on condition, approximately half of the participants ($n\approx75$) will receive RL-based personalization. The IM factor may affect the intervention context and relevant state variables, but its role is outside the scope of this paper.

Before online RL begins, participants pass through two initial phases: a baseline phase and an initial micro-randomization phase. The baseline phase lasts approximately one week but may be extended until sufficient Fitbit data are collected. During this phase, participants complete a baseline check-in containing the 12-item CAE measure and the 2-item perceived-utility measure. Let $Y_0$ denote the mean of the 12 CAE items; $Y_0$ is fully observed for all participants. Physical-activity data are also collected once participants begin wearing the Fitbit.

For notational convenience, we refer to the baseline phase as week 0 and the initial micro-randomization phase as week 1, although their calendar durations may be longer. The baseline phase ends after the participant accumulates at least seven days with at least eight hours of Fitbit wear per day. The participant then enters week 1, during which activity-suggestion actions are randomized with probability $0.5$.

\subsection{ADAPT MRT Design}
The ADAPT micro-randomized trial (MRT) evaluates components of the HeartSteps mobile health intervention for supporting physical activity in cardiac rehabilitation patients. Participants receive the intervention for approximately three months while wearing a Fitbit activity tracker and interacting with the HeartSteps app.

Several intervention components are repeatedly randomized during the MRT. Activity suggestions are randomized twice daily with probability $0.5$, while morning salience messages and planning prompts are each randomized once daily with probability $0.5$. Other intervention components, including weekly reflection and planning activities, continuously available in-app features, and daily and weekly ecological momentary assessments, are delivered without micro-randomization.

The MRT collects longitudinal data on physical activity, intervention exposure, psychosocial measures, survey completion, and app engagement. These data are particularly relevant to the present work because the activity-suggestion decision schedule and many of the measured variables closely correspond to those used by the RL system in the forthcoming trial. We therefore use the MRT data both to estimate the data-generating models underlying the pre-deployment simulation environment and to construct informative priors for participant-specific RLSVI.


\subsection{MRT-to-Factorial Trial Variable Correspondence}
Some variables required by the future factorial trial RL system were measured differently in the MRT. Table~\ref{table:data} summarizes the primary mappings from future trial target variables to their MRT-based surrogates used for simulator construction and RL initialization.

\begin{table}[htbp]
\centering
\begin{tabularx}{\textwidth}{|X|X|} 
\hline
\textbf{Factorial Trial Variable} & \textbf{MRT Surrogate} \\ [0.5ex] 
\hline\hline
Short weekly survey measuring affective association of physical activity:
\begin{enumerate}
    \item ``For me, being active is \ldots'' with response options from unenjoyable (1) to enjoyable (7);
    \item ``Exercise leaves me feeling \ldots'' with response options from exhausted (1) to energized (7);
    \item ``For me, exercise is \ldots'' with response options from stressful (1) to relaxing (7).
\end{enumerate}
&
Corresponding items from the Core Affective Experiences (CAE) survey:
\begin{enumerate}
    \item ``Exercise feels \ldots'' with response options from very terrible (1) to wonderful (7);
    \item the CAE item corresponding to the energized--exhausted dimension;
    \item the CAE item corresponding to the relaxing--stressful dimension.
\end{enumerate}
\\
\hline
Daily morning check-in survey:
\begin{enumerate}
    \item whether the participant was active yesterday (this can also be reported elsewhere);
    \item anticipated affect: ``Next time I am active, I expect it will feel \ldots''
\end{enumerate}
&
The MRT end-of-day survey on day $d$ is used as the surrogate for the factorial trial morning check-in on day $d+1$. In particular:
\begin{enumerate}
    \item the MRT end-of-day item ``Were you active at least once today?'';
    \item the MRT end-of-day anticipated-affect item.
\end{enumerate}
\\
\hline
Active minutes:
\begin{enumerate}
    \item Prior 2 hour active minutes
    \item Next 4 hour active minutes
\end{enumerate}
&
Step count:
\begin{enumerate}
    \item Prior 2 hour step count
    \item Next 4 hour step count
\end{enumerate}
\\
\hline
\end{tabularx}
\caption{Mapping Factorial Trial Target Variables to MRT Surrogates}
\label{table:data}
\end{table}

\subsection{Data Preprocessing and Standardization}
As of July 26, 2026, the ADAPT MRT had 43 completed participants, defined as participants with an active-phase span of at least 83 days (the planned 84-day window, allowing a one-day shortfall). Eight of these 43 participants were excluded from the analysis sample using a priori availability rules: four (112, 117, 138, and 219) lacked sufficient wearable data for the primary proximal outcome (fewer than 20 wear-valid decision windows or fewer than 20 non-missing next-four-hour step counts), and five (82, 112, 251, 257, and 259) had no observed weekly surveys, and/or no observed daily end-of-day surveys; one participant (112) met both criteria. An additional seven participants were excluded based on prespecified CAE ceiling criteria: their minimum weekly CAE over the 12-week active phase was at least 6, or their week-1 CAE was 7. 
After these exclusions, 28 participants remained for fitting the environment models.

To improve numerical stability and place continuous predictors on comparable scales, we transformed the merged longitudinal panel prior to fitting the simulation models. Calendar variables, such as study week, were linearly rescaled into centered, unitless covariates ranging approximately within $[-1,1]$. The time-of-day indicator is coded as $t_n=0$ for the morning decision time and $t_n=1$ for the afternoon decision time.
Count-valued variables (step counts) were log-transformed via $x^\star = \log(x+1)$ to reduce skewness and the influence of extreme values. Page views were also treated as counts but were modeled using hurdle models because of their high proportion of zeros.

Continuous variables were subsequently standardized as $z = (x^\star-\mu)/\sigma$, where $\mu$ and $\sigma$ denote the mean and standard deviation computed from the analysis sample for the corresponding variable. The derived centering and scaling constants, along with the empirical support limits, were retained to ensure reproducibility, facilitate clipping of simulated values when necessary, and enable online standardization during the forthcoming trial.

For ordinal survey variables coded on a fixed $1$--$7$ scale, we use scale-based normalization rather than empirical z-score standardization, mapping to $[-1,1]$ via $u = \frac{x-4}{3}$. This centers the variable at the midpoint of the response scale and places it on a scale comparable to other standardized predictors.


The fitted simulation models distinguish between missingness in outcome variables and missingness in predictor variables. When a variable is used as an outcome, observations with missing outcome values are excluded from the corresponding model-fitting procedure. When a variable is used as a predictor, missing values are imputed with their mean before model fitting.

\section{Causal-DAG-Guided Online RL Design}
\label{app:rl_design}
This section describes how the working causal DAG is translated into the RL design, including four components: estimation of the partially observed weekly CAE and perceived-utility states, construction of a parsimonious decision-time state, representation of weekly bottleneck structure, and the use of intermediate mediators for delayed credit assignment.

\subsection{Partial Observability}
\subsubsection{CAE Belief-State Model}
\label{subsec:cae-belief}
Let $Y_w$ denote the weekly CAE state available for the Monday-night update in week $w$. It summarizes CAE over the preceding intervention week and is queried on Sunday evening at the end of week $w-1$. It can be latent because the sometimes short weekly CAE survey is sent or the user does not complete the survey.
Specifically, the weekly CAE-related observation is
\begin{align*}
O_w^Y =
\begin{cases}
Y_w, & J_w=1,\ I_w=1,\\[2pt]
\widetilde Y_w, & J_w=1,\ I_w=0,\\[2pt]
\emptyset, & J_w=0,
\end{cases}
\end{align*}
where $I_w$ indicates whether the full or short CAE instrument is requested, $J_w$ indicates engagement with the weekly intervention with a check-in survey, $Y_w$ is the mean of the 12 full-CAE items, and $\widetilde Y_w$ is the mean of the three items in the short instrument.

Consistent with the working DAG, weekly survey completion is modeled through perceived utility rather than as a direct measurement of CAE. Specifically, we assume
\begin{align*}
p\!\left(
J_w=1
\mid
Y_w,E_w,H_{w-1}^{-}
\right)
=
p\!\left(
J_w=1
\mid
E_w,H_{w-1}^{-}
\right),
\end{align*}
where $H_{w-1}^{-}$ denotes the information available before the weekly CAE observation $O^Y_w$ is realized. Thus, conditional on perceived utility and the observed history, survey completion provides no additional information about latent CAE. This is a working observation-model assumption rather than a consequence of the DAG itself. We do not condition on $I_w$ in this assumption because the weekly check-in occurs at the end of the weekly intervention: once a participant engages in the weekly Sunday night intervention, the completion of the survey by the individuals does not depend on whether the survey is the full or short CAE survey (i.e. the value of $I_w$).  

Because $Y_w$ may not be directly observed, we maintain the filtering distribution
\begin{align*}
b_w(y)
=
p\!\left(Y_w=y\mid\mathcal H_w^{\mathrm{obs}}\right)
\end{align*}
using particle filtering, where $H_w^{\mathrm{obs}} = H_{w-1}^- \cup \{I_w,J_w,O^Y_w\} $.

\paragraph{Working particle-filtering models.}
The particle filter combines a weekly CAE transition model, an observation model for the short CAE instrument, and action-conditional models for next-four-hour step count and anticipated affect: 
\begin{align}
Y_w
&=
\theta_0^{Y}
+
\theta_1^{Y} \overline{M}^{Y,SC}_{w-1}
+ 
\theta_2^{Y} \overline{M}^{Y,AA}_{w-1}
+
\theta_3^{Y}Y_{w-1}
+
\epsilon_w^Y,
\nonumber \\[0.25em]
\widetilde Y_w
&=
\theta_0^{\widetilde Y}
+
\theta_1^{\widetilde Y}Y_w
+
\epsilon_w^{\widetilde Y},
\nonumber \\[0.25em]
M_{w,d,t}^{Y,SC}
&=
\theta_0^{MY,SC}
+
\theta_1^{MY,SC}Y_w
+
\theta_2^{MY,SC}E_w
+
(\boldsymbol\theta_3^{MY,SC})^\top C_{w,d,t}
+
(\boldsymbol\theta_4^{MY,SC})^\top M_{w,d,t-1}^{Y,SC}
\nonumber\\
&\quad+
A_{w,d,t}
\left[
\theta_5^{MY,SC}
+
\theta_6^{MY,SC}Y_w
+
\theta_7^{MY,SC}E_w
+
(\boldsymbol\theta_8^{MY,SC})^\top C_{w,d,t}
\right]
+
\epsilon_{w,d,t}^{MY,SC},
\nonumber\\[0.25em]
M_{w,d}^{Y,AA}
&=
\theta_0^{MY,AA}
+
\theta_1^{MY,AA}Y_w
+
\theta_2^{MY,AA}E_w
+
(\boldsymbol\theta_3^{MY,AA})^\top M_{w,d-1}^{Y}
+
(\boldsymbol\theta_4^{MY,AA})^\top C_{w,d-1}
\nonumber\\
&\quad+
A_{w,d-1,1}
\left[
\theta_5^{MY,AA}
+
\theta_6^{MY,AA}Y_w
+
\theta_7^{MY,AA}E_w
+
(\boldsymbol\theta_8^{MY,AA})^\top C_{w,d-1}
\right]
\nonumber\\
&\quad+
A_{w,d-1,2}
\left[
\theta_9^{MY,AA}
+
\theta_{10}^{MY,AA}Y_w
+
\theta_{11}^{MY,AA}E_w
+
(\boldsymbol\theta_{12}^{MY,AA})^\top C_{w,d-1}
\right]
+
\epsilon_{w,d}^{MY,AA}.
\label{eq:pf-working-models}
\end{align}
Here, $\overline{M}_{w-1}^{Y,SC}$ and $\overline{M}_{w-1}^{Y,AA}$ summarize next-four-hour step count and anticipated affect, respectively, during week $w-1$. The lagged mediator terms use the most recent available value within the same week, with zero used at the beginning of a week. The Gaussian error terms have model-specific variances. 

Let $m\in\{SC;\ AA;\ Y;\ \widetilde Y\}$ index the four working models, with $\boldsymbol\theta^m \sim N(\nu_0^m,\Gamma_0^m)$. 
Conditional on particle $j$, latent CAE values in the cumulative design matrices are replaced by the corresponding particle trajectory, while latent perceived utility is replaced by its current approximation $\widehat e_w$. Using data available through week $w-2$, conjugate Bayesian linear-regression updates yield
\begin{align}
\Gamma_{w-2}^{m,(j)}
&=
\left[
(\Gamma_0^m)^{-1}
+
\frac{
(\mathbf X_{w-2}^{m,(j)})^\top
\mathbf X_{w-2}^{m,(j)}
}{
(\sigma^m)^2
}
\right]^{-1},
\nonumber\\
\nu_{w-2}^{m,(j)}
&=
\Gamma_{w-2}^{m,(j)}
\left[
(\Gamma_0^m)^{-1}\nu_0^m
+
\frac{
(\mathbf X_{w-2}^{m,(j)})^\top
\mathbf y_{w-2}^{m,(j)}
}{
(\sigma^m)^2
}
\right].
\label{eq:particlelearning}
\end{align}

Algorithm~\ref{alg:belief.state.all} combines these parameter updates with the week-$w-1$ mediator likelihood and the available weekly CAE observation to update $b_w$. The resulting weighted particles define
\[
\widehat b_w
=
\sum_{j=1}^J
\widehat v_w^{(j)}
\widehat y_w^{(j)},
\qquad
\widetilde b_w
=
\left[
\sum_{j=1}^J
\widehat v_w^{(j)}
\left(
\widehat y_w^{(j)}-\widehat b_w
\right)^2
\right]^{1/2}.
\]
These posterior mean and standard-deviation summaries enter the RL state, and $\widehat b_w$ provides the weekly CAE learning signal in the baseline algorithm.

\begin{algorithm}[htbp]
    \caption{Estimating Belief State $b_{w}(y)$ on Monday Night of Week $w$ ($w\ge 2$)}
    \label{alg:belief.state.all}
    \textbf{Input}: History $H_{w-1}$; particles $\{ \widehat{y}_{1:w-1}^{(j)} \}_{j=1}^J$ with weights $\{\widehat{v}_{w-1}^{(j)} \}_{j=1}^J$;
    prior $\boldsymbol{\theta}^{m} \sim \mathcal{N}(\nu_0^{m}, \Gamma_0^{m})$ for each model $m \in \{MY,DT; \ MY,D; \ Y; \ \widetilde{Y}\}$; current variance estimates for the particle-filtering models.

\textbf{Output:} Updated particles $\{ \widehat{y}_{1:w}^{(j)} \}_{j=1}^J$, updated weights $\{\widehat{v}_{w}^{(j)}\}_{j=1}^J$, 
    
    \begin{algorithmic}[1] 
    \STATE Update the empirical-Bayes variance estimates for the particle-filtering models:
\[
\widehat{\mathcal S}_{w}^{\mathrm{PF}}
\leftarrow
\textsc{UpdateEBVariancesPF}(H_{w-1}),
\]
where
\[
\widehat{\mathcal S}_{w}^{\mathrm{PF}}
=
\left\{
(\widehat\sigma_{w}^{Y})^2,
(\widehat\sigma_{w}^{\widetilde Y})^2,
(\widehat\sigma_{w}^{MY,DT})^2,
(\widehat\sigma_{w}^{MY,D})^2
\right\}.
\]
        \FOR{$j \in \{1:J\}$}
    \STATE Draw model parameters
        $\{\widehat{\boldsymbol\theta}_w^{m,(j)}\}$
        from their posterior distributions in \eqref{eq:particlelearning}.
\STATE Compute the mediator likelihood contribution from week \(w-1\): 
\vspace{-0.8em}
\begin{align*} 
L_{M,w-1}^{(j)} &= \prod_{d=1}^{6}\prod_{t=1}^{2} \phi\left( M_{w-1,d,t}^{Y}; \left(X_{w-1}^{MY,DT,(j)}(d,t)\right)^\top \widehat{\boldsymbol\theta}_{w}^{MY,DT,(j)}, (\widehat \sigma_w^{MY,DT})^2 \right), \\
&\qquad \prod_{d=2}^{7} \phi\left( M_{w-1,d}^{Y}; \left(X_{w-1}^{MY,D,(j)}(d,t)\right)^\top \widehat{\boldsymbol\theta}_{w}^{MY,D,(j)}, (\widehat \sigma_w^{MY,D})^2 \right)
\end{align*}
\vspace{-1.8em}
    \IF{$I_{w} = 1$ and $J_w^{\mathrm{week}}=1$}
        \STATE Set the new particle equal to the observed full CAE outcome:
        $
            y_{w}^{(j)} = Y_{w} 
        $
        \STATE Set the weight:
        $
            \widetilde{v}_{w}^{(j)} = \widehat{v}_{w-1}^{(j)} \cdot L_{M,w-1}^{(j)} \cdot \phi (Y_w; (X^{Y,(j)}_{w-1})^\top\widehat{\btheta}_w^{Y,(j)},(\widehat \sigma_w^{Y})^2).
        $ 
    \ELSIF{$I_{w} = 0$ and $J_w^{\mathrm{week}}=1$}
    \STATE Propagate the latent state: 
        $ y_w^{(j)} \sim N\left(\frac{1}{\widehat \sigma^2_{Y}} \left(X_{w-1}^{Y,(j)}\right)^\top \widehat{\boldsymbol\theta}_{w}^{Y,(j)} + \frac{1}{\widehat \sigma^2_{\widetilde Y}} \theta_1^{\widetilde Y} (\widetilde Y_w-\theta_0^{\widetilde Y}), \left(\frac{1}{\widehat \sigma^2_{Y}} + \frac{\theta_1^{\widetilde Y}}{\widehat \sigma^2_{\widetilde Y}}\right)^{-1} \right). $
        \STATE Set the  weight:
        $
            \widetilde{v}_{w}^{(j)} = \widehat{v}_{w-1}^{(j)} \cdot L_{M,w-1}^{(j)}\cdot  \phi (\widetilde{Y}_w; (X^{\widetilde{Y},(j)}_{w-1})^\top\widehat{\btheta}_w^{\widetilde{Y},(j)},(\widehat \sigma_w^{\widetilde{Y}})^2).
       $
    \ELSE
    \STATE Propagate the latent state: 
    $ y_w^{(j)} \sim N\left( \left(X_{w-1}^{Y,(j)}\right)^\top \widehat{\boldsymbol\theta}_{w}^{Y,(j)}, (\widehat \sigma_w^{Y})^2 \right). $
        \STATE Set the weight:
        $
            \widetilde{v}_{w}^{(j)} = \widehat{v}_{w-1}^{(j)} \cdot L_{M,w-1}^{(j)}.
        $
    \ENDIF
    \ENDFOR
        \STATE Normalize particle weight:
        $
            v_{w}^{(j)} = \frac{\widetilde{v}_{w}^{(j)}}{\sum_{i=1}^J \widetilde{v}_{w}^{(i)}}
        $
        for $j \in \{1:J\}$.
    \STATE Calculate the effective sample size 
        $
        \text{ESS} := \left[\sum_{j=1}^J (v_{w}^{(j)})^2\right]^{-1}.
        $
        \IF{$\mathrm{ESS}<0.5J$}
        \STATE Resample
        $\{\widehat y_{1:w-1}^{(j)},y_w^{(j)}\}_{j=1}^J$
        with weights $\{v_w^{(j)}\}_{j=1}^J$ to obtain
        $\{\widehat y_{1:w}^{(j)}\}_{j=1}^J$, and set
        $\widehat v_w^{(j)}=1/J$ for all $j$.  
    \ELSE
        \STATE Set
        $\widehat y_{1:w}^{(j)}=(\widehat y_{1:w-1}^{(j)},y_w^{(j)})$
        and $\widehat v_w^{(j)}=v_w^{(j)}$ for all $j$.
    \ENDIF
        \STATE \textbf{return}
    $\{\widehat y_{1:w}^{(j)},\widehat v_w^{(j)}\}_{j=1}^J$.
    \end{algorithmic}
\end{algorithm}

\subsubsection{Perceived-Utility Estimation}
\label{subsec:pu}
Perceived utility $E_w$ is also latent. 
Rather than maintaining a second online filtering distribution, we construct an MRT-trained recursive approximation $\widehat e_w$ to its filtered mean, mirroring the structure of the fitted state-space model: an autoregressive carry-over of last week's estimate (the transition), the current week's engagement-mediator averages (the mediator pathway), and the check-in survey bundle (completion indicator $J_w$ and responses $U_{w,1},U_{w,2}$).
Define the weekly averages  
\begin{align*}
\overline M_{w-1}^{E,PJ}
&=
\frac{1}{6}
\sum_{d=1+1}^{6}
M_{w-1,d}^{E,PJ},\\
\overline M_{w-1}^{E,FW}
&=
\frac{1}{6}
\sum_{d=1}^{6}
M_{w-1,d+1}^{E,FW},\\
\overline M_w^{E,PV} 
&= \frac{1}{12}\sum_{d=1}^{6}\sum_{t=1}^{2} M_{w,d,t}^{E,PV}.
\end{align*}
The approximation is
\begin{align}
\widehat e_w
=
\beta_0
+
\beta_E\,\widehat e_{w-1}
+
\beta_{PV}\overline M_{w-1}^{E,PV}
+
\beta_{FW}\overline M_{w-1}^{E,FW}
+
\beta_{PJ}\overline M_{w-1}^{E,PJ}
+
\beta_{J} J_w^{\mathrm{week}}
+
\beta_{U} J_w^{\mathrm{week}}\frac{U_{w,1}+U_{w,2}}{14},
\label{eq:ew}
\end{align}
initialized at $\widehat e_1 = 2$, the pin used for $E_1$ during estimation. Because $U_{w,1}$ and $U_{w,2}$ are each coded from 1 to 7, their summed score divided by 14 lies in $[1/7,1]$ when observed.

The coefficients in \eqref{eq:ew} are estimated offline from the historical MRT and held fixed during the factorial trial.   The regression target is the filtered latent trajectory $\widehat E_{iw}$ from the fitted state-space models of Section~\ref{subsec:sim-pu}.  To remove between-participant level differences, outcome and predictors are centered within participant and a
common slope vector is estimated under a nonnegativity constraint,
\begin{align*}
\widehat{\bm\beta}
=
\arg\min_{\bm\beta\ge0}
\sum_{i,w}
\left(
\widetilde E_{iw}
-
\widetilde{\bm X}_{iw}^{\mathsf T}\bm\beta
\right)^2,
\end{align*}
after which the intercept is calibrated on the original scale, $\widehat\beta_0=\overline E-\overline{\bm X}^{\mathsf T}\widehat{\bm\beta}$. Because the autoregressive predictor $\widehat e_{w-1}$ is itself an output of the fitted map, estimation alternates between fitting $\widehat{\bm\beta}$ and regenerating the recursive predictions $\{\widehat e_{iw}\}$ under the current fit, iterating to a fixed point. At deployment the same normalized averages are computed over the trial's Monday--Saturday schedule (12 decision slots, 6 days).

Under the nonnegativity constraint, the fitted weights of the two weekly check-in terms are zero (Table~\ref{tab:pu-weights}): the daily check-in, page-view, and wear signals, together with the autoregressive term, absorb the information the weekly items would otherwise provide. 

\begin{table}[H]
\centering
\caption{Estimated weights of the perceived-utility approximation
$\widehat e_w$ in \eqref{eq:ew}.}
\label{tab:pu-weights}
\begin{tabular}{lc}
\hline
\textbf{Component} & \textbf{Estimated weight} \\
\hline
Intercept & $0.915$ \\
$\widehat e_{w-1}$ & $0.119$ \\
$\overline M_{w-1}^{E,PV}$ & $0.318$ \\
$\overline M_{w-1}^{E,FW}$ & $0.232$ \\
$\overline M_{w-1}^{E,PJ}$ & $0.905$ \\
$J_w^{\mathrm{week}}$ & $0.000$ \\
$J_w^{\mathrm{week}}\bigl[U_{w,1}+U_{w,2}\bigr]/14$ & $0.000$ \\
\hline
\end{tabular}
\end{table}

\subsection{State Construction}
\label{subsec:state}

At decision time $(w,d,t)$, the RL state combines estimates of the weekly latent states, summaries of within-week mediator history, and currently available contextual information. Let $\mathcal M_{w,d,t}^{\mathrm{pre}}$ denote all mediator measurements observed strictly before decision time $(w,d,t)$. Only information available at the time of decision is used in constructing the state.

Because mediator histories grow over the week, each mediator stream is compressed into a recency-weighted summary. If $x_1,\ldots,x_k$ are the observations from mediator stream $Z$ available before $(w,d,t)$, define
\[
\overline M_{w,d,t}^{Z}
=
\frac{
\sum_{i=1}^k \gamma_k^{\,k-i}x_i
}{
\sum_{i=1}^k \gamma_k^{\,k-i}
},
\qquad
\gamma_k=1-\frac{1}{k},
\]
for $k\geq 1$, and set $\overline M_{w,d,t}^{Z}=0$ when $k=0$. Thus, more recent observations receive greater weight.


We summarize five mediator streams motivated by the working causal DAG:
\begin{itemize}
    \item $\overline M^{Y,AA}$: anticipated affect, assessed in the daily morning check-in;
    \item $\overline M^{Y,SC}$: physical activity following an activity-suggestion decision, measured using Fitbit activity during the subsequent four-hour window;
    \item $\overline M^{E,PV}$: unprompted app page views during the four hours following a decision;
    \item $\overline M^{E,FW}$: Fitbit wear status on the following morning; and
    \item $\overline M^{E,PJ}$: completion of the following morning check-in.
\end{itemize}
The first two streams represent hypothesized pathways to future CAE, whereas the remaining three represent hypothesized pathways to future perceived utility. Daily measurements, such as anticipated affect, Fitbit wear, and morning check-in completion, may reflect both activity-suggestion decisions from the preceding day. Because the morning check-in window can overlap with the next morning's decision time, a response is included in the state only after it has actually been observed; measurements arriving after a decision are available only for subsequent decisions.

The resulting decision-time state is
\begin{align}
S_{w,d,t}
=
\Big[
&\widehat b_w,\,
\widetilde b_w,\,
\widehat e_w,\,
\overline M_{w,d,t}^{Y,AA},\,
\overline M_{w,d,t}^{Y,SC},\,
\overline M_{w,d,t}^{E,PV},\,
\overline M_{w,d,t}^{E,FW},\,
\overline M_{w,d,t}^{E,PJ},
\nonumber\\
& C_{w,d,t}^{\top},\,
d_n,\,
t_n
\Big]^\top,
\label{eq:state}
\end{align}
where $d_n\in\{0,1\}$ indicates day type (0 is weekday, 1 is weekend) and $t_n\in\{0,1\}$ indicates the morning or afternoon decision time. The context vector $C_{w,d,t}$ contains:
\begin{itemize}
    \item steps during the two hours preceding the decision;
    \item steps on the previous day;
    \item recent intervention burden, summarized by an exponentially weighted average of activity suggestions delivered over the previous seven days;
    \item the fraction of delivered activity suggestions with which the participant interacted over the previous 14 decision times; and
    \item the fraction of the previous seven days classified as active.
\end{itemize}

The linear Q-function uses both state main effects and a parsimonious set of prespecified action-interaction features:
\begin{align}
\phi(S_{w,d,t},a)
&=
\left[
1,\,
d_n,\,
t_n,\,
\widehat e_w,\,
\widehat b_w,\,
\widetilde b_w
\right]
\nonumber\\
&\quad\frown
\left[
\overline M_{w,d,t}^{Y,AA},\,
\overline M_{w,d,t}^{Y,SC},\,
\overline M_{w,d,t}^{E,PV},\,
\overline M_{w,d,t}^{E,FW},\,
\overline M_{w,d,t}^{E,PJ},\,
C_{w,d,t}^{\top}
\right]
\nonumber\\
&\quad\frown
a
\left[
1,\,
\widehat e_w,\,
\widehat b_w,\,
\widetilde b_w,\,
C_{w,d,t}^{\top},\,
d_n,\,
t_n
\right],
\label{eq:feature-map}
\end{align}
where $\frown$ denotes vector concatenation. The first two blocks encode the baseline state, whereas the final block contains the prespecified effect-modification features. These action-interaction terms are chosen for parsimony and scientific relevance; they are not identified by the DAG.

\begin{algorithm}[H]
    \caption{Feature Construction Subroutine}
    \label{alg:feature.construction}

    \textbf{Input}: Week $w$; decision time $(d,t)$; candidate action $a\in\{0,1\}$; perceived-utility estimate $\widehat e$; belief-state summaries $(\widehat b,\widetilde b)$; raw mediator history $\mathcal M_{w,d,t}^{\mathrm{pre}}$; context $C_{w,d,t}$.

    \textbf{Output}: Feature vector $\phi(S_{w,d,t},a)$.

    \begin{algorithmic}[1]

    \STATE Construct the decision-time state $S_{w,d,t}$ using \eqref{eq:state}

    \STATE Construct the action-specific feature vector $\phi(S_{w,d,t},a)$ using \eqref{eq:feature-map}.

    \STATE \textbf{return} $\phi(S_{w,d,t},a)$.

    \end{algorithmic}
\end{algorithm}

\subsection{DAG-Guided Delayed Credit Assignment}
\label{subsec:dag-credit}

The working DAG highlights two distinct sources of delayed credit. First, the 12 activity-suggestion decisions within a week share a single weekly CAE learning signal. Second, effects transmitted through perceived utility may influence CAE only after an additional week. We consider complementary mechanisms that exploit two forms of structure encoded in the DAG: low-dimensional weekly bottleneck states and intermediate mediators along the benefit and burden pathways.

\subsubsection{Week-Level Bottleneck Value}
\label{subsec:bottleneck}
The baseline RLSVI update propagates information from one week to the next through the first decision-time state of the following week. The bottleneck variant instead introduces an explicit week-level value function based on the low-dimensional weekly state
\begin{align*}
S_{w,0}
=
[\widehat e_w,\widehat b_w,\widetilde b_w]^\top.
\end{align*}
We approximate the bottleneck value and decision-time action-value function by
\begin{align*}
V(S_{w,0}) = f(S_{w,0})^\top \alpha,
\qquad
Q(S_{w,d,t},a)
=
\phi(S_{w,d,t},a)^\top \beta.
\end{align*}
and estimate them jointly through
\begin{align*}
\theta =
\begin{bmatrix}
\alpha\\
\beta
\end{bmatrix},
\qquad
\theta\sim N(\mu_0,\Sigma_0).
\end{align*}

The corresponding TD loss links the week-level value to the first decision-time value, uses standard within-week TD backups for intermediate decisions, and replaces the usual cross-week decision-time target at the final decision with
\[
\widehat b_{w'+1}
+
\gamma
V(S_{w'+1,0}).
\]
The full loss is given in Equation~\eqref{eq:bottleneck-loss}. This formulation changes the value-function factorization without changing the scientific reward and shortens the path through which information must propagate across weeks.

For ensemble member $b$, we freeze the greedy action using the previous ensemble parameter,
\begin{align*}
a_{w',d,t}^{(b)}
=
\arg\max_{a\in\{0,1\}}
\phi(S_{w',d,t},a)^\top \beta_{w-2}^{(b)}.
\end{align*}
The resulting bottleneck-state TD objective is
\begin{align}
L_{\mathrm{BN}}^{(b)}(\alpha,\beta\mid H_{w-1})
&=
\frac{1}{\sigma_Q^2}
\sum_{w'=1}^{w-1}
\left[
\phi(S_{w',1,1},a_{w',1,1}^{(b)})^\top\beta
-
f(S_{w',0})^\top\alpha
\right]^2
\nonumber\\
&\quad+
\frac{1}{\sigma_Q^2}
\sum_{w'=1}^{w-1}
\sum_{(d,t)<(6,2)}
\left[
y_{w',d,t}^{(b)}
-
\phi(S_{w',d,t},A_{w',d,t})^\top\beta
\right]^2
\nonumber\\
&\quad+
\frac{1}{\sigma_Q^2}
\sum_{w'=1}^{w-1}
\left[
\widehat b_{w'+1}
+
 \gamma f(S_{w'+1,0})^\top\alpha
-
\phi(S_{w',6,2},A_{w',6,2})^\top\beta
\right]^2
\nonumber\\
&\quad+
(\theta-\mu_0)^\top\Sigma_0^{-1}(\theta-\mu_0),
\label{eq:bottleneck-loss}
\end{align}
where, for nonterminal decision times $(d,t)<(6,2)$,
\[
y_{w',d,t}^{(b)}
=
\phi(S_{w',\operatorname{next}(d,t)},
a_{w',\operatorname{next}(d,t)}^{(b)})^\top
\beta_{w-2}^{(b)}.
\]
The bottleneck-state loss couples the week-level value model and the decision-time $Q$-function through the shared parameter vector $\theta$, thereby inducing correlated posterior perturbations across the two components.

\subsubsection{Weekly Reward Construction}
\label{subsec:weekly-reward}

The baseline weekly reward is $R_{w+1}^{\mathrm{base}}
= \widehat b_{w+1}$. Because perceived utility lies on a more delayed pathway to CAE, we additionally consider an alternative weekly learning signal.
which employs potential-based shaping,
\begin{align}
R_{w+1}^{\mathrm{PB}}
=
\widehat b_{w+1}
+
\gamma\widehat e_{w+1}
-
\widehat e_w.
\end{align}
This corresponds to the potential $\Phi(S_w)=\widehat e_w$.
Under the working belief-state MDP and the standard terminal-potential condition, potential-based shaping preserves the optimal policy while bringing information about changes in perceived utility closer to the week in which they occur \citep{ng1999policy}.

\subsubsection{Within-Week Reward Redistribution}
\label{subsec:redistribution}
Changing the weekly reward does not by itself resolve the within-week credit-assignment problem, because the learning signal still arrives only once per week. We therefore also consider reward redistribution, which allocates a chosen weekly reward across the 12 within-week decisions using mediator information represented in the working DAG.

\paragraph{Stage 1: decomposition of once-daily mediators.}
The daily mediators $Z \in \{AA, FW, PJ\}$ are observed once per day and are affected by both actions $A_{w,d,1}$ and $A_{w,d,2}$. Because their contributions cannot be directly attributed to either decision from the observed daily measurement alone, we first construct a model-based decomposition in the spirit of RUDDER \citep{arjona2019rudder}.

Define the decision-time feature vector
\begin{align}
\psi(S_{w,d,t},a)
&=
\left[
1,\,
\widehat e_{w},\,
\widehat b_{w},\,
\widetilde b_{w}
\right]
\nonumber\\
&\quad \frown
\left[
\overline{M}^{Y,AA}_{w,d,t},\,
\overline{M}^{Y,SC}_{w,d,t},\,
\overline{M}^{E,PV}_{w,d,t},\,
\overline{M}^{E,FW}_{w,d,t},\,
\overline{M}^{E,PJ}_{w,d,t},\,
C_{w,d,t}^{\top}
\right]
\nonumber\\
&\quad \frown\;
a
\left[
1,\,
t_n,\,
\widehat e_{w},\,
\widehat b_{w},\,
\widetilde b_{w},\,
C_{w,d,t}^{\top}
\right].
\label{eq:daily-mediator-features}
\end{align}
where $\frown$ denotes vector concatenation. A state-only slot indicator $t_n$ is omitted because it is collinear with the intercept; the action--slot interaction is identified because the two within-day actions vary independently, so the decomposition attributes \emph{action} effects to morning versus afternoon while splitting the non-action baseline evenly.

At the Monday-night update in week $w$, each mediator-specific parameter $\eta^Z$ is estimated using all completed weeks $w'=1,\ldots,w-1$:
\begin{align}
    L_{\mathrm{day},w}^{Z}(\eta^Z)
    &=
    \frac{1}{\sigma_{\mathrm{day}}^{2}}
    \sum_{w'=1}^{w-1}\sum_{d=1}^{6}
    \left[
    \sum_{t=1}^{2}
    \psi\!\left(S_{w',d,t},A_{w',d,t}\right)^{\top}\eta^{Z}
    -
    Z_{w',d+1}
    \right]^{2}
    \nonumber\\
    &\quad+
    (\eta^{Z}-\eta^{Z}_{0})^{\top}
    \Sigma_{\eta^{Z},0}^{-1}
    (\eta^{Z}-\eta^{Z}_{0}),
    \label{eq:daily-mediator-loss}
\end{align}
with prior mean $\eta^{Z}_{0}$ and prior covariance
$\Sigma_{\eta^{Z},0}$. 
Let $\widehat\eta_w^Z$ denote the minimizer. For a completed decision time $(w',d,t)$, define its fitted contribution to the daily mediator by
\begin{align}
\widehat Z_{w,d,t}
=
\psi^{Z}\!\left(S_{w,d,t},A_{w,d,t}\right)^{\top}\widehat\eta^{Z}_{w},
\qquad t \in \{1,2\},
\quad Z \in \{AA, FW, PJ\}.
\label{eq:daily-mediator-share}
\end{align}

Although $M^{Y,SC}$ is observed at every decision time, its value mixes the action effect with baseline activity.  We therefore regress each of the twelve weekly slot values on baseline covariates ($d_n$, $t_n$, $\widehat e_w$, $\widehat b_w$, $\widetilde b_w$, $C_{w,d,t}$) and an action block $a[1,\,t_n]$, in the same Gaussian-prior ridge form as \eqref{eq:daily-mediator-loss}. The fitted action-attributable contribution $\widehat{SC}_{w,d,t}$ replaces the realized step count in the Stage-2 features below. Page views are not decomposed: their fitted action effect is weak, and their realized value functions as a readout of the latent engagement state rather than as attributable credit, so they enter Stage~2 directly.

\paragraph{Stage 2: within-week redistribution.}
We next redistribute the selected weekly reward $R_{w+1}$ across the 12 decisions in each completed week $w$. 

At the week-$w$ update, define
\begin{align}
\chi_{w,d,t}
&=
\left[
1,\,
d_{n},\,
\widehat e_{w},\,
\widehat b_{w},\,
\widetilde b_{w}
\right]
\nonumber\\
&\quad \frown
\left[
\overline{M}^{Y,AA}_{w,d,t},\,
\overline{M}^{Y,SC}_{w,d,t},\,
\overline{M}^{E,PV}_{w,d,t},\,
\overline{M}^{E,FW}_{w,d,t},\,
\overline{M}^{E,PJ}_{w,d,t},\,
C_{w,d,t}^{\top}
\right]
\nonumber\\
&\quad \frown
\left[
M^{E}_{w,\mathrm{next}(d,t)},\,
\widehat{SC}_{w,d,t},\,
\widehat{AA}_{w,d,t},\,
\widehat{FW}_{w,d,t},\,
\widehat{PJ}_{w,d,t}
\right],
\label{eq:redistribution-features}
\end{align}
where $M^{Y}_{w,\mathrm{next}(d,t)}$ and $M^{E}_{w,\mathrm{next}(d,t)}$ are the decision-level mediators observed after action $A_{w',d,t}$, and the remaining three terms $\widehat{AA}_{w,d,t}$, $\widehat{FW}_{w,d,t}$, $\widehat{PJ}_{w,d,t}$ are Stage-1 contributions defined in \eqref{eq:daily-mediator-share}. The realized step count is omitted in favor of its fitted action-attributable contribution $\widehat{SC}_{w,d,t}$. 

At the Monday-night update in week $w$, the redistribution parameter $\eta$ is estimated by matching the sum of the decision-level predictions within each completed week to the corresponding weekly reward:
\begin{align}
L_{\mathrm{sh},w}(\eta)
&=
\frac{1}{\sigma_{\mathrm{sh}}^2}
\sum_{w'=1}^{w-1}
\left[
\sum_{d=1}^{6}\sum_{t=1}^{2}
\chi_{w',d,t}^{\top}\eta
-
R_{w'+1}
\right]^2
\nonumber\\
&\quad+
(\eta-\eta_0)^{\top}\Sigma_{\eta,0}^{-1}(\eta-\eta_0),
\label{eq:reward-shaping-loss}
\end{align}
with prior mean $\eta_0$ and prior covariance $\Sigma_{\eta,0}$. Let $\widehat\eta_w$ denote the minimizer. The redistributed reward for decision $(w,d,t)$ is
\begin{align}
R_{w,d,t}^{\mathrm{sh}}
=
\chi_{w,d,t}^{\top}\widehat \eta_w .
\label{eq:preliminary-shaped-reward}
\end{align}


\section{Participant-Specific Online Learning with RLSVI}
\label{app:rlsvi}

Given the DAG-guided state and reward representations described in Section~\ref{app:rl_design}, we use randomized least-squares value iteration (RLSVI) to learn a participant-specific activity-suggestion policy. Historical MRT data initialize the learner, while subsequent updates use only data accumulated from the trial participant.

\subsection{Linear Q function}
Under policy $\pi$, the decision-time action-value function is
\begin{align}
Q^\pi(S_{w,d,t},a)
=
\mathbb{E}^{\pi}
\left[
\sum_{w'=w}^{\infty}
\gamma^{\,w'-w}
R_{w'+1}
\;\middle|\;
S_{w,d,t}, A_{w,d,t}=a
\right].
\label{eq:q-function}
\end{align}
Within-week transitions are undiscounted, while $\gamma$ discounts rewards across weekly boundaries.

The decision-time action-value function is approximated by
\begin{align*}
  Q(S,a;\beta)
=
\phi(S,a)^\top\beta,
\qquad
\beta\sim N(\mu_0,\Sigma_0),  
\end{align*}
where $\phi(S,a)$ is defined in Equation~\eqref{eq:feature-map}. The linear representation limits model complexity in the participant-specific, data-scarce setting.

\subsection{MRT-Informed Prior}
\label{subsec:rlsvi-prior}

MRT data are used to construct informative priors for the particle-filtering working models, the Q-function, and the credit-assignment (Stage-1/Stage-2) models. All priors are Gaussian, $\xi \sim N(\mu_{\xi,0}, \Sigma_{\xi,0})$, and are built by one generic two-stage empirical-Bayes scheme; only the response and design differ across models.

\paragraph{Generic construction.}
First, all participants' MRT rows are stacked and a single ridge regression is fit, yielding the pooled coefficient vector $\theta^{\mathrm{pool}}$, which becomes the prior mean. Second, the same regression is fit separately to each participant's rows, yielding per-participant estimates $\theta_u$ and residual variances $\sigma^2_u$. The prior covariance is diagonal, with each element equal to the across-participant variance of the corresponding coordinate of $\theta_u$ — so coefficients that vary strongly between people are weakly constrained, and vice versa — and the noise variance is initialized at the average per-participant residual variance. 
One exception to diagonality: the joint $(\alpha,\beta)$ prior of the bottleneck variant uses the full sample covariance of the per-participant $(\alpha_u, \beta_u)$, preserving cross-block correlations.

For the four particle-filter working models (four-hour step count, anticipated affect, weekly CAE, short CAE) the regressions are the working models themselves on the MRT panel. For the Stage-1 and Stage-1b decompositions, the regressions are \eqref{eq:daily-mediator-loss} and its slot-level analogue on the MRT weeks. For the Stage-2 redistribution model, the response is the shaped weekly target and the design is the week-summed features of \eqref{eq:redistribution-features}. Q-function priors cannot be read off a supervised fit; they are constructed by offline fitted-Q iteration (25 iterations per participant) on the MRT trajectories, run separately for each discount setting ($\bar\gamma \in \{0.5, 0.9\}$) and for each reward construction, so that every RL variant receives a prior matched to the objective it optimizes: the weekly-CAE target for the base variants, the potential-shaped target for V4, and the redistributed slot rewards for the V5 variant.

\paragraph{Deployment regime.}
At evaluation, the policy-side prior means (Q-function and Stage-1 decompositions) are re-centered at zero with covariance inflated tenfold (diagonal floored at $0.1$), while the particle-filter and Stage-2 priors retain their MRT-informed centers. 

\paragraph{Held-out evaluation.}
During pre-deployment evaluation on the testbed, all priors supplied to a simulated participant are refit on the MRT data excluding that participant (28 leave-one-out folds), so no participant's own historical data informs their priors. For the eventual trial, all eligible MRT data will be used.

\subsection{Action Selection}
\label{subsec:action-selection}

RLSVI maintains an ensemble of $B$ randomized Q-functions,
\begin{align*}
Q(S,a;\bbeta^{(b)})=\phi(S,a)^\top \bbeta^{(b)},
\qquad b=1,\ldots,B.
\end{align*}
At each decision time $(w,d,t)$, the subroutine takes as input the two action-specific feature vectors
\begin{align*}
\phi_1 = \phi(S_{w,d,t},1),
\qquad
\phi_0 = \phi(S_{w,d,t},0),
\end{align*}
constructed using Algorithm~\ref{alg:feature.construction}. The fraction of ensemble members that prefer action 1 is
\begin{align}
\widehat\pi_{w,d,t}
=
\frac{1}{B}
\sum_{b=1}^{B}
\mathbb I
\left\{
\phi_1^\top\beta^{(b)}
>
\phi_0^\top\beta^{(b)}
\right\}.
\label{eq:ensemble-vote-prob}
\end{align}
To maintain randomized exploration and facilitate after-study analyses,
\begin{align}
\pi_{w,d,t}
=
\min\{1-\epsilon_0,\max(\widehat\pi_{w,d,t},\epsilon_0)\}.
\label{eq:clipped-prob}
\end{align}
and
\begin{align}
A_{w,d,t}\sim \operatorname{Bernoulli}(\pi_{w,d,t}).
\label{eq:action-sampling}
\end{align}

\begin{algorithm}[H]
    \caption{Action-Selection Subroutine}
    \label{alg:select.action}

    \textbf{Input}: Feature vectors $\phi_1=\phi(S_{w,d,t},1)$ and $\phi_0=\phi(S_{w,d,t},0)$; RLSVI ensemble $\{\bbeta^{(b)}\}_{b=1}^B$; clipping parameter $\epsilon_0$.

    \textbf{Output}: Activity-suggestion action $A_{w,d,t}$ and assignment probability $\pi_{w,d,t}$.

    \begin{algorithmic}[1]

    \STATE Compute the ensemble voting probability $\widehat\pi_{w,d,t}$ using \eqref{eq:ensemble-vote-prob}.

    \STATE Compute the clipped assignment probability $\pi_{w,d,t}$ using \eqref{eq:clipped-prob}.

    \STATE Sample the activity-suggestion action $A_{w,d,t}$ using \eqref{eq:action-sampling}.

    \STATE \textbf{return} $(A_{w,d,t},\pi_{w,d,t})$.

    \end{algorithmic}
\end{algorithm}

\subsection{Weekly RLSVI Update}
\label{subsec:rlsvi-update}

The participant-specific ensemble is updated each Monday night using all completed transitions through week $w-1$. Let $W$ denote the study horizon.

Let $\operatorname{next}(d,t)$ denote the next decision time within the same week. For ensemble member $b$, TD targets use the previous ensemble parameter:
\begin{align}
y_{w',d,t}^{(b)}
=
\max_{a\in\{0,1\}}
\phi(S_{w',\operatorname{next}(d,t)},a)^\top
\beta_{w-2}^{(b)},
\qquad
(d,t)\neq(6,2).
\label{eq:td-target-nonterminal}
\end{align}
At the week boundary,
\begin{align}
y_{w',6,2}^{(b)}
=
R_{w'+1}
+
 \gamma
\max_{a\in\{0,1\}}
\phi(S_{w'+1,1,1},a)^\top
\beta^{(b)}_{w-2}.
\label{eq:td-target-terminal}
\end{align}
For the baseline algorithm,
$R_{w'+1}=\widehat b_{w'+1}^{Y}$;
the alternatives in Section~\ref{app:variants} replace this reward and, when applicable, the corresponding TD structure.

For each ensemble member, the regularized TD objective is
\begin{align}
L^{(b)}(\beta\mid H_{w-1})
&=
\frac{1}{\sigma_Q^2}
\sum_{w'=1}^{w-1}
\sum_{d=1}^6
\sum_{t=1}^2
\left[
y_{w',d,t}^{(b)}
-
\phi(S_{w',d,t},A_{w',d,t})^\top\beta
\right]^2
\nonumber\\
&\quad+
(\beta-\mu_0)^\top\Sigma_0^{-1}(\beta-\mu_0).
\label{eq:rlsvi-loss}
\end{align}

Let $\mathbf X_{w-1}$ stack the realized state-action feature vectors and let $\mathbf y_{w-1}^{(b)}$ stack the corresponding TD targets:
\begin{align}
\bX_{w-1}
=
\begin{bmatrix}
\phi(S_{1,1,1}^{\dagger},A_{1,1,1})^\top\\
\phi(S_{1,1,2}^{\dagger},A_{1,1,2})^\top\\
\vdots\\
\phi(S_{w-1,6,2}^{\dagger},A_{w-1,6,2})^\top
\end{bmatrix},
\qquad
\by_{w-1}^{(b)}
=
\begin{bmatrix}
y_{1,1,1}^{(b)}\\
y_{1,1,2}^{(b)}\\
\vdots\\
y_{w-1,6,2}^{(b)}
\end{bmatrix}.
\label{eq:rlsvi-design-target}
\end{align}
The posterior covariance and ensemble-specific posterior mean are
\begin{align}
\Sigma_{w-1}
&=
\left(
\Sigma_0^{-1}
+
\frac{1}{\sigma_Q^2}\bX_{w-1}^\top\bX_{w-1}
\right)^{-1},
\nonumber\\
\mu_{w-1}^{(b)}
&=
\Sigma_{w-1}
\left(
\Sigma_0^{-1}\mu_0
+
\frac{1}{\sigma_Q^2}\bX_{w-1}^\top\by_{w-1}^{(b)}
\right).
\label{eq:rlsvi-posterior}
\end{align}

We use temporally correlated perturbations across weekly updates. Initialize
\begin{align}
z_0^{(b)}\sim N(0,\Sigma_0),
\qquad b=1,\ldots,B,
\label{eq:init-perturbation}
\end{align}
and for $w\geq 2$, sample
\begin{align}
z_{w-1}^{(b)}
\sim
N
\left(
\gamma z_{w-2}^{(b)},
(1-\gamma^2)\Sigma_{w-1}
\right).
\label{eq:correlated-perturbation}
\end{align}
The updated ensemble parameter is
\begin{align}
\bbeta_{w-1}^{(b)}
=
\mu_{w-1}^{(b)}
+
z_{w-1}^{(b)}.
\label{minimizer}
\end{align}

\begin{algorithm}[H]
    \caption{Weekly RLSVI Update Subroutine}
    \label{alg:rlsvi.update}

    \textbf{Input}: Update week $w$; history through transitions ending in week $w-1$; prior parameters $(\mu_0,\Sigma_0)$; noise variance estimate $\sigma_{Q,w-1}^2$; discount factors $ \gamma$; previous perturbations $\{z_{w-2}^{(b)}\}_{b=1}^B$; previous ensemble parameters $\{\bbeta_{w-2}^{(b)}\}_{b=1}^B$.

    \textbf{Output}: Updated ensemble parameters $\{\bbeta_{w-1}^{(b)}\}_{b=1}^B$ and perturbations $\{z_{w-1}^{(b)}\}_{b=1}^B$.

    \begin{algorithmic}[1]

    \STATE Construct the realized feature matrix $\bX_{w-1}$ using \eqref{eq:rlsvi-design-target}.

    \STATE Compute the posterior covariance $\Sigma_{w-1}$ using \eqref{eq:rlsvi-posterior}.

    \FOR{$b=1,\ldots,B$}

        \STATE Construct the temporal-difference targets $\by_{w-1}^{(b)}$ using \eqref{eq:td-target-nonterminal} for nonterminal decision times and \eqref{eq:td-target-terminal} for final decision times.

        \STATE Compute the posterior mean $\mu_{w-1}^{(b)}$ using \eqref{eq:rlsvi-posterior}.

        \STATE Sample the temporally correlated perturbation $z_{w-1}^{(b)}$ using \eqref{eq:correlated-perturbation}.

        \STATE Set the updated ensemble parameter $\bbeta_{w-1}^{(b)}$ using \eqref{minimizer}.

    \ENDFOR


    \STATE \textbf{return} $\{\bbeta_{w-1}^{(b)},z_{w-1}^{(b)}\}_{b=1}^B$.

    \end{algorithmic}
\end{algorithm}

\subsection{Empirical-Bayes Noise-Variance Updates}
\label{subsec:eb-variance}
Each Gaussian working model contains a scalar noise-variance parameter. These variances are initialized from the MRT and updated each Monday night using empirical Bayes. The same generic procedure is used across model classes; only the response vector and design matrix differ.

For a generic Bayesian linear model $m$, let
\[
\mathbf y^m
=
\mathbf X^m\theta^m+\epsilon^m,
\qquad
\epsilon_i^m \overset{\mathrm{ind}}{\sim} N(0,\sigma_m^2),
\qquad
\theta^m\sim N(\mu_0^m,\Sigma_0^m),
\]
where $\sigma_m^2$ is scalar. Conditional on the fixed prior hyperparameters
$(\mu_0^m,\Sigma_0^m)$, we estimate $\sigma_m^2$ by maximizing the marginal likelihood obtained after integrating out $\theta^m$:
\[
p(\mathbf y^m\mid \mathbf X^m,\sigma_m^2)
=
\int
p(\mathbf y^m\mid \mathbf X^m,\theta^m,\sigma_m^2)
p(\theta^m)
\,d\theta^m.
\]
Equivalently,
\[
\mathbf y^m\mid \mathbf X^m,\sigma_m^2
\sim
N\left(
\mathbf X^m\mu_0^m,\,
\sigma_m^2 I_{n_m}+\mathbf X^m\Sigma_0^m(\mathbf X^m)^\top
\right),
\]
where $n_m$ is the number of observations for model $m$. 

Therefore, the marginal log-likelihood is
\begin{align}
\ell_m(\sigma_m^2)
&=
-\frac{1}{2}
\Bigg[
n_m\log(2\pi)
+
\log\left|
\sigma_m^2 I_{n_m}
+
\mathbf X^m\Sigma_0^m(\mathbf X^m)^\top
\right|
\nonumber\\
&\qquad\quad+
\left(\mathbf y^m-\mathbf X^m\mu_0^m\right)^\top
\left(
\sigma_m^2 I_{n_m}
+
\mathbf X^m\Sigma_0^m(\mathbf X^m)^\top
\right)^{-1}
\left(\mathbf y^m-\mathbf X^m\mu_0^m\right)
\Bigg].
\label{eq:marginal-loglik-variance}
\end{align}

For computational stability, we use an equivalent precision-form expression. Let $\rho_m=1/\sigma_m^2$. Up to constants that do not depend on $\rho_m$, the marginal log-likelihood can be written as
\begin{align}
\widetilde \ell_m(\rho_m)
&=
-\frac{1}{2}
\log\left|
(\Sigma_0^m)^{-1}
+
\rho_m(\mathbf X^m)^\top\mathbf X^m
\right|
+
\frac{n_m}{2}\log \rho_m
-
\frac{\rho_m}{2}(\mathbf y^m)^\top\mathbf y^m
\nonumber\\
&\quad+
\frac{1}{2}
\left\{
(\Sigma_0^m)^{-1}\mu_0^m
+
\rho_m(\mathbf X^m)^\top\mathbf y^m
\right\}^\top
\left\{
(\Sigma_0^m)^{-1}
+
\rho_m(\mathbf X^m)^\top\mathbf X^m
\right\}^{-1}
\nonumber\\
&\quad\quad\times
\left\{
(\Sigma_0^m)^{-1}\mu_0^m
+
\rho_m(\mathbf X^m)^\top\mathbf y^m
\right\}.
\label{eq:precision-marginal-loglik}
\end{align}

Algorithm~\ref{alg:generic-eb-variance} gives the generic scalar variance update. 
In implementation, we use the precision-form marginal log-likelihood in 
\eqref{eq:precision-marginal-loglik}, but we optimize over the log-variance
\[
r_m=\log(\sigma_m^2),
\qquad
\rho_m=\frac{1}{\sigma_m^2}=\exp(-r_m).
\]
Thus, for a candidate value $r_m$, the objective is evaluated as
$
-\widetilde\ell_m\left(\exp(-r_m)\right)
$.

We use a bounded one-dimensional search centered at the empirical residual scale around the prior mean. Specifically,
\[
s_m^2
=
\max\left\{
\frac{1}{n_m}
\left\|
\mathbf y^m-\mathbf X^m\mu_0^m
\right\|_2^2,
10^{-12}
\right\},
\qquad
c_m=\log(s_m^2).
\]
Given a prespecified log-scale half-width $h_\sigma = 8$ and lower variance bound $\sigma_{\min}^2 = 10^{-4}$, the optimizer searches over
\[
r_{m,\min}
=
\max\left\{
c_m-h_\sigma,
\log(\sigma_{\min}^2)
\right\},
\qquad
r_{m,\max}
=
c_m+h_\sigma.
\]
The empirical-Bayes estimate is then
\[
\widehat r_m
=
\arg\min_{r_m\in[r_{m,\min},r_{m,\max}]}
\left\{
-\widetilde\ell_m\left(\exp(-r_m)\right)
\right\},
\qquad
\widehat\sigma_m^2
=
\exp(\widehat r_m).
\]
This one-dimensional bounded optimization is performed using 
\texttt{scipy.optimize.minimize\_scalar}. 

\begin{algorithm}[H]
    \caption{Generic Empirical-Bayes Variance Update}
    \label{alg:generic-eb-variance}

    \textbf{Input}: Design matrix $\mathbf X$; response vector $\mathbf y$; prior mean $\mu_0$; prior covariance $\Sigma_0$; log-scale half-width $h_\sigma$=8; lower variance bound $\sigma_{\min}^2=10^{-4}$.

    \textbf{Output}: Empirical-Bayes variance estimate $\widehat\sigma^2$.

    \begin{algorithmic}[1]

    \STATE Define the precision-form marginal log-likelihood
    $\widetilde\ell(\rho)$ as in \eqref{eq:precision-marginal-loglik}.

    \STATE Compute the prior-predictive residual scale:
    \vspace{-0.8em}
    \[
    s^2
    =
    \max\left\{
    \frac{1}{n}
    \left\|
    \mathbf y-\mathbf X\mu_0
    \right\|_2^2,
    10^{-12}
    \right\},
    \qquad
    c=\log(s^2).
    \]
    \vspace{-2em}

    \STATE Define the bounded search interval for the log-variance:
    \vspace{-0.8em}
    \[
    r_{\min}
    =
    \max\left\{
    c-h_\sigma,
    \log(\sigma_{\min}^2)
    \right\},
    \qquad
    r_{\max}
    =
    c+h_\sigma.
    \]
    \vspace{-2em}

    \STATE Compute
    \vspace{-0.8em}
    \[
    \widehat r
    =
    \arg\min_{r\in[r_{\min},r_{\max}]}
    \left\{
    -\widetilde\ell\left(\exp(-r)\right)
    \right\}.
    \]
    \vspace{-1.5em}

    \STATE Set
    \vspace{-0.4em}
    \[
    \widehat\sigma^2=\exp(\widehat r).
    \]
    \vspace{-2em}


    \STATE \textbf{return} $\widehat\sigma^2$.

    \end{algorithmic}
\end{algorithm}

Importantly, the algorithm does not maintain particle-specific or ensemble-specific variance parameters. For the particle-filtering working models, the empirical-Bayes variance update is performed once per working model. At the Monday-night update in week $w$, before propagating particles to $Y_w$, latent CAE columns in the cumulative particle-filtering designs are filled using the smoothed posterior mean CAE trajectory under the current pre-update particle approximation,
\[
\bar y_{\ell\mid w-1}
=
\sum_{j=1}^J
\widehat v_{w-1}^{(j)}
\widehat y_{\ell}^{(j)},
\qquad
\ell=1,\ldots,w-1.
\]
The resulting collapsed design matrix and response vector are supplied to Algorithm~\ref{alg:generic-eb-variance}, yielding one scalar estimate for each particle-filtering working-model variance:
\[
(\widehat\sigma_{w}^{Y})^2,\quad
(\widehat\sigma_{w}^{\widetilde Y})^2,\quad
(\widehat\sigma_{w}^{MY,DT})^2,\quad
(\widehat\sigma_{w}^{MY,D})^2.
\]
These variances are shared across particles in the subsequent belief-state update.

For the RLSVI regression, the TD targets are ensemble-specific
because they depend on $\bbeta^{(b)}$. Ensemble members share the design
matrix and differ only through the exploration perturbations, so estimating
$\sigma_Q^2$ separately per member would count exploration spread as
observation noise. We therefore apply
Algorithm~\ref{alg:generic-eb-variance} once, to the ensemble-averaged
target vector $\bar{\mathbf y}_{w-1} = B^{-1}\sum_{b=1}^{B}\mathbf
y_{w-1}^{(b)}$, and the resulting $\widehat\sigma_{Q,w-1}^{2}$ is shared
across ensemble members in the subsequent posterior update.

For the bottleneck-state variant, the design matrix itself is ensemble-specific, because the bottleneck-consistency rows use member-specific greedy actions. In that case Algorithm~\ref{alg:generic-eb-variance} is applied to each member's stacked three-block regression (bottleneck consistency, nonterminal TD, and terminal TD), and the resulting estimates are averaged over members. The Stage-1 and Stage-2 redistribution regressions use fixed noise variances estimated from the MRT; only the working-model and TD variances are updated online.

\subsection{Weekly Update Schedule}
\label{subsec:weekly-schedule}

Week 1 is an initial micro-randomization phase in which activity suggestions are assigned independently with probability $0.5$. The first online RL update occurs on Monday night of week 2. Thereafter, each week follows the same schedule.

The two Monday activity-suggestion decisions occur before the weekly survey response window closes and therefore use the most recently available perceived-utility estimate, CAE belief summaries, and RLSVI ensemble together with the current real-time context. After the Monday-night update, the newly estimated weekly states and RLSVI ensemble are used for the remaining decision times that week.

At the Monday-night update, the algorithm:
\begin{enumerate}
    \item constructs the perceived-utility approximation $\widehat e_w$;
    \item updates the CAE belief state and computes
    $(\widehat b_w,\widetilde b_w)$;
    \item updates the required empirical-Bayes variance estimates;
    \item updates any redistribution models required by the selected variant; and
    \item updates the RLSVI ensemble.
\end{enumerate}

\section{RL Variants Evaluated in Pre-Deployment Experiments}
\label{app:variants}

\subsection{Baseline Configuration}
Algorithm~\ref{alg:rlsvi1} gives the base online RL algorithm. It calls several subroutines defined in subsequent sections. Specifically, \textsc{UpdateBeliefState} denotes the particle-filtering update for the CAE belief state, \textsc{ConstructFeature} constructs the action-specific feature vector from the current summaries, mediator history, and context, \textsc{SelectAction} computes the clipped assignment probability and samples the activity-suggestion action, and \textsc{UpdateRLSVI} performs the weekly RLSVI parameter update. Fixed hyperparameters listed in the input of Algorithm~\ref{alg:rlsvi1}, are passed implicitly to the relevant subroutines.

\begin{algorithm}[htbp]
    \caption{Online RL Algorithm for the ADAPT Factorial Trial}
    \label{alg:rlsvi1}
    \textbf{Input}: Weekly discount factors $\gamma$; ensemble size $B$;
    prior parameters $N(\mu_0,\Sigma_0)$; 
    clipping parameter $\epsilon_0$; 
    baseline CAE value $Y_1$; 
    initial perceived-utility estimate $\widehat e_1=2$;
    inputs for Algorithm~\ref{alg:belief.state.all}.

    \begin{algorithmic}[1]

    \STATE Initialize the belief state using the baseline CAE value:
    $\widehat b_1=Y_1$ and $\widetilde b_1=0$.
    \STATE Initialize posterior perturbations:
    $
    z_0^{(b)}\sim N(0,\Sigma_0),\quad
    \bbeta_0^{(b)}=\mu_0+z_0^{(b)},\quad b=1,\ldots,B.
    $
    
    \item[] \textit{Uniform randomization in week 1 before RL updating.}
    
    \STATE Assign activity-suggestion actions by
    $A_{1,d,t}\sim \operatorname{Bernoulli}(0.5)$ for all
    $d\in\{1,\ldots,6\}$ and $t\in\{1,2\}$.

    \FOR{week $w=2,3,\ldots$}
    
        \item[] \textit{Sunday-night weekly survey query decision.}
        
         \STATE Set $I_w=1$ if $w=2$, and sample $I_w\sim\operatorname{Bernoulli}(0.5)$ if $w\geq 3$.

        \item[] \textit{Monday decisions before weekly updating.}
        
        \FOR{$(d,t)\in\{(1,1),(1,2)\}$}
            \STATE Construct the two action-specific feature vectors using Algorithm~\ref{alg:feature.construction}:
            \vspace{-0.8em}
            \begin{align*}
             &\phi_1^-
            \leftarrow
            \textsc{ConstructFeature}
            (w,d,t,1;\widehat e_{w-1},\widehat b_{w-1},\widetilde b_{w-1}, \mathcal M_{w,d,t}^{\mathrm{pre}}, C_{w,d,t}),   \\
            &\phi_0^-
            \leftarrow
            \textsc{ConstructFeature}
            (w,d,t,0;\widehat e_{w-1},\widehat b_{w-1}, \widetilde b_{w-1},\mathcal M_{w,d,t}^{\mathrm{pre}}, C_{w,d,t}).
            \end{align*}
            \vspace{-2em}
            \STATE Select the activity-suggestion action using Algorithm~\ref{alg:select.action}:
            \vspace{-0.8em}
            \begin{align*}
            (A_{w,1,t},\pi_{w,1,t})
            \leftarrow
            \textsc{SelectAction}
            \left(
            \phi_1^-,
            \phi_0^-,
            \{\bbeta_{w-2}^{(b)}\}_{b=1}^B,
            \epsilon_0
            \right).
            \end{align*}
            \vspace{-2em}
        \ENDFOR

        \item[] \textit{Monday-night update.}
        \STATE Compute the approximated perceived-utility $\widehat e_w$ according to Equation \eqref{eq:ew}.

        \STATE Update the belief state for $Y_w$ using Algorithm~\ref{alg:belief.state.all}:
        \vspace{-0.8em}
        \begin{align*}
            \left(
        \{\widehat y_{1:w}^{(j)}\}_{j=1}^J,
        \{\widehat v_w^{(j)}\}_{j=1}^J
        \right)
        \leftarrow
        \textsc{UpdateBeliefState}(H_{w-1},\{\widehat y_{1:w-1}^{(j)}\}_{j=1}^J,
\{\widehat v_{w-1}^{(j)}\}_{j=1}^J).
        \end{align*}
        \vspace{-2em}
        \STATE Compute the belief-state summaries:
        \vspace{-0.8em}
        \begin{align*}
          \widehat b_w
        =
        \sum_{j=1}^J
        \widehat v_w^{(j)}\widehat y_w^{(j)},
        \qquad
        \widetilde b_w
        =
        \left[
        \sum_{j=1}^J
        \widehat v_w^{(j)}
        \left(\widehat y_w^{(j)}-\widehat b_w\right)^2
        \right]^{1/2}.  
        \end{align*}
        \vspace{-1.8em}
        \STATE Update the empirical-Bayes variance estimates of RLSVI:
        \vspace{-0.8em}
        \begin{align*}
        \widehat \sigma_{Q,w-1}^2    
        \leftarrow
        \textsc{UpdateEBVariancesRL}
        \left(
        H_{w-1},
        \widehat e_{1:w-1},
        \widehat b_{1:w},
        \widetilde b_{1:w}
        \right),
        \end{align*}
        \vspace{-2em}
        \STATE Update the RLSVI ensemble using Algorithm~\ref{alg:rlsvi.update}:
        \vspace{-0.8em}
        \begin{align*}
            \left(
        \{\bbeta_{w-1}^{(b)}\}_{b=1}^B,
        \{z_{w-1}^{(b)}\}_{b=1}^B
        \right)
        \leftarrow
        \textsc{UpdateRLSVI}
        \left(
        w,
        H_{w-1},
        \{\bbeta_{w-2}^{(b)}\}_{b=1}^B,
        \{z_{w-2}^{(b)}\}_{b=1}^B,
        \widehat \sigma_{Q,w-1}^2  
        \right).
        \end{align*}
        \vspace{-2em}

        \item[] \textit{Remaining within-week decisions after weekly updating.}
        \FOR{$(d,t)\in\{(2,1),(2,2),\ldots,(6,1),(6,2)\}$}
                \STATE Construct the two action-specific feature vectors using Algorithm~\ref{alg:feature.construction}:
                \vspace{-0.8em}
                \begin{align*}
                &\phi_1
                \leftarrow
                \textsc{ConstructFeature}
                (w,d,t,1;\widehat e_w,\widehat b_w,\widetilde b_w,\mathcal M_{w,d,t}^{\mathrm{pre}}, C_{w,d,t}), \\
                &\phi_0
                \leftarrow
                \textsc{ConstructFeature}
                (w,d,t,0;\widehat e_w,\widehat b_w,\widetilde b_w,\mathcal M_{w,d,t}^{\mathrm{pre}}, C_{w,d,t}).
                \end{align*}
                \vspace{-2em}

                \STATE Select the activity-suggestion action using Algorithm~\ref{alg:select.action}:
                \vspace{-0.8em}
                \begin{align*}
                    (A_{w,d,t},\pi_{w,d,t})
                \leftarrow
                \textsc{SelectAction}
                \left(
                \phi_1,
                \phi_0,
                \{\bbeta_{w-1}^{(b)}\}_{b=1}^B,
                \epsilon_0
                \right).
                \end{align*}
                \vspace{-2em}
        \ENDFOR

    \ENDFOR
    \end{algorithmic}
\end{algorithm}

\subsection{Combined Variants}
Rather than fully crossing all design choices, we organize the
experiments around a base configuration and vary one component at a time. The base RL algorithm uses $\gamma = 0.9$, no bottleneck-state TD loss, and no reward shaping. First, variants 1 is rerun with
$\gamma = 0.5$ as a sensitivity analysis (variants 2).
We then vary (i) whether the bottleneck-state TD loss in \eqref{eq:bottleneck-loss} is added, and (ii) which of the two reward designs is used (variants 4--5).   The two baseline policies--never sending, and sending independently with probability $0.5$ at each decision time--provide reference points for assessing the added value of personalization and learning.

\begin{table}[H]
\centering
\caption{Summary of RL variants and baseline policies. Reward designs: V1 = return-invariant shaping; V2 = return-invariant shaping with redistribution.}
\label{tab:rl-variants}
\begin{tabular}{c c c c c}
\hline
\textbf{Variant} & \textbf{Policy class} & $\gamma$ &
\textbf{Bottleneck TD} & \textbf{Reward design} \\
\hline
1 & RL        & $0.9$ & No  & None \\
2 & RL               & $0.5$ & No  & None \\
3 & RL               & $0.9$ & Yes & None \\
4 & RL               & $0.9$ & No  & V1 \\
5 & RL               & $0.9$ & No  & V2 \\

\hline
6 & Never send             & -- & -- & -- \\
7 & Randomized, $p=0.5$    & -- & -- & -- \\
\hline
\end{tabular}
\end{table}

\section{ADAPT Simulation Environment}
\label{app:simulation_models}

We construct a closed-loop simulation testbed from the historical ADAPT MRT to evaluate candidate online RL algorithms before deployment in the ADAPT factorial trial. 

The environment contains three interacting components. Section~\ref{subsec:sim-cae} models the CAE-related process, including activity-related mediators and weekly CAE. Section~\ref{subsec:sim-pu} models latent perceived utility together with engagement-related mediators and emissions. Section~\ref{subsec:sim-other} models additional time-varying state variables. The fitted components are then combined sequentially to generate trajectories under a specified activity-suggestion policy, as described in Section~\ref{subsec:generate-testbed-data}.

\subsection{Variables to Model}
\label{subsec:variablesmodel}

The simulation environment generates the following variables.

\paragraph{CAE-related mediators.}
\begin{itemize}
    \item $M_{w,d,t}^{Y,SC}$: log-transformed step count during the four hours following decision time $t$ on day $d$ of week $w$;
    \item $M_{w,d}^{Y,AA}$: anticipated affect measured once daily.
\end{itemize}

\paragraph{Perceived-utility-related mediators.}
\begin{itemize}
    \item $M_{w,d,t}^{E,PV}$: unprompted app page views following decision time $(w,d,t)$;
    \item $M_{w,d}^{E,FW}$: indicator of Fitbit wear for the corresponding daily wear window;
    \item $M_{w,d}^{E,PJ}$: indicator of daily check-in completion.
\end{itemize}

\paragraph{Weekly latent states and outcomes.}
\begin{itemize}
    \item $E_w$: latent weekly perceived utility of the app;
    \item $Y_w$: CAE, defined as the mean of the 12 CAE items, each coded from 1 to 7.
\end{itemize}

\paragraph{Weekly emissions.}
\begin{itemize}
    \item $\widetilde Y_w$: short weekly CAE, constructed from three CAE items on the same 1--7 scale;
    \item $J_w$: indicator of weekly check-in completion;
    \item $U_{w,1}$ and $U_{w,2}$: weekly ratings of app helpfulness and pleasantness, respectively.
\end{itemize}

\paragraph{Additional time-varying state variables.}
\begin{itemize}
    \item $C_{w,d,t}^{SC}$: log-transformed step count during the two hours preceding decision time $(w,d,t)$;
    \item $C_{w,d,t}^{AS}$: indicator of interaction with a delivered activity suggestion;
    \item $C_{w,d}^{AD}$: indicator of an active day.
\end{itemize}

Throughout, $A_{w,d,t}\in\{0,1\}$ indicates whether an activity suggestion is delivered. The historical MRT allowed activity-suggestion decisions on all seven days, so $d\in\{1,\ldots,7\}$ when fitting MRT-based models. The forthcoming trial uses the six-day Monday--Saturday decision schedule described previously.

For each conditional model, $S$ denotes the history available immediately before the modeled variable is generated. We use $g(S)$ for baseline features and $f(S)$ for action-effect moderators. This notation is model-specific: the exact feature sets are given below.

\subsection{CAE-Related Process}
\label{subsec:sim-cae}

The models for next-four-hour step count $M^{Y,SC}$, anticipated affect $M^{Y,AA}$, weekly CAE $Y$, and short CAE $\widetilde Y$ are fit jointly across participants using Bayesian hierarchical linear models. The hierarchical specification allows the simulator to retain participant-specific dynamics while borrowing information across participants when individual MRT trajectories are sparse.

For a generic CAE-related outcome $Z$, let $\theta_i^Z$ denote participant $i$'s coefficient vector and $x_{i,j}^Z$ the corresponding design vector for observation $j$. We use
\begin{align}
Z_{i,j} 
&= (x_{i,j}^Z)^\top \theta_i^Z + \epsilon_{i,j}^Z,
&
\epsilon_{i,j}^Z &\sim \mathcal{N}\!\left(0, \sigma_Z^2\right),
\nonumber \\
\theta_i^Z &= \theta^{Z} + b_i^Z,
&
b_i^Z &\sim \mathcal{N}\!\left(0,
\mathrm{diag}\!\left((\tau_1^Z)^2, \ldots, (\tau_{p_Z}^Z)^2\right)\right),
\label{eq:hier-glm}
\end{align}
with a residual variance $\sigma_Z^2$ common to all participants and participant-specific random effects on the regression coefficients.

At the population level,
\begin{align}
\theta^Z
&\sim
\mathcal N\!\left(
\mu_0^Z,
\operatorname{diag}\left(
(s_{0,1}^Z)^2,\ldots,(s_{0,p_Z}^Z)^2
\right)
\right),
\nonumber\\
\sigma_Z
&\sim
\operatorname{HalfNormal}(1),
\nonumber\\
\tau_k^Z
&\sim
\operatorname{Gamma}
\left(
\text{shape}=2,\,
\text{mode}=m_k^Z
\right),
\qquad
k=1,\ldots,p_Z.
\label{eq:hier-glm-priors}
\end{align}
The shape-2 Gamma prior assigns zero density at $\tau_k^Z=0$ and therefore discourages collapse to the boundary in which all participants have identical coefficients, while still allowing small between-participant variability when supported by the data. We do not use the classical non-informative inverse-Gamma $(\epsilon,\epsilon)$ distribution on the variance: because when $\epsilon \rightarrow0$, it leads to an improper posterior, and is sensitive to the value of $\epsilon$ \citep{gelman2006prior}; the shape-2 Gamma is the boundary-avoiding prior of \citet{chung2013nondegenerate}, with mode $m_k^Z$ directly interpretable as the most plausible between-participant scale for each coefficient. 
The residual scale, unlike the random-effect scales, is informed by every
observation rather than by the $28$ participant-level deviations, so its
likelihood concentrates away from zero and no boundary-avoiding prior is
needed; we therefore use a weakly informative $\operatorname{HalfNormal}(1)$,
whose scale matches the standardized outcomes.

Most population-level coefficients receive weakly informative priors $N(0,\sqrt{10})$. For selected coefficients whose direction is specified by the working causal model, we use weak directional regularization by shifting the population-level prior mean and reducing its variance. 

Posterior inference uses the No-U-Turn sampler \citep{hoffman2014no} with a non-centered parameterization of the random effects, implemented via \texttt{PyMC} \citep{abril2023pymc}. For $M^{Y,SC}$, $Y$, and $\widetilde Y$, we use four chains with $3{,}000$ warm-up iterations followed by $2{,}000$ retained draws per chain and target acceptance $0.99$. For $M^{Y,AA}$, we use $1{,}500$ warm-up iterations followed by $1{,}000$ retained draws per chain and target acceptance $0.95$. We require no divergent transitions and maximum $\widehat R<1.01$. The simulator uses posterior-mean participant-specific coefficients; residual noise is generated by the autoregressive residual bootstrap of Section \ref{subsec:generate-testbed-data} from the fitted residuals, not from the Gaussian residual model.

\paragraph{Next-four-hour step count.}
For the log-transformed step count during the four hours following each activity-suggestion decision,
\begin{align}
M_{w,d,t,i}^{Y,SC}
&=
g^{Y,SC}(S_{w,d,t,i})^\top \alpha_i^{Y,SC}
+
A_{w,d,t,i}\,
f^{Y,SC}(S_{w,d,t,i})^\top \beta_i^{Y,SC}
+
\epsilon_{w,d,t,i}^{Y,SC},
\qquad
\epsilon_{w,d,t,i}^{Y,SC}
\sim
\mathcal{N}(0, \sigma_{Y,SC}^{2}),
\label{eq:genmysc}
\end{align}
The baseline feature vector contains an intercept, the previous decision slot's outcome, yesterday's step count, a exponentially weighted average of same-slot four-hour step counts over the past seven days, prior two-hour step count, recent active-day fraction, recent burden, recent unprompted page views, recent Fitbit wear, recent interaction with delivered suggestions, yesterday's anticipated affect, weekday/weekend, morning/afternoon, previous-week CAE, and previous-week perceived utility.

The moderator vector contains yesterday's step count, prior two-hour step count, recent burden, recent page views, Fitbit wear, suggestion interaction, yesterday's anticipated affect, decision time, previous-week CAE, and previous-week perceived utility.

Unless otherwise specified, coefficients random-effect-SD mode $0.15$. Recent burden receives a mildly negative prior
$(\mu_0=-0.05,s_0=0.03,m=0.05)$,
and its interaction with the activity suggestion receives
$(\mu_0=-0.05,s_0=0.04,m=0.05)$.
Previous-week perceived utility, previous-week CAE, the suggestion main effect, and suggestion interactions with previous-week perceived utility and CAE receive weak positive prior centers
$\mu_0=0.03$ with $s_0=0.08$.
These directional priors encode working population-level assumptions rather than fixed treatment effects.

\paragraph{Daily anticipated affect.}

For anticipated affect observed after the day's activity-suggestion decisions,
\begin{align}
M_{w,d,i}^{Y,AA}
&=
g^{Y,AA}(S_{w,d,i})^\top \alpha_i^{Y,AA}
+
\sum_{t=1}^{2}
A_{w,d,t,i}\,
f^{Y,AA}(S_{w,d,i})^\top \beta_{t,i}^{Y,AA}
+
\epsilon_{w,d,i}^{Y,AA},
\qquad
\epsilon_{w,d,i}^{Y,AA}
\sim
\mathcal{N}(0, \sigma_{Y,AA}^{2}),
\end{align}
The baseline feature vector contains an intercept, yesterday's anticipated affect, recent active-day fraction, weekday/weekend, previous-week perceived utility, previous-week CAE, and recent burden. Each decision-specific moderator vector contains an intercept, previous-week perceived utility, previous-week CAE, recent burden, and recent active-day fraction.

The prior specification parallels that for $M^{Y,SC}$. Recent burden and its suggestion interactions receive weak negative prior centers, whereas previous-week perceived utility, previous-week CAE, the two suggestion main effects, and selected suggestion interactions receive weak positive prior centers. All remaining coefficients receive weakly informative zero-centered priors.

\paragraph{Weekly CAE.}

Weekly CAE is generated from
\begin{align}
Y_{w,i}
=
g^{Y}(S_{w,i})^\top \alpha_i^{Y}
+
\epsilon_{w,i}^{Y},
\qquad
\epsilon_{w,i}^Y
\sim
\mathcal{N}(0, \sigma_Y^{2}).
\end{align}
The weekly feature vector contains
\begin{enumerate}
    \item an intercept;
    \item last week's CAE;
    \item week in study;
    \item exponentially weighted average of the within-week four-hour
          step-count mediators $\{M_{w,d,t}^{Y,SC}\}$;
    \item exponentially weighted average of the within-week daily
          anticipated-affect mediators $\{M_{w,d}^{Y,AA}\}$.
\end{enumerate}

The within-week summaries use normalized exponentially decaying weights with
$\gamma=13/14$ across the 14 MRT decision slots and
$\gamma=6/7$ across the seven daily anticipated-affect measurements.
When some values needed for a weekly summary are unavailable, missing values are replaced using the prespecified within-week imputation rule before the weighted summary is computed.

The intercept and week-in-study keep random-effect-SD mode $0.1$.
The remaining three components each have a positive prior mean and a tight
$\sigma_0 = 0.15$ (mode $0.15$): the autoregressive term is centered at
$0.4$, and the step-count and anticipated-affect averages are centered at $0.15$ and $0.3$, respectively; the two averages are highly correlated in the training data ($\rho \approx 0.9$), and these mild centers prevent the fit from arbitrarily reallocating their shared signal.


\paragraph{Short CAE.}

The short CAE measurement is generated conditionally on full weekly CAE:

\begin{align}
\widetilde{Y}_{w,i}
=
g^{\widetilde{Y}}(S_{w,i})^\top \alpha_i^{\widetilde{Y}}
+
\epsilon_{w,i}^{\widetilde{Y}},
\qquad
\epsilon_{w,i}^{\widetilde{Y}}
\sim
\mathcal{N}(0, \sigma_{\widetilde{Y}}^{2}),
\end{align}
where $g^{\widetilde Y}(S_w)$
contains an intercept and $Y_w$. Both coefficients receive weakly informative priors.

\subsection{Perceived-Utility Process}
\label{subsec:sim-pu}
We represent perceived utility using a participant-specific latent state-space model. The latent state $E_w$ evolves jointly with engagement-related mediators and weekly emissions: unprompted page views $M^{E,PV}$, Fitbit wear $M^{E,FW}$, daily check-in completion $M^{E,PJ}$, weekly check-in completion $J_w^{\mathrm{week}}$, and the two weekly perceived-utility items $U_{w,1}$ and $U_{w,2}$.

The final state-space model is fit independently for each MRT participant; no final model coefficient is shared across participants. A pooled fit is used only to initialize the participant-specific optimization described below. We suppress the participant index in this subsection. The initial latent state is fixed at $E_1=2$, which anchors the location of the participant-specific latent process.

\paragraph{Latent transition.}
For $w = 2, \ldots, W$,
\begin{align}
E_w
=
g^{E}(S_{w-1}^{E})^\top \alpha^{E}
+
\epsilon_w^E,
\qquad
\epsilon_w^E
\sim
\mathcal{N}(0, \sigma_E^2),
\label{eq:ssm-ew}
\end{align}
where $g^{E}(S_{w-1}^{E})$ stacks an intercept, the previous state $E_{w-1}$, and the realized weekly mediator summaries $PV_{w-1}$, $FW_{w-1}$, and $PJ_{w-1}$.
Writing $\alpha^{E} = (a_0, a_1, a_2, a_3, a_4)^\top$, the coefficient $a_1$ captures week-to-week persistence and $a_2, a_3, a_4$ capture feedback through realized engagement. 
The weekly summaries are
\begin{align}
PV_w
=
\tfrac{1}{14}
\sum_{d=1}^{7} \sum_{t=1}^{2} M_{w,d,t}^{E,PV},
\qquad
FW_w
=
\tfrac{1}{7}
\sum_{d=1}^{7} M_{w,d}^{E,FW},
\qquad
PJ_w
=
\tfrac{1}{7}
\sum_{d=1}^{7} M_{w,d}^{E,PJ}.
\label{eq:pu-weekly-summaries}
\end{align}

\paragraph{Weekly emissions.}
For $w = 1, \ldots, W$,
\begin{align}
J_w \mid E_w
&\sim
\operatorname{Bernoulli}\!\bigl(\pi_w^{J}\bigr),
\qquad
\operatorname{logit}\bigl(\pi_w^{J}\bigr) = b_0 + b_1 E_w,
\nonumber\\
U_{w,j} \mid E_w,\, J_w = 1
&\sim
\mathcal{N}\!\bigl(c_{0,j} + c_{1,j} E_w,\; \sigma_{U_j}^2\bigr),
\qquad j \in \{1, 2\}.
\label{eq:pu-weekly-emissions}
\end{align}
The two perceived-utility items are observed only when the participant completes the weekly check-in. Because weekly completion is itself modeled conditional on $E_w$, the joint likelihood explicitly represents engagement-related survey non-response under the working state-space model.

\paragraph{Carry-over from the weekly check-in.}
Participation in the weekly check-in immediately preceding the current week may affect subsequent engagement. Each within-week engagement model therefore includes the carry-over term
\begin{align}
q^{k}\bigl(E_w, B_{w,d,t}\bigr)
=
J_w
\bigl(
q_0^{k} + q_E^{k} E_w + q_B^{k} B_{w,d,t}
\bigr),
\label{eq:pu-jlag}
\end{align}
where $B_{w,d,t}$ denotes recent intervention burden and $k$ indexes the relevant mediator model. 

\paragraph{Decision-time page views.}
Page views are modeled using a two-part hurdle model. 
Let $O_{w,d,t} = \mathbb{I}\{\text{page-view count} > 0\}$ denote occurrence and
let $M_{w,d,t}^{E,PV}$ denote the standardized log page-view value, with
zero-count slots placed at a fixed floor below the smallest positive value.
Occurrence follows a logistic model and, conditional on occurrence, intensity
follows a Gaussian model on the standardized log scale:
\begin{align}
O_{w,d,t} \mid E_w
&\sim
\operatorname{Bernoulli}\!\bigl(\pi_{w,d,t}^{O}\bigr),
&
\operatorname{logit}\bigl(\pi_{w,d,t}^{O}\bigr)
&=
\eta_{w,d,t}^{O},
\nonumber\\
M_{w,d,t}^{E,PV} \mid E_w,\, O_{w,d,t} = 1
&\sim
\mathcal{N}\!\bigl(\mu_{w,d,t}^{PV},\, \sigma_{PV}^2\bigr),
&
\mu_{w,d,t}^{PV}
&=
\eta_{w,d,t}^{I},
\label{eq:pu-pv-model}
\end{align}
where both linear predictors share the same functional form,
\begin{align}
\eta_{w,d,t}^{k}
=
g^{k}(S_{w,d,t})^\top \alpha^{k}
+
A_{w,d,t}\, f^{k}(S_{w,d,t})^\top \beta^{k}
+
q^{k}(E_w, S_{w,d,t}),
\qquad k \in \{O, I\},
\end{align}
with separate coefficients for occurrence ($k = O$) and intensity ($k = I$).
The baseline vector $g^{k}$ contains an intercept, the current state $E_w$, the
morning/afternoon indicator, the weekday/weekend indicator, recent burden, and
the previous decision slot's standardized page-view value (a first-order
autoregressive term); the moderator vector $f^{k}$ contains an intercept and
$E_w$, so that the effect of a walking suggestion on engagement may depend on
current perceived utility.

\paragraph{Daily mediators.}
Fitbit wear and daily check-in completion follow participant-specific logistic models:
\begin{align}
M_{w,d}^{E,k} \mid E_w
\sim
\operatorname{Bernoulli}\!\bigl(\pi_{w,d}^{k}\bigr),
\qquad
\operatorname{logit}\bigl(\pi_{w,d}^{k}\bigr)
=
g^{k}(S_{w,d})^\top \alpha^{k}
+ \sum_{t=1}^{2}
A_{w,d,t}\, f_t^{k}(S_{w,d})^\top \beta_t^{k}
+
q^{k}(E_w, S_{w,d}),
\label{eq:pu-daily-models}
\end{align}
for $k \in \{FW, PJ\}$. The baseline vectors contain an intercept, $E_w$, the
weekday/weekend indicator, recent burden, and the previous day's value of the
same mediator (a first-order autoregressive term); each moderator vector
$f_t^{k}$ contains an intercept and $E_w$.

\subsubsection{Estimation}
\label{sssec:pu-estimation}
Because $E_w$ is one-dimensional, we evaluate the state-space likelihood by deterministic quadrature. Filtering is performed on a fixed grid
$\mathcal{G} = \{e_1, \ldots, e_K\} \subset [-4, 4]$
with trapezoidal weights. Week~1 contributes its emission likelihood at the
fixed initial state $E_1 = 2$. For each subsequent week, the filtered distribution is propagated through the Gaussian transition \eqref{eq:ssm-ew}, multiplied by the current week's emission likelihood, and renormalized.

Estimation proceeds in two stages. Stage 1 maximizes the summed penalized log-likelihood across participants using a shared parameter vector and a coarse grid ($K=161$); this fit is used only to obtain a stable common initialization. Stage 2 fits each participant separately using a finer grid ($K=321$), initialized at the Stage-1 solution.

In both stages we use the weighted quadratic penalty $\lambda \sum_k w_k (\theta_k - c_k)^2$ with $\lambda = 0.5$ and $w_k \in \{0, 1\}$. Intercepts and log residual noise scales are unpenalized. In Stage 1, all penalized coefficients are centered at zero. In Stage 2, selected coefficients receive weak directional centers of magnitude $0.05$ based on the working causal model: the three transition feedback coefficients, the $E_w$ slopes in the page-view, wear, and check-in models, and the suggestion-by-$E_w$ interactions are centered in the positive direction, whereas the suggestion main effects in the engagement models are centered in the negative direction. All remaining penalized coefficients are shrunk toward zero.

Two additional penalties are used to discourage unstable latent dynamics. First, the direct persistence coefficient is constrained to $a_1 \in (-0.98, 0.98)$. Second, we monitor the first-order compound feedback gain
\begin{align}
g
=
a_1
+ a_2\, \widehat{\partial PV_w / \partial E_w}
+ a_3\, \widehat{\partial FW_w / \partial E_w}
+ a_4\, \widehat{\partial PJ_w / \partial E_w}.
\end{align}
The derivatives are local linearizations of the fitted emission models; logistic derivatives are conservatively bounded using the maximum Bernoulli slope $1/4$, and the page-view derivative combines the occurrence and positive-intensity components of the hurdle model. The gain is evaluated under the extreme suggestion and weekly-check-in configurations ($A \in \{0, 1\} \times J_{w-1}^{\mathrm{week}} \in \{0, 1\}$) used in the stability analysis below.

A finite barrier
\[
A_g
\frac{(g^E)^2}
{\max\{1-(g^E)^2,\varepsilon\}},
\qquad
A_g=3,\quad
\varepsilon=10^{-3},
\]
strongly penalizes parameter configurations approaching or exceeding unit feedback gain. A lighter
$\operatorname{arctanh}^2(a_1)$
penalty is retained to discourage $a_1$ from approaching its box constraint. The residual scales
$\sigma_E,\sigma_{U_1},\sigma_{U_2},\sigma_{PV}$
are parameterized on the log scale and constrained to $[0.03,10]$.

\subsection{Additional Time-Varying Variables}
\label{subsec:sim-other}
The remaining time-varying variables are fit separately for each participant using $\ell_2$-penalized regression. The penalty is selected by cross-validation over a common prespecified grid; when a participant has insufficient observations for cross-validation, a fixed unit penalty is used.

For prior two-hour step count,
\begin{align}
C_{w,d,t}^{SC}
=
g^{SC}(S_{w,d,t})^\top \alpha^{SC}
+
\epsilon_{w,d,t}^{SC},
\qquad
\epsilon_{w,d,t}^{SC}
\sim
\mathcal{N}(0, \sigma_{SC}^{2}),
\end{align}
where the features include an intercept, an exponentially weighted
average of same-slot two-hour step counts over the past seven days, weekday/weekend indicator, and morning/afternoon indicator.

For interaction with a delivered activity suggestion,
\begin{align}
C_{w,d,t}^{AS}
\sim
\operatorname{Bernoulli}(\mu_{w,d,t}^{AS}),
\qquad
\operatorname{logit}(\mu_{w,d,t}^{AS})
=
g^{AS}(S_{w,d,t})^\top \alpha^{AS},
\end{align}
where $g^{AS}$ contains an intercept, the fraction of delivered walking
suggestions interacted with over the past seven days, weekday/weekend
indicator, and morning/afternoon indicator.

Daily active status is modeled as
\begin{align}
C_{w,d}^{AD}
\sim
\operatorname{Bernoulli}(\mu_{w,d}^{AD}),
\qquad
\operatorname{logit}(\mu_{w,d}^{AD})
=
g^{AD}(S_{w,d})^\top \alpha^{AD},
\end{align}
where $g^{AD}$ contains an intercept, the fraction of active days over the
past seven days, and weekday/weekend indicator.

\subsection{Stability Diagnostics}
\label{subsec:pu-stability}
The fitted MRT models are used to simulate trajectories substantially longer than those observed during model fitting. We therefore perform local stability diagnostics to identify parameterizations for which feedback between weekly states and within-week mediators would cause the simulated process to be dominated by its clipping boundaries.

\paragraph{Perceived utility.}
For perceived utility, the state $E_w$ influences $E_{w+1}$ directly through the persistence coefficient $a_1$ and indirectly through the week-$w$ mediators, whose weekly summaries enter the transition \eqref{eq:ssm-ew}. Its local gain is
\[
g^{E}
=
a_1
+ a_2\, \widehat{\partial PV_w / \partial E_w}
+ a_3\, \widehat{\partial FW_w / \partial E_w}
+ a_4\, \widehat{\partial PJ_w / \partial E_w}.
\]
We evaluate this quantity under the extreme never-suggest and always-suggest regimes and under both values of the weekly-check-in carry-over indicator. Each fitted participant model is required to satisfy $|g^E|<1$ for all evaluated configurations.

\paragraph{Weekly CAE.}
For weekly CAE, last week's outcome $Y_{w}$ enters the CAE model
directly and through the week-$w$ step-count and anticipated-affect
mediators, each of which includes $Y_{w}$ (and its suggestion interactions)
as a covariate. The corresponding gain is
\[
g^{Y}
=
\alpha_1^{Y}
+ \lambda_{SC}\,
\widehat{\partial \overline{M}_w^{Y,SC} / \partial Y_{w}}
+ \lambda_{AA}\,
\widehat{\partial \overline{M}_w^{Y,AA} / \partial Y_{w}},
\]
where $\alpha_1^{Y}$ is the autoregressive coefficient and $\lambda_{SC}$,
$\lambda_{AA}$ are the loadings on the two within-week mediator summaries. The mediator derivatives incorporate both baseline and action-interaction terms and are evaluated under never-suggest and always-suggest. We require $\lvert g^{Y} \rvert < 1$ under both regimes for every fitted participant model.

The same diagnostics are repeated after any treatment-effect calibration used to construct alternative simulation environments.

\subsection{Generating Data from the Testbed}
\label{subsec:generate-testbed-data}

A simulated participant is associated with one of the fitted MRT participant models. Experimental sampling of participant types is described separately in Section~\ref{app:experiments}. Conditional on the selected participant type, variables are generated sequentially according to the temporal ordering of the fitted models, with activity-suggestion actions supplied by the policy being evaluated.

For a continuous variable $Z$ with fitted conditional mean
$\widehat\mu_Z(S,A)$
and simulated residual $\epsilon^Z$, we generate
\[
Z^\ast
=
\widehat\mu_Z(S,A)+\epsilon^Z
\]
and clip the result to the prespecified support
$[\ell_Z,u_Z]$
derived from the MRT training data:
\[
Z
=
\min\left\{
u_Z,\,
\max\{\ell_Z,Z^\ast\}
\right\}.
\]
with limits taken from the observed range of the training data. 
The latent perceived-utility state is generated analogously using
\[
\epsilon_w^E
\sim
\mathcal N(0,\widehat\sigma_E^2)
\]
and clipped to $[-4,4]$, matching the support used for quadrature during estimation.

The weekly perceived-utility items are generated on the fitted continuous scale and then mapped to the allowable 1 - 7 response scale. Binary variables, including weekly and daily check-in completion, Fitbit wear, active-day status, suggestion interaction, and page-view occurrence, are generated directly from their fitted Bernoulli probabilities. No additional residual perturbation is added to a fitted Bernoulli probability.

Page views are generated from the fitted hurdle model. Occurrence is first sampled from the Bernoulli component. If no page view occurs, the mediator is assigned the prespecified standardized zero-count value. Conditional on occurrence, positive intensity is generated from the fitted intensity model with a bootstrapped continuous residual and is bounded below by the standardized value corresponding to one page view.

For continuous observed streams, residual variation is generated using a participant-specific AR(1) residual bootstrap. Let $\{\widehat\epsilon_j\}_{j=1}^{n}$
be the fitted residuals for participant $i$ and stream $Z$, with empirical mean $\mu$ and lag-one autocorrelation $\rho$, truncated to
$|\rho|\le0.95$.
The simulated residual sequence follows
\begin{align}
\epsilon_t
=
\mu
+
\rho(\epsilon_{t-1}-\mu)
+
\sqrt{1-\rho^2}\,z_t,
\label{eq:residual-bootstrap}
\end{align}
where the initial value is sampled from the participant's residual pool and
$z_t$
is sampled with replacement from the mean-centered residual pool
$\{\widehat\epsilon_j-\mu\}$.
This construction approximately preserves the participant-specific marginal residual distribution and first-order serial dependence without replaying the observed residual sequence deterministically.

For the four streams most directly related to the primary outcome---weekly CAE, short CAE, anticipated affect, and next-four-hour step count---participants with limited residual information use a partially pooled innovation distribution that combines their participant-specific residual pool with a population residual pool. Population residuals are re-centered to the participant-specific residual mean before sampling. This prevents extremely sparse participant-specific residual pools from producing unrealistically low or unstable innovation variance while retaining participant-specific location and, when estimable, serial dependence.

Initial states are taken from each participant's observed MRT history at study entry, and the perceived-utility state is initialized at $E_1 = 2$, matching the estimation convention.

In addition to serial residual dependence, the generator explicitly represents two cross-component relationships. First, AM next-four-hour step count is linked to same-day PM prior-two-hour step count. Let $Z_{i,s,\mathrm{AM}}^{F}$ denote the centered and scaled AM next-four-hour step-count residual after generation and $Z_{i,s,\mathrm{PM}}^{P,0}$ the corresponding baseline PM prior-two-hour step residual. We generate the adjusted PM residual as
\[
\epsilon_{i,s,\mathrm{PM}}^{P}
=
\bar r_i^{P}
+
s_i^{P}
\left[
\lambda Z_{i,s,\mathrm{AM}}^{F}
+
\sqrt{1-\lambda^{2}}
Z_{i,s,\mathrm{PM}}^{P,0}
\right],
\]
where $\bar r_i^{P}$ and $s_i^{P}$ are the participant-specific residual mean and standard deviation. We set $\lambda=0.30$, based on the pooled fitted-residual correlation of $0.2997$ across 1,709 aligned MRT pairs.

Second, suggestion interaction is linked to page-view occurrence. At occasions with a delivered suggestion, interaction is generated first and the page-view occurrence probability is then adjusted according to
\[
\operatorname{logit}\bigl(p_{i,w,d,t}^{O,*}\bigr)
=
\operatorname{logit}\bigl(p_{i,w,d,t}^{O}\bigr)
+
a_t
+
b_t
\bigl(C_{i,w,d,t}^{AS}-p_{i,w,d,t}^{AS}\bigr).
\]
Separate AM and PM offset-logistic calibrations give $(a_t,b_t)=(0.037,3.901)$ for AM and then $(-0.163,2.323)$ for PM. Conditional on page-view occurrence, positive intensity is generated from its fitted conditional mean and continuous residual. Other cross-component dependence propagates through the evolving simulated history and the shared CAE and perceived-utility states rather than through additional residual coupling.

\subsection{Standardized Testbed Effect}
\label{subsec:STE}
To characterize the strength of the activity-suggestion signal relative to within-participant stochastic variation, we define a standardized testbed effect (STE). This is a calibration metric for the simulation environment, not a measure of the performance of the online RLSVI algorithm.

Let the fitted participant-specific simulators define an empirical population of simulator types indexed by $i=1,\ldots,N_0$.
For a policy $\pi$, define the undiscounted cumulative CAE over a $W$-week trajectory by
\[
G_i^\pi
=
\sum_{w=1}^{W}Y_{i,w}^{\pi}.
\]
Let $\pi^0$ denote the never-suggest policy.

For STE calculation only, the weekly check-in channel is held fixed across policies so that the comparison isolates the activity-suggestion pathway. In particular, the full weekly CAE is made available at every weekly update under both the reference policy and $\pi^0$.

\paragraph{Reference-policy effect.}
For each simulator type $i$, we construct a high-performing reference policy
$\pi_i^{\mathrm{ref}}$ using the simulator-specific procedure described below. Define
\begin{align*}
\Delta_i
=
\mathbb E
\left[
G_i^{\pi_i^{\mathrm{ref}}}
-
G_i^{\pi^0}
\right],
\qquad
\sigma_i^2
=
\operatorname{Var}
\left(
G_i^{\pi^0}
\right),
\end{align*}
where the expectation and variance are over simulator randomness. 

The average STE is 
\[
\operatorname{STE}_{\mathrm{pop}}
=
\frac{1}{N_0} \sum_{i=1}^{N_0} \frac{\Delta_i}{\sigma_i}.
\]
Thus, each simulator type's improvement is standardized by its own trajectory-level variation under never-suggest before averaging across types. This choice targets intervention signal relative to the stochastic variation faced within an individual trajectory rather than between-participant heterogeneity.

The STE should be interpreted as an upper-reference signal metric. The reference policy has access to complete weekly CAE observations and is not subject to the online algorithm's clipping constraint on randomization probabilities. It therefore has an informational and policy-class advantage over the deployed RLSVI learner.

\paragraph{Construction of the reference policy.}
Because the optimal policy of each nonlinear simulator is not available analytically, we approximate a high-performing simulator-specific policy using offline RL. For each simulator type $i$, we generate $B_{\mathrm{train}}$ trajectories of length $W+1$ under a behavior policy that independently sends each activity suggestion with probability $0.5$. Weekly CAE is assigned as the reward at the final controlled decision time of each week. We then fit discrete conservative Q-learning (CQL), using discount factor one within each week and $0.9$ across weekly boundaries.
The fitted deterministic reference policy is
\[
\widehat\pi_i^{\mathrm{ref}}(s)
=
\mathbb I
\left\{
\widehat Q_i(s,1)
-
\widehat Q_i(s,0)
>
0
\right\}.
\]
Training, checkpoint selection, and final evaluation use disjoint simulator
seeds. During training, candidate checkpoints are scored periodically on an
independent selection set, and the checkpoint with the largest paired
improvement over never-suggest is retained. Because this maximization is
optimistically biased, the reported STE is computed on a separate,
independent test set that plays no role in training or checkpoint selection.

On the test set, the reference and never-suggest policies are evaluated using common random numbers. For episode
$b=1,\ldots,B_{\mathrm{eval}}$,
the same simulator seed is used for both policies, yielding paired cumulative outcomes
$G_{i,b}^{\widehat\pi_i^{\mathrm{ref}}}$
and
$G_{i,b}^{\pi^0}$.
We estimate
\[
\widehat\Delta_i
=
\max\Bigl\{0,\;
\frac{1}{B_{\mathrm{eval}}}\sum_{b=1}^{B_{\mathrm{eval}}}
\bigl(G_{i,b}^{\widehat\pi_i^{\mathrm{ref}}} - G_{i,b}^{\pi^0}\bigr)
\Bigr\},
\]
and
\[
\widehat\sigma_i^2
=
\frac{1}{B_{\mathrm{eval}}-1}
\sum_{b=1}^{B_{\mathrm{eval}}}
\left(
G_{i,b}^{\pi^0}
-
\overline G_i^{\pi^0}
\right)^2.
\]
The estimated population STE is
\[
\widehat{\operatorname{STE}}_{\mathrm{pop}}
=
\frac{1}{N_0}
\sum_{i=1}^{N_0}
\frac{\widehat\Delta_i}{\widehat\sigma_i}.
\]

\paragraph{Implementation Details.}

The CQL Q-function uses two hidden layers with 128 and 64 units and a fixed 20-dimensional state representation containing calendar features, the agent-computable perceived-utility proxy, the previous week's realized CAE, within-week mediator summaries, and decision-time context. Training uses $10^5$ gradient steps, batch size $256$, learning rate $10^{-4}$, conservative penalty $\alpha=0.1$, and a target-network update every $500$ gradient steps, based on $B_{\mathrm{train}}=5000$ behavior-policy trajectories. Checkpoint selection uses 200 independent episodes, and final evaluation uses
$B_{\mathrm{eval}}=500$ paired episodes per simulator type.

\subsection{Simulation-Environment Variants}
\label{ssec:ste-tuning}

The fitted MRT models define the \emph{vanilla} testbed ($\widehat{\operatorname{STE}}_{\mathrm{pop}} = 0.479$). To evaluate the algorithms under weaker and stronger intervention signal, we construct variants by rescaling fitted coefficients along prespecified causal pathways, leaving all other model components unchanged. 

The \emph{low-signal} testbed strengthens the burden chain $A_{w,d,t} \to M_{w,\cdot}^{E} \to E_{w+1} \to M_{w+1,\cdot}^{Y} \to Y_{w+2}$ with a single tuning constant $\kappa$: the main action effects on the three engagement mediators ($M^{E,PV}$ occurrence and intensity, $M^{E,FW}$, $M^{E,PJ}$) are shifted by $-\kappa$; the mediator-to-state transition loadings ($a_2, a_3, a_4$) and the state-to-mediator legs ($E_w \to M^{Y,SC}$, $E_w \to M^{Y,AA}$) are shifted by $+\kappa$, so that suggestions suppress engagement, suppressed engagement lowers the next week's state, and the lowered state reaches CAE. 

The \emph{high-signal} testbed strengthens the benefit chain $A_{w,d,t} \to M_{w,\cdot}^{Y} \to Y_{w+1}$: the full action blocks of the step-count and affect models --- main effects and all fitted action--state interactions --- are  multiplied by $\kappa$, preserving each participant's moderation structure, and the step-count-to-CAE loading receives a small compensating shift, $c(\kappa-1)$ with $c=0.05$. 

Each variant's $\kappa$ is calibrated by root finding against a fast Monte-Carlo proxy of the STE and then confirmed with the full reference-policy procedure of
Section~\ref{subsec:STE}, giving $\widehat{\operatorname{STE}}_{\mathrm{pop}} = 0.275$ (low) and $0.813$ (high).
Priors supplied to the online algorithms are not refit on the tuned environments, mirroring deployment, where working models are calibrated on historical data from a source environment.


\section{Simulator Validation}
\label{app:validation}

We validate the vanilla simulator in two steps. We first examine fitted-model residuals to identify temporal and cross-variable dependence remaining after model fitting, and then assess whether the connected simulator reproduces the observed MRT trajectories. 

\subsection{Component-Model Diagnostics}
\label{subsec:validation-component}

The fitted component models are specified in Section~\ref{app:simulation_models}. Here, we assess whether their conditional means and probabilities reproduce the observed MRT data and examine temporal and cross-variable dependence remaining after model fitting. These diagnostics identify residual structure that is carried into the stochastic generator used to construct participant trajectories.

\paragraph{CAE-related Gaussian model fit.}
For each CAE-related Gaussian component $Z$, we examine the participant-conditional residual
\(
\widehat\epsilon_{i,j}^{Z}
=
Z_{i,j}
-
(x_{i,j}^{Z})^\top\widehat\theta_i^{Z},
\)
where $(x_{i,j}^{Z})^\top\widehat\theta_i^{Z}$ is the fitted conditional mean for participant $i$ at observation $j$. Thus, the residual represents variation remaining after accounting for the participant-specific history, actions, and other predictors included in the fitted model. Because these conditional models form the deterministic backbone of the simulator, structure remaining in their residuals indicates stochastic variation or dependence that must be represented separately during simulation.

\begin{figure}[p]
\centering
\includegraphics[width=\textwidth]{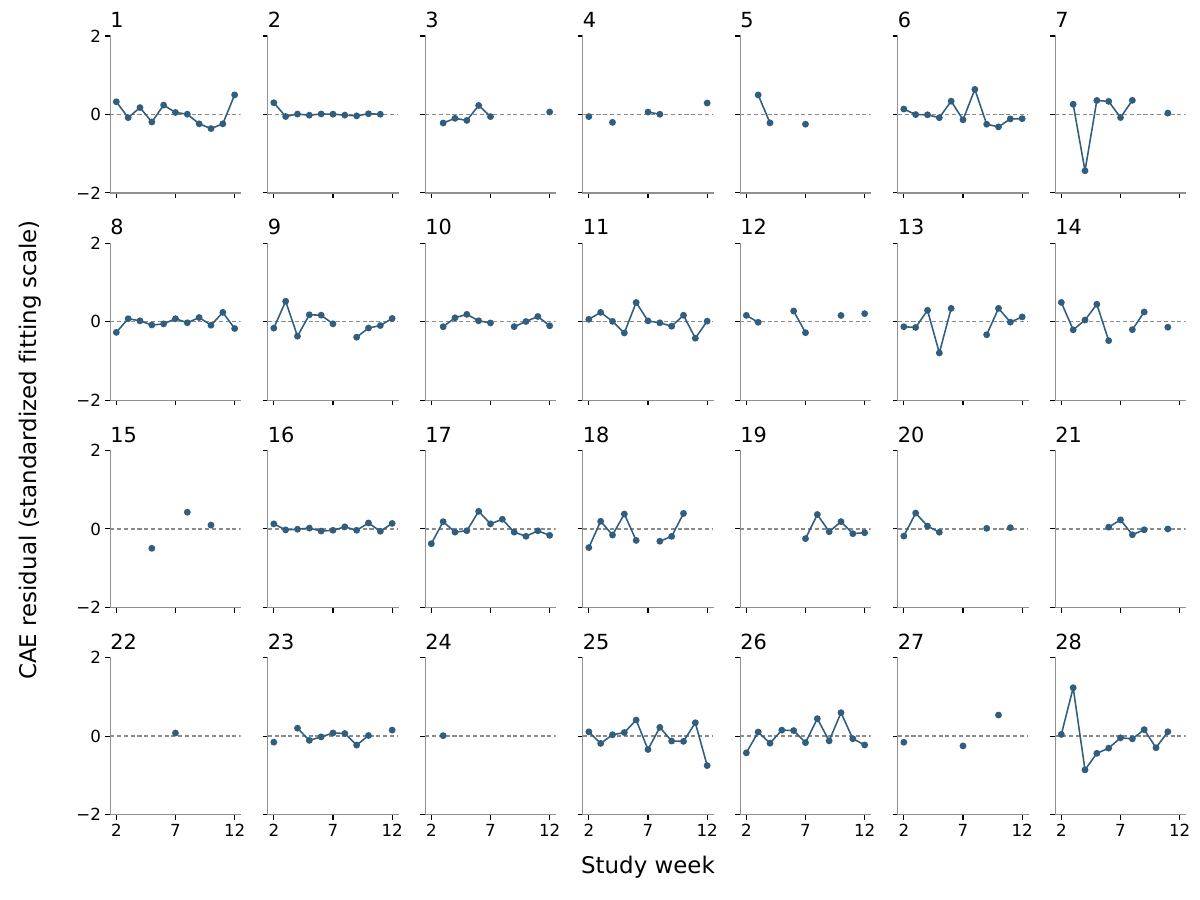}
\caption{Participant-conditional CAE residuals for the 28 fitted participant models. Residuals are expressed on the standardized fitting scale and plotted against study week. Points denote observed assessments, lines connect consecutive observed weeks, and dashed horizontal lines mark zero.}
\label{fig:cae-residuals}
\end{figure}

\begin{figure}[p]
\centering
\includegraphics[width=\textwidth]{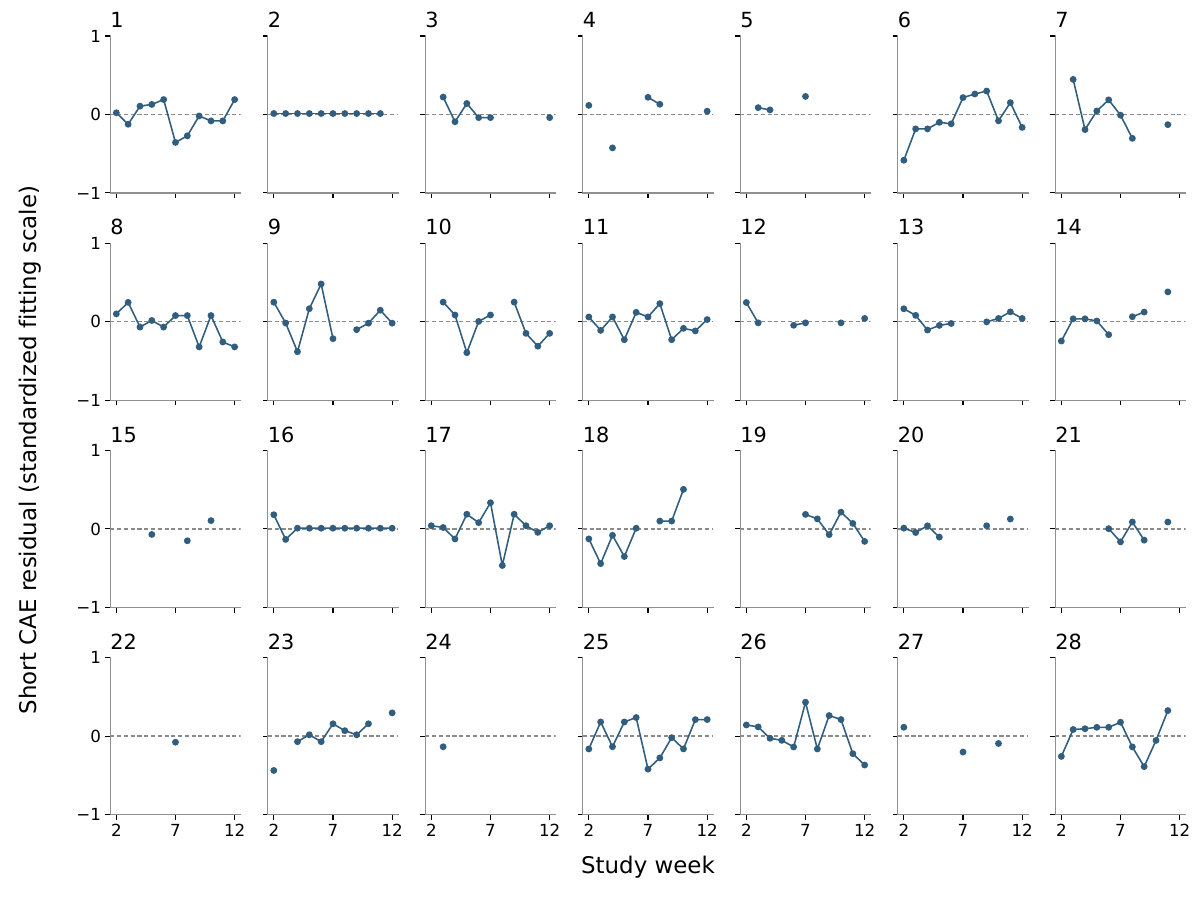}
\caption{Participant-conditional short CAE residuals for the 28 fitted participant models. Residuals are expressed on the standardized short CAE fitting scale. Points denote observed assessments, lines connect consecutive observed weeks, and dashed horizontal lines mark zero. All panels use common axes and the participant order established for CAE.}
\label{fig:short-cae-residuals}
\end{figure}

\begin{figure}[p]
\centering
\includegraphics[width=\textwidth]{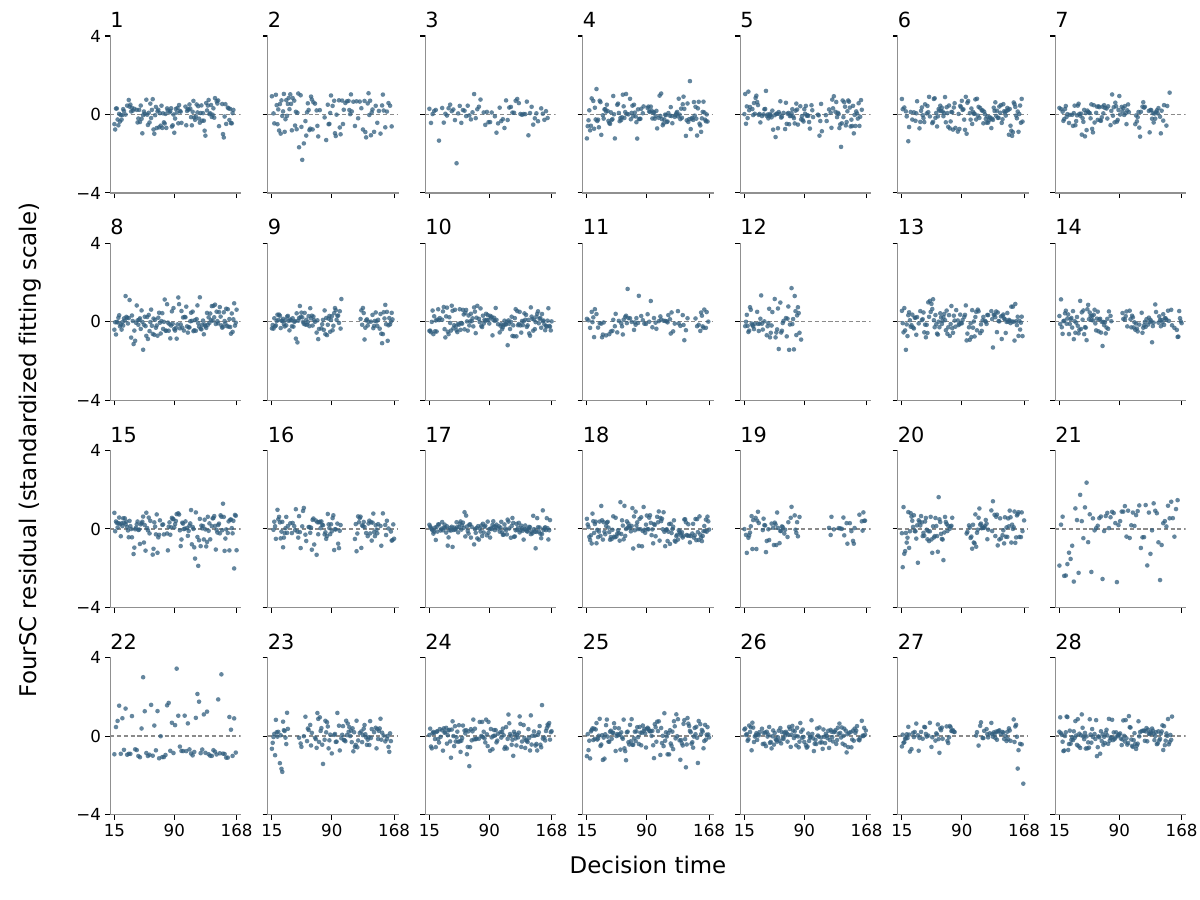}
\caption{Participant-conditional FourSC residuals for the 28 fitted participant models. Points show observed residuals on the standardized log-step-count fitting scale, and dashed lines mark zero. All panels use common axes.}
\label{fig:foursc-residuals}
\end{figure}

\begin{figure}[p]
\centering
\includegraphics[width=\textwidth]{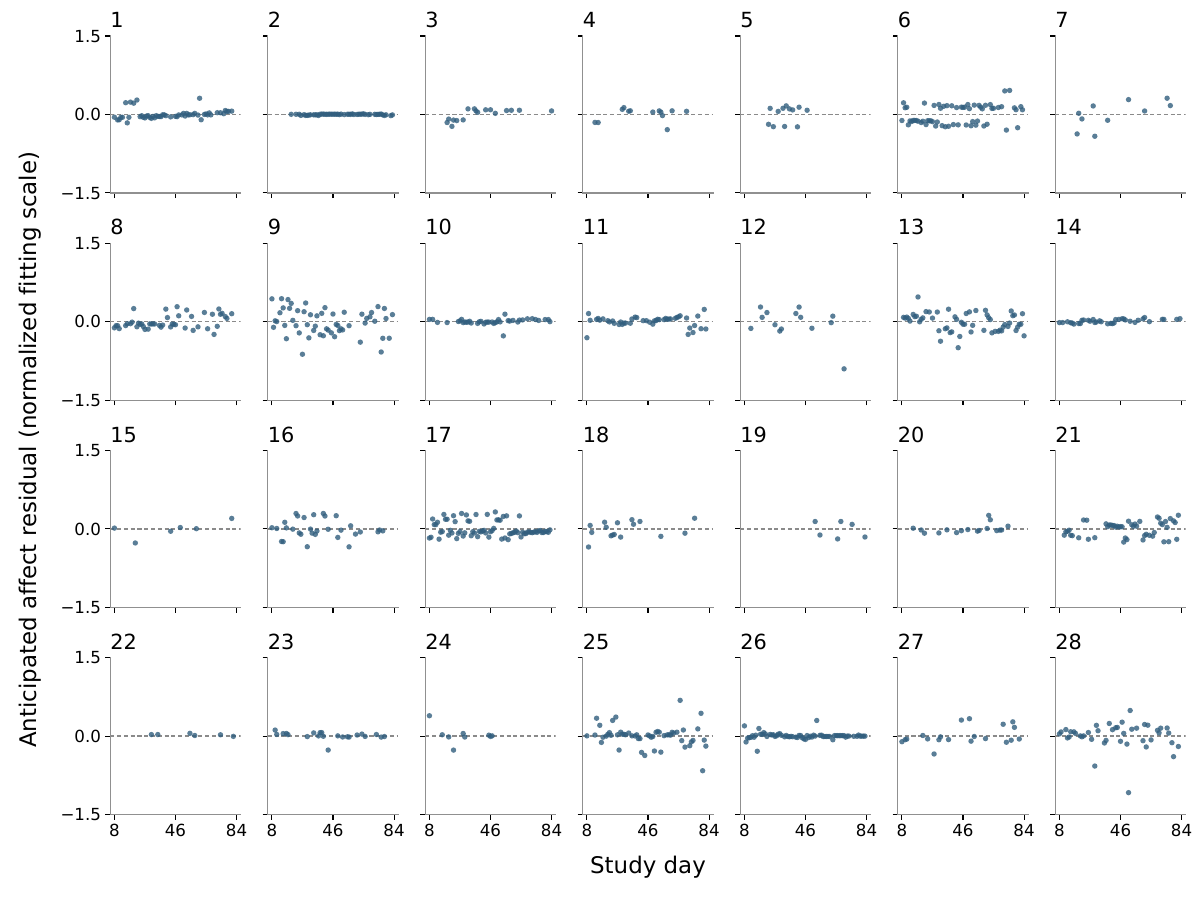}
\caption{Participant-conditional anticipated-affect residuals for the 28 fitted participant models. Points show observed residuals on the normalized fitting scale, and dashed lines mark zero. All panels use common axes.}
\label{fig:anticipated-affect-residuals}
\end{figure}

Figures~\ref{fig:cae-residuals}--\ref{fig:anticipated-affect-residuals} show the residuals for weekly CAE, short CAE, next-four-hour step count, and anticipated affect. Across the four components, residuals generally fluctuate around zero, suggesting little systematic level discrepancy after conditioning on the fitted predictors. Their dispersion and occasional larger deviations nevertheless differ across participants and components. We therefore next examine whether this remaining variation also contains temporal or cross-variable dependence before specifying the stochastic generator.

\paragraph{Bernoulli component fit.}
For each Bernoulli component $k$, we compare the observed outcome $z_{i,j}^k$ with its participant-specific fitted probability $\widehat p_{i,j}^k$ and define the residual as $\widehat e_{i,j}^k=z_{i,j}^k-\widehat p_{i,j}^k$. We assess mean calibration by comparing the observed event rate $\bar z_i^k$ with the mean fitted probability $\bar p_i^k$, and probabilistic accuracy using the Brier score \citep{glenn1950verification}, $\mathrm{BS}_i^k=(n_i^k)^{-1}\sum_j(z_{i,j}^k-\widehat p_{i,j}^k)^2$, where smaller values indicate more accurate probability predictions.

Table~\ref{tab:bernoulli-fit} reports these diagnostics for page-view occurrence, Fitbit wear, daily and weekly check-in completion, active-day status, and suggestion interaction. The observed rates and mean fitted probabilities are closely aligned for the three perceived-utility-related mediators: the largest participant-level absolute differences are $0.011$ for page-view occurrence, $0.010$ for Fitbit wear, and $0.018$ for daily check-in completion. Suggestion interaction is also closely reproduced, with a largest discrepancy of $0.046$. Larger differences occur for weekly check-in completion and active-day status, primarily for participants with limited variation or few observed outcomes. Overall, the fitted Bernoulli models reproduce participant-specific event rates well for the engagement processes most directly entering the perceived-utility dynamics, while leaving residual variation to be examined for temporal and cross-variable dependence.

\paragraph{Auxiliary continuous component fit.}
For prior-two-hour steps, positive page-view intensity, helpfulness, and pleasantness, we assess participant-specific fit using residuals defined as the transformed observed outcome minus its fitted conditional mean. We summarize residual location by the participant-specific mean residual and residual magnitude by the root mean squared error (RMSE).

Table~\ref{tab:auxiliary-decision-fit} reports these diagnostics for prior-two-hour steps and positive page-view intensity. Mean residuals remain close to zero across participants, ranging from $-0.067$ to $0.038$ for prior-two-hour steps and from $-0.026$ to $0.034$ for positive page-view intensity, although residual dispersion varies across participants. The smallest page-view errors occur for some participants with very few positive page-view observations, so they provide limited information about fit across a broader range of intensities. Table~\ref{tab:utility-items-fit} shows a similar pattern for helpfulness and pleasantness, with participant-specific mean residuals ranging from $-0.019$ to $0.059$ and from $-0.022$ to $0.041$, respectively. Overall, these models show little systematic mean discrepancy while retaining participant-specific residual variation for the subsequent dependence diagnostics.

\paragraph{Temporal residual dependence.}
After assessing conditional fit, we examine whether the fitted residuals retain serial dependence at the time scale of each process. For weekly components, we calculate participant-specific correlations between residuals at consecutive assessment weeks. For daily components, we examine lags of one and seven calendar days, corresponding to adjacent days and the same weekday in successive weeks. For decision-time components, we examine AM-to-same-day-PM and PM-to-next-day-AM residual correlations. Tables~\ref{tab:weekly-residual-correlation}, \ref{tab:daily-residual-correlation}, and \ref{tab:ampm-residual-correlation} report these correlations together with the number of available residual pairs.

The diagnostics show residual temporal structure beyond the fitted conditional means and probabilities. Most notably, weekly CAE residual correlations are negative for all 21 participants with reportable values, ranging from $-0.728$ to $-0.017$. The other weekly components show more heterogeneous dependence across participants. At the daily scale, residual dependence is also evident for some participants, including a lag-one Fitbit-wear correlation of $0.554$ for participant 21 and a lag-seven daily-check-in correlation of $0.370$ for participant 11. Decision-time residual dependence is generally more modest for next-four-hour step count, but stronger associations remain for some auxiliary and engagement components. Overall, the direction and magnitude of residual dependence vary across participants and processes, indicating that independent residual draws would not adequately represent the remaining within-participant dynamics and motivating participant- and component-specific serial dependence in the stochastic generator.

\paragraph{Cross-variable residual dependence.}
We next examine whether residuals from different component models remain associated after conditioning on their fitted predictors. Weekly residuals are aligned by assessment week, daily residuals by calendar day, and decision-time residuals by AM or PM occasion. We additionally compare AM next-four-hour step-count residuals with same-day PM prior-two-hour step-count residuals because their measurement windows overlap. Tables~\ref{tab:weekly-joint-continuous}--\ref{tab:daily-joint-residual-correlation} and Tables~\ref{tab:ampm-joint-steps}--\ref{tab:ampm-joint-engagement} report the participant-specific correlations and available pair counts.

The weekly and daily associations vary substantially in sign and magnitude across participants, without a common cross-variable pattern. Two decision-time relationships are more systematic. AM next-four-hour step-count and same-day PM prior-two-hour step-count residuals are positively correlated for 25 of the 27 participants with available pairs. Suggestion-interaction and page-view-occurrence residuals are positive for all 27 participants with reportable AM correlations, while the PM associations are more heterogeneous. These recurring relationships indicate cross-component dependence not captured by the fitted conditional models and motivate the corresponding dependence mechanisms in the stochastic generator.

\paragraph{Residual dependence in the stochastic generator.} The residual diagnostics above inform the stochastic component of the simulator. For continuous streams, participant-specific residual sequences are generated with an AR(1)-style bootstrap that carries first-order serial dependence into simulated trajectories, with partial pooling when participant-specific residual information is limited. The two systematic cross-component relationships identified above are also represented explicitly during generation: AM next-four-hour step count informs same-day PM prior-two-hour step count, and suggestion interaction informs page-view occurrence. Other dependence propagates through the evolving simulated history and the shared CAE and perceived-utility states. Full generation procedures are provided in Section~\ref{subsec:generate-testbed-data}.

\subsection{Simulator-Level Validation}

At the simulator level, we assess whether the fitted components, when linked through the evolving simulated history, reproduce the participant-specific trajectories observed in the MRT. For each of the 28 participant models, we generate 100 trajectories over MRT study weeks 2--12 using the participant's recorded activity-suggestion sequence and matched initial history. Fitted parameters are held fixed, while generated values recursively update subsequent states and history summaries. We compare the observed MRT measurements with pointwise 95\% simulation intervals defined by the 2.5th and 97.5th percentiles of the 100 simulated trajectories. 

\begin{figure}[p]
    \centering
    \includegraphics[width=\textwidth]{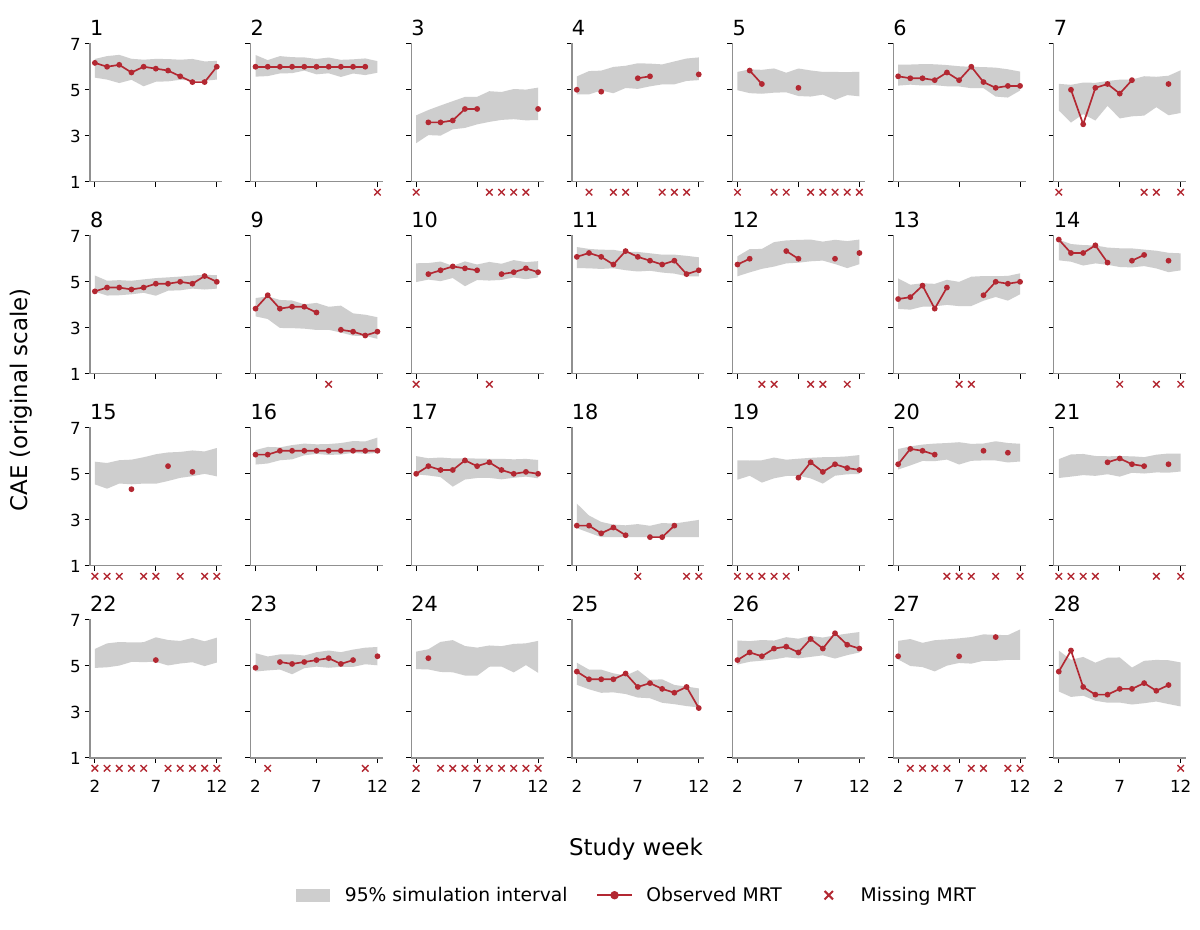}
    \caption{Observed and simulated CAE for the 28 participants under their recorded MRT action histories. Gray bands show pointwise 95\% simulation intervals, red points and lines show observed CAE, and crosses mark missing MRT assessments. All panels use common axes.}
    \label{fig:cae-trajectories}
\end{figure}

\paragraph{Main trajectory reproduction.}
Figures~\ref{fig:cae-trajectories}, \ref{fig:foursc-trajectories}, and \ref{fig:anticipated-affect-trajectories} compare the observed and simulated trajectories for weekly CAE, next-four-hour step count, and anticipated affect, while Figures~\ref{fig:pageview-occurrence-trajectories}, \ref{fig:fitbit-wear-trajectories}, and \ref{fig:daily-completion-trajectories} show the three perceived-utility-related mediators. To summarize agreement across participants, we first calculate within each participant the proportion of observed MRT measurements that fall within the pointwise 95\% simulation interval and then average these proportions equally across the 28 participants. The resulting participant-averaged coverage is 92.7\% for weekly CAE, 89.9\% for next-four-hour step count, and 93.8\% for anticipated affect. The corresponding coverages for page-view occurrence, Fitbit wear, and daily check-in completion are 97.4\%, 97.4\%, and 98.4\%, respectively, giving an average of 97.7\% across the three perceived-utility-related mediators.

The trajectory plots also show that the simulator retains substantial participant heterogeneity in outcome and mediator levels. For example, the CAE trajectories preserve the lower level of participant 18, the higher levels of participants 2 and 16, and broad declines for participants 9 and 25. The step-count and anticipated-affect trajectories similarly retain large between-participant differences in their typical levels. Agreement is less uniform for some localized changes and extreme observations, indicating that the simulator reproduces the broader participant-specific trajectory structure more consistently than every individual fluctuation.

\begin{figure}[p]
    \centering
    \includegraphics[width=\textwidth]{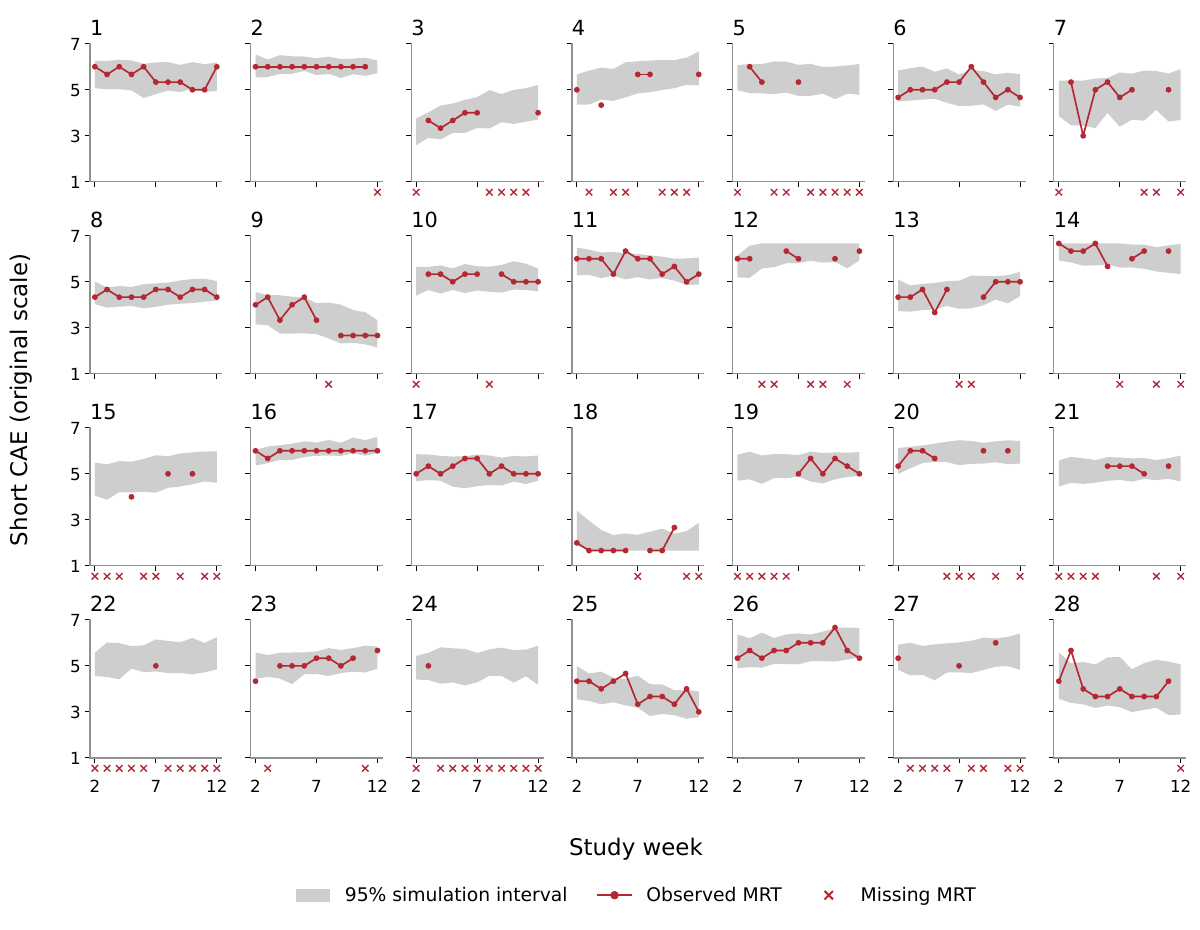}
    \caption{
        Observed and simulated short CAE for the 28 participants under
        their recorded MRT action histories. Gray bands show pointwise
        95\% simulation intervals, and red points and lines show observed
        assessments. Crosses below each panel mark missing assessments,
        which are not joined by lines. All panels use the original
        1--7 scale and a common study-week axis.
    }
    \label{fig:short-cae-trajectories}
\end{figure}

\begin{figure}[p]
    \centering
    \includegraphics[width=\textwidth]{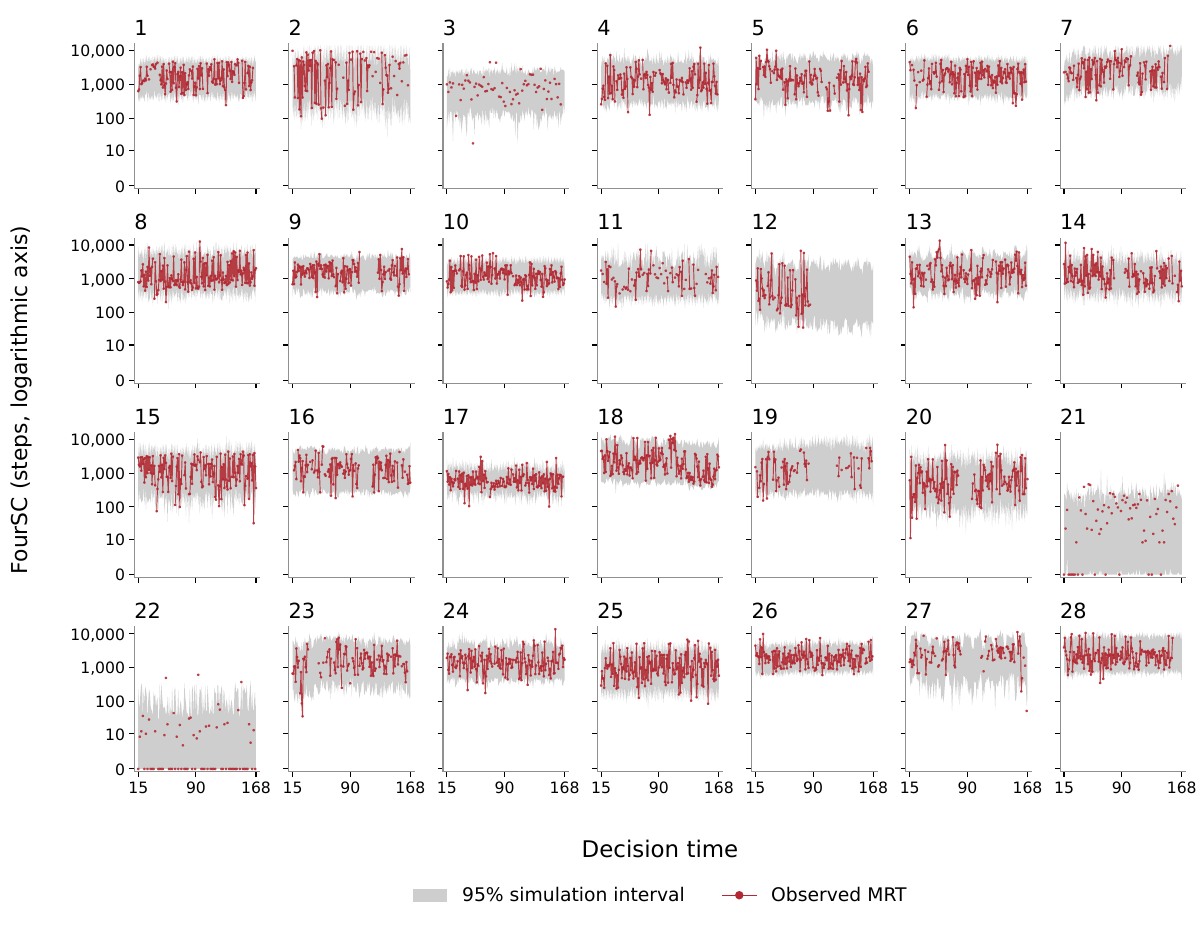}
    \caption{Observed and simulated FourSC for the 28 participants under their recorded MRT action histories. Gray bands show pointwise 95\% simulation intervals, and red points and lines show observed step counts. All panels use a common $\log(1+\mathrm{steps})$ axis labeled in step counts. Missing observations are not joined.}
    \label{fig:foursc-trajectories}
\end{figure}

\begin{figure}[p]
    \centering
    \includegraphics[width=\textwidth]{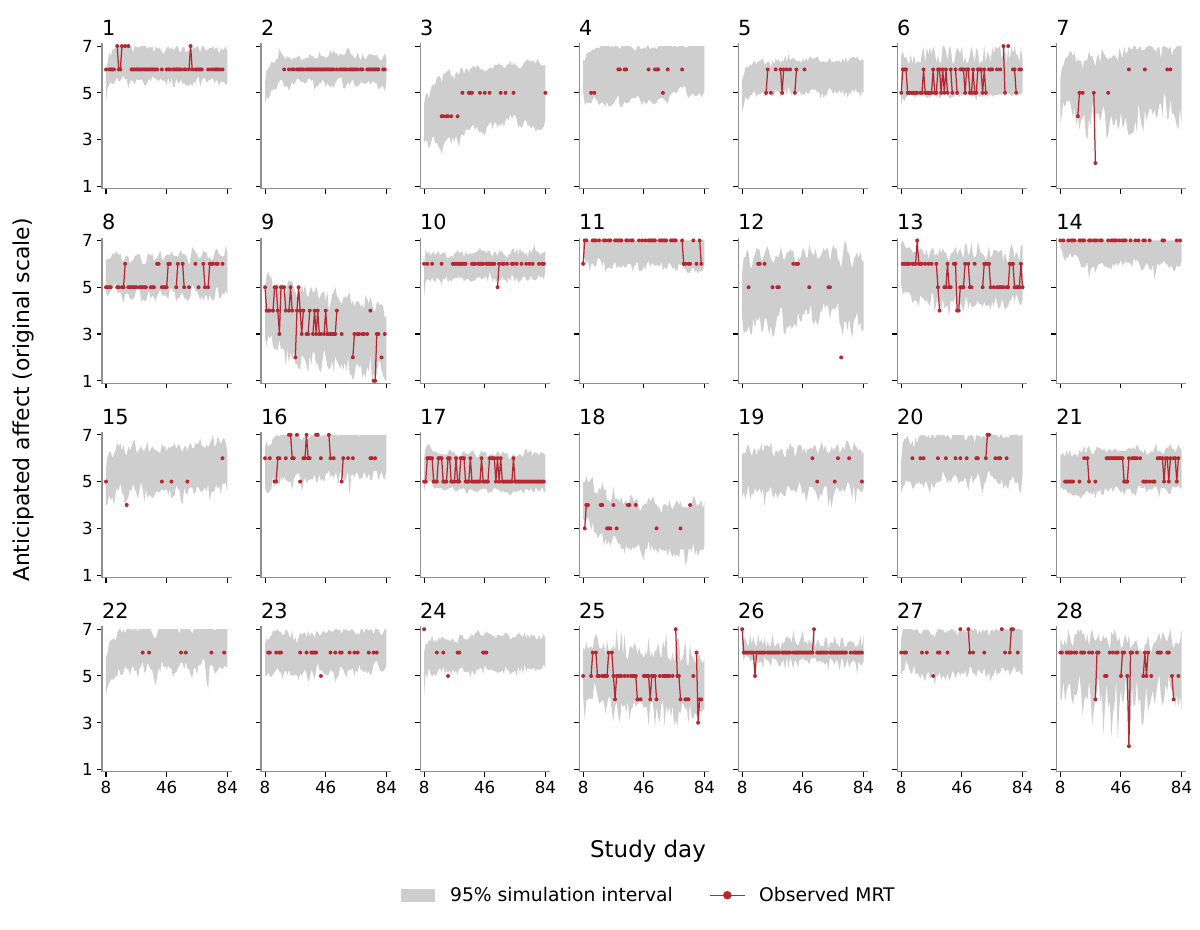}
    \caption{Observed and simulated anticipated affect for the 28 participants under their recorded MRT action histories. Gray bands show pointwise 95\% simulation intervals, and red points and lines show recorded responses. All panels use the original 1--7 scale and a common study-day axis. Missing responses are not joined.}
    \label{fig:anticipated-affect-trajectories}
\end{figure}

\begin{figure}[p]
    \centering
    \includegraphics[width=\textwidth]{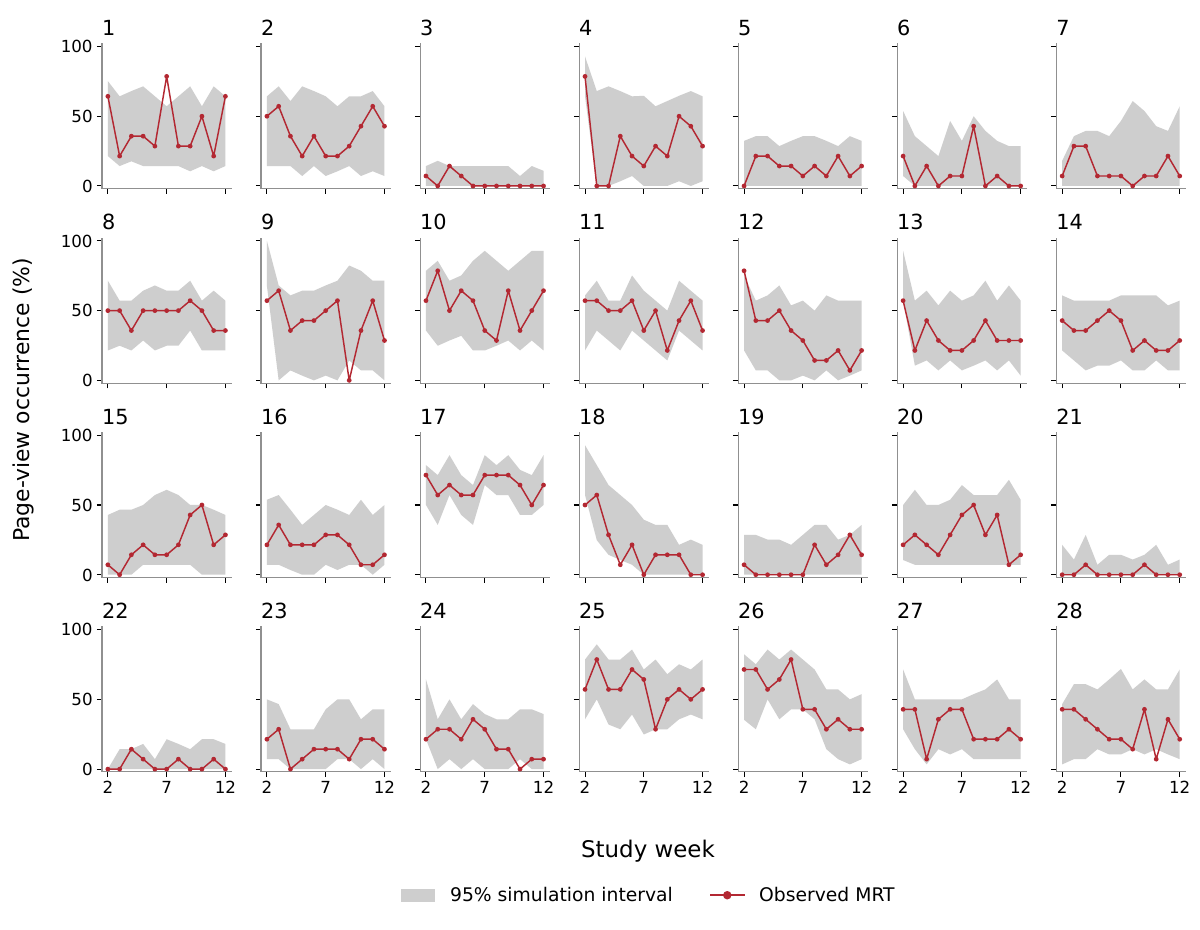}
    \caption{
        Observed and simulated weekly page-view occurrence for the
        28 participants under their recorded MRT action histories.
        Red points and lines show the percentage of decision occasions
        with at least one page view. Gray bands show pointwise 95\%
        simulation intervals calculated after weekly aggregation within
        each trajectory. All panels share the same axes.
    }
    \label{fig:pageview-occurrence-trajectories}
\end{figure}

\begin{figure}[p]
\centering
\includegraphics[width=\textwidth]{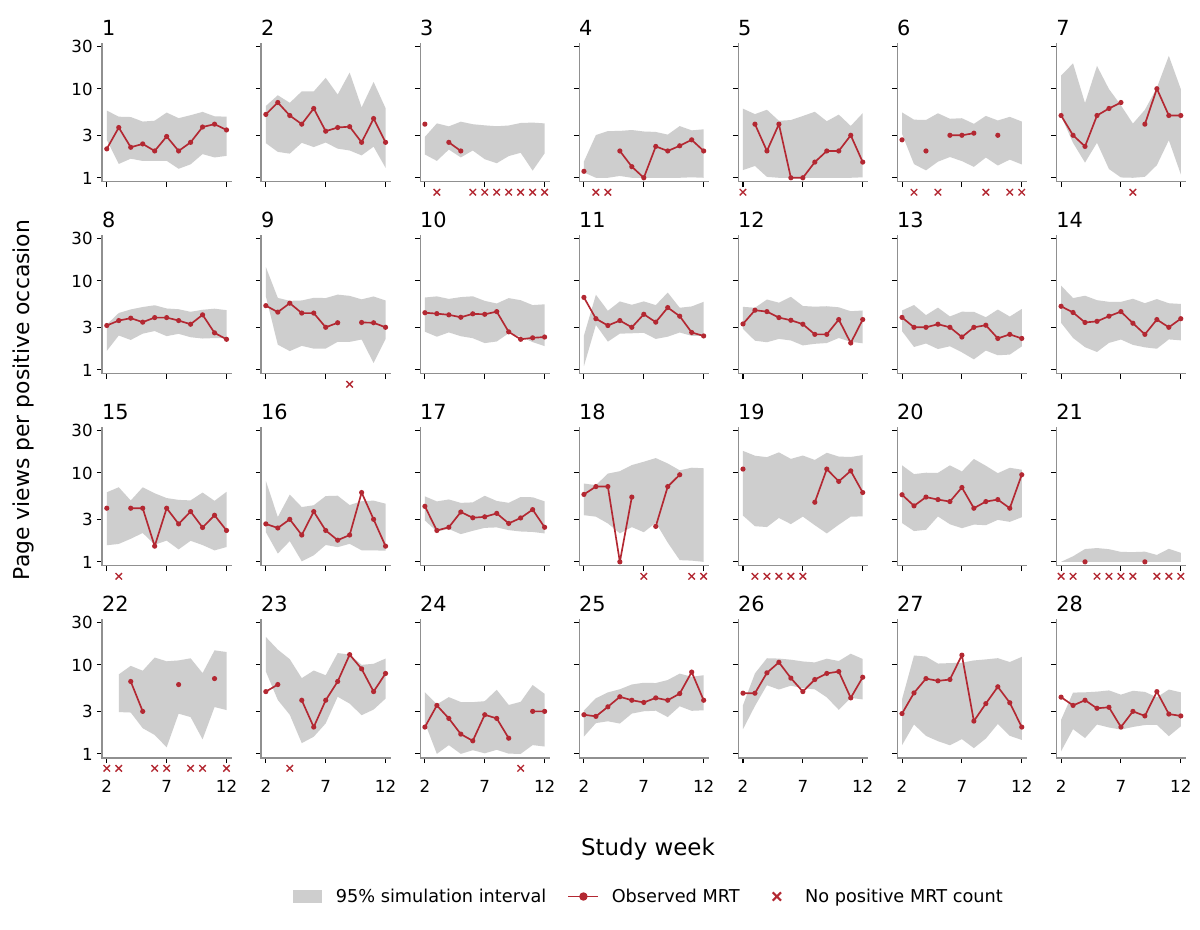}
\caption{Observed and simulated mean page-view counts among positive occasions in each study week. Gray bands show pointwise 95\% simulation intervals and red points show observed means. Crosses mark MRT weeks with no positive count, rather than missing logging data. All panels share a logarithmic count axis.}
\label{fig:positive-pageview-trajectories}
\end{figure}

\begin{figure}[p]
\centering
\includegraphics[width=\textwidth]{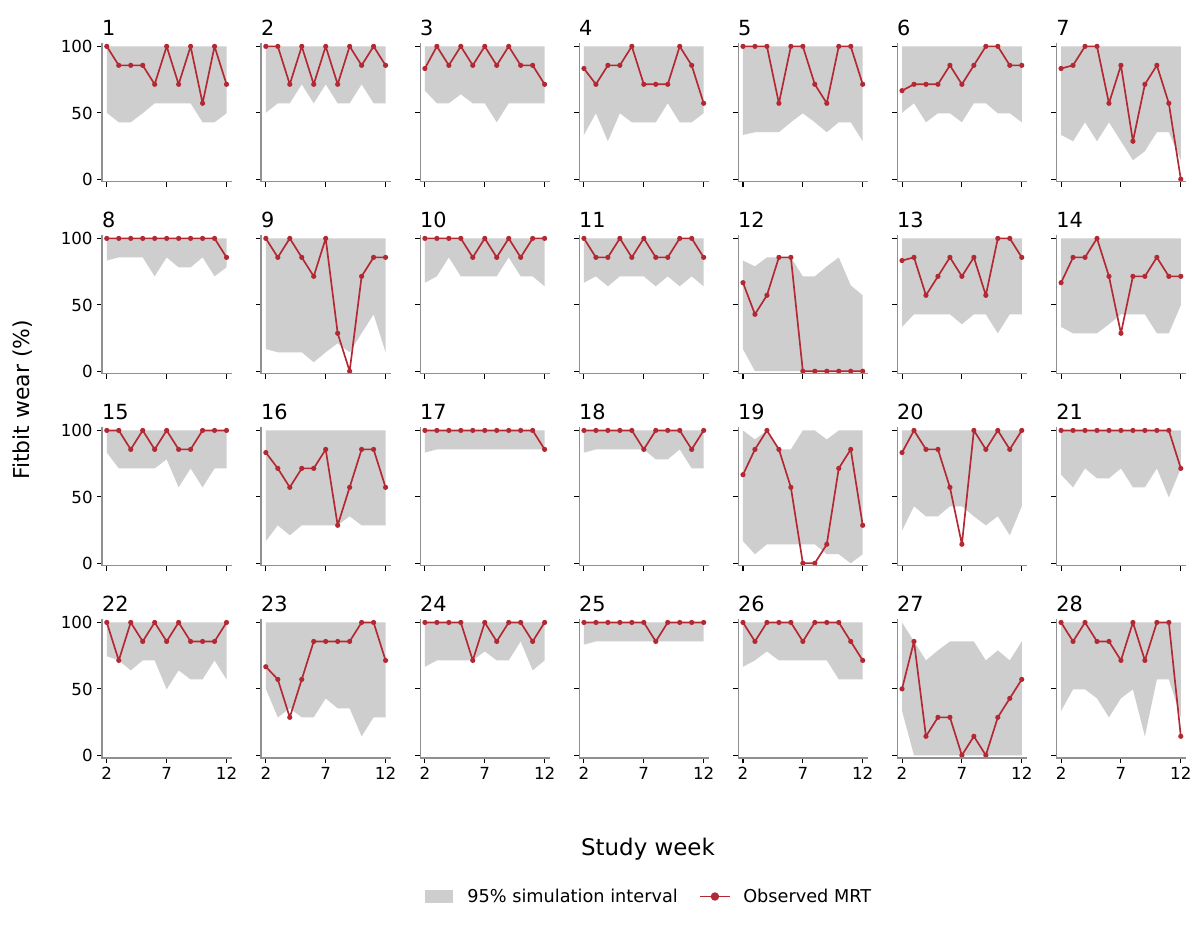}
\caption{Observed and simulated Fitbit-wear proportions by study week. Gray bands show pointwise 95\% simulation intervals and red points and lines show MRT proportions. Outcomes are aligned to their actual morning dates. All panels share the same percentage axis.}
\label{fig:fitbit-wear-trajectories}
\end{figure}

\begin{figure}[p]
\centering
\includegraphics[width=\textwidth]{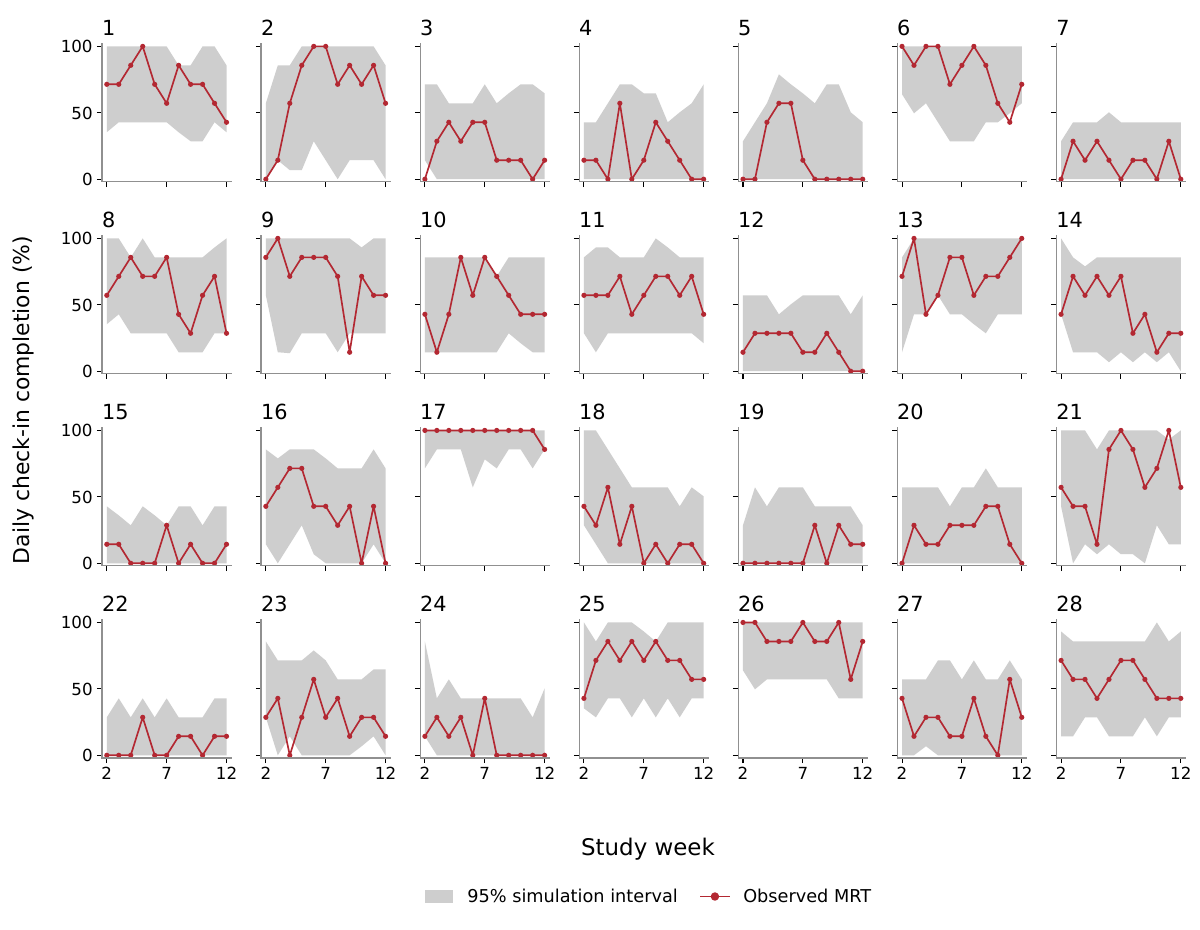}
\caption{Observed and simulated daily check-in completion proportions by study week. Each proportion uses seven daily opportunities. Gray bands show pointwise 95\% simulation intervals and red points and lines show observed proportions. Noncompletion remains an observed zero.}
\label{fig:daily-completion-trajectories}
\end{figure}

\begin{figure}[p]
\centering
\includegraphics[width=\textwidth]{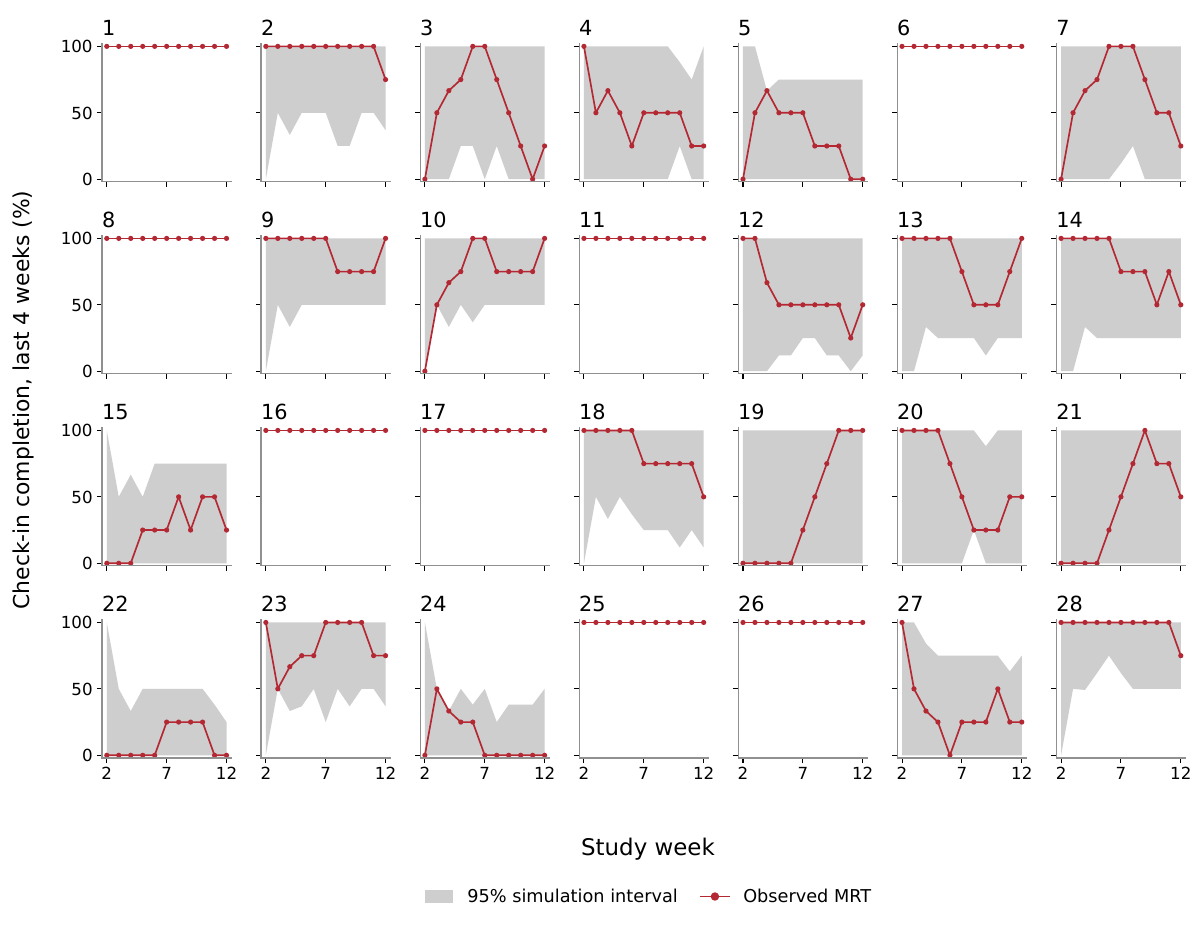}
\caption{Observed and simulated assessment-completion proportions over the most recent four study weeks. The initial points use the shorter available history. Gray bands show pointwise 95\% simulation intervals and red points and lines show the corresponding MRT summaries. Adjacent plotted points share weeks.}
\label{fig:weekly-completion-trajectories}
\end{figure}

\begin{figure}[p]
\centering
\includegraphics[width=\textwidth]{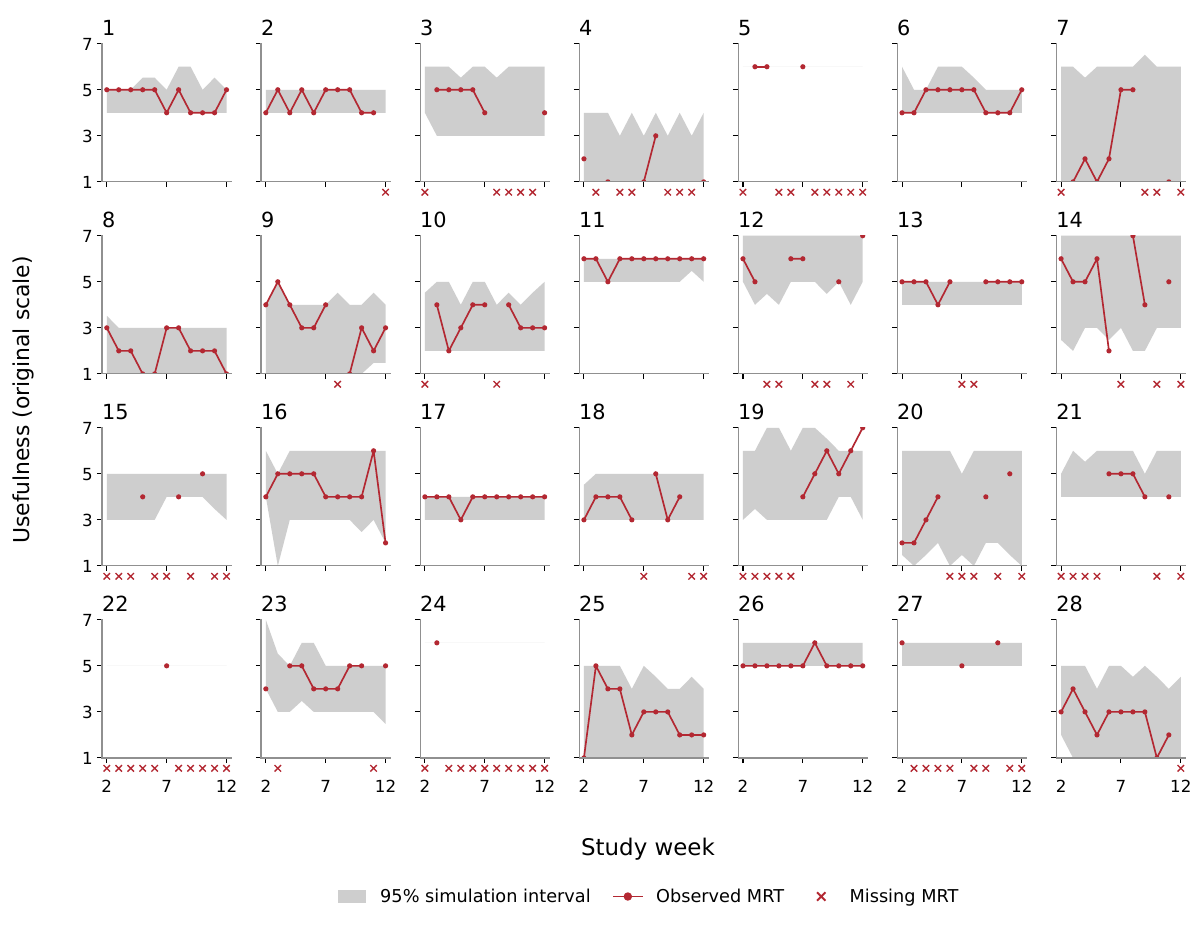}
\caption{Observed and simulated usefulness responses. Gray bands show pointwise 95\% intervals from all generated responses and red points show MRT responses. Missing assessments remain gaps and are marked by crosses below the panels. All panels share the original 1--7 scale.}
\label{fig:usefulness-trajectories}
\end{figure}

\begin{figure}[p]
\centering
\includegraphics[width=\textwidth]{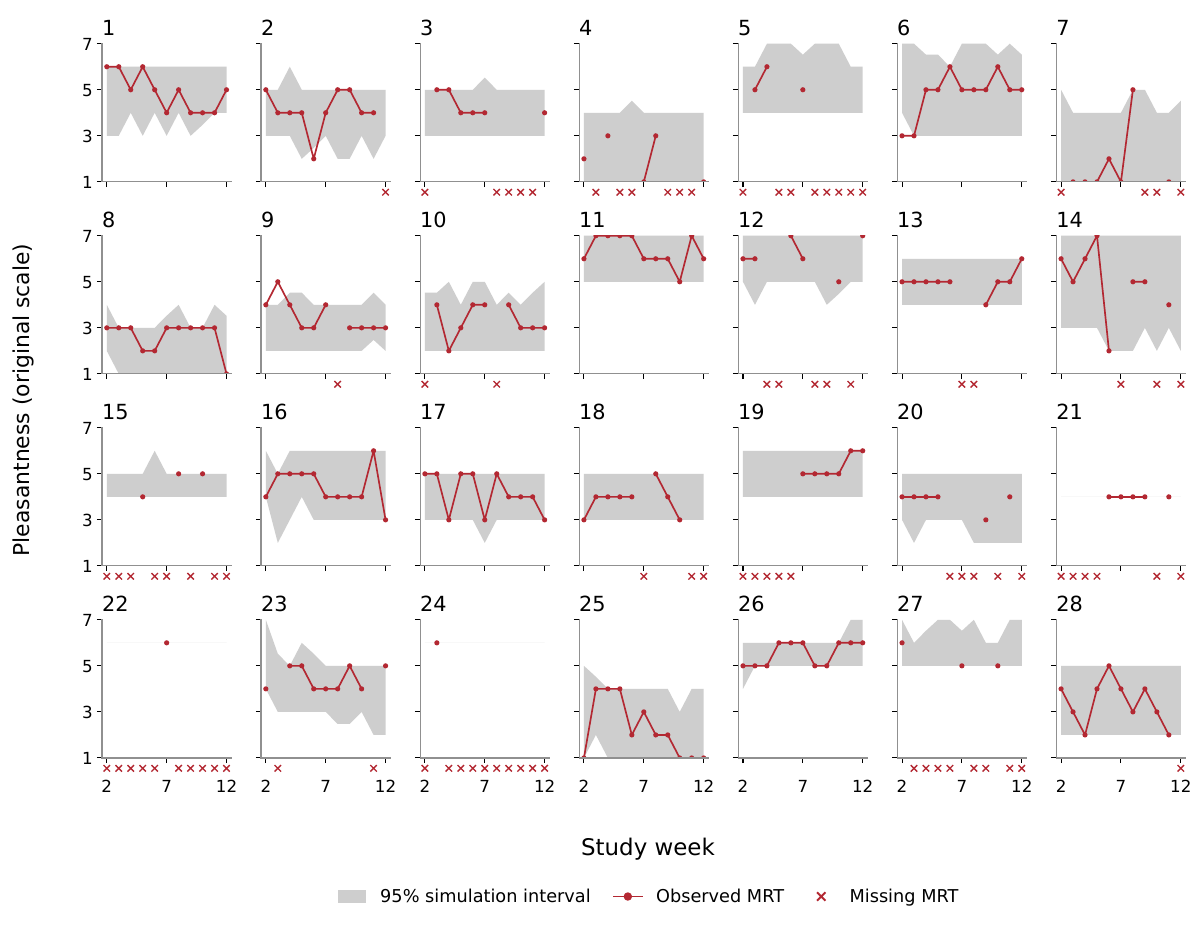}
\caption{Observed and simulated pleasantness responses. Gray bands show pointwise 95\% intervals from all generated responses and red points show MRT responses. Missing assessments remain gaps and are marked by crosses below the panels. All panels share the original 1--7 scale.}
\label{fig:pleasantness-trajectories}
\end{figure}

\paragraph{Supporting trajectory reproduction.} The remaining trajectory diagnostics provide complementary checks on the observation and engagement processes. Figure~\ref{fig:short-cae-trajectories} shows that short CAE largely follows the participant-level patterns reproduced for full CAE, as expected from its conditional generation model. Figure~\ref{fig:positive-pageview-trajectories} shows that the simulator retains substantial heterogeneity in positive page-view intensity, although these comparisons are less informative for participants with few positive page-view occasions. Figures~\ref{fig:weekly-completion-trajectories}--\ref{fig:pleasantness-trajectories} similarly preserve broad differences across participants in weekly assessment completion, usefulness, and pleasantness. As with the main processes, agreement is stronger for participant-specific levels and broad longitudinal patterns than for every local change over time.

\begin{figure}[p]
\centering
\includegraphics[width=\textwidth]{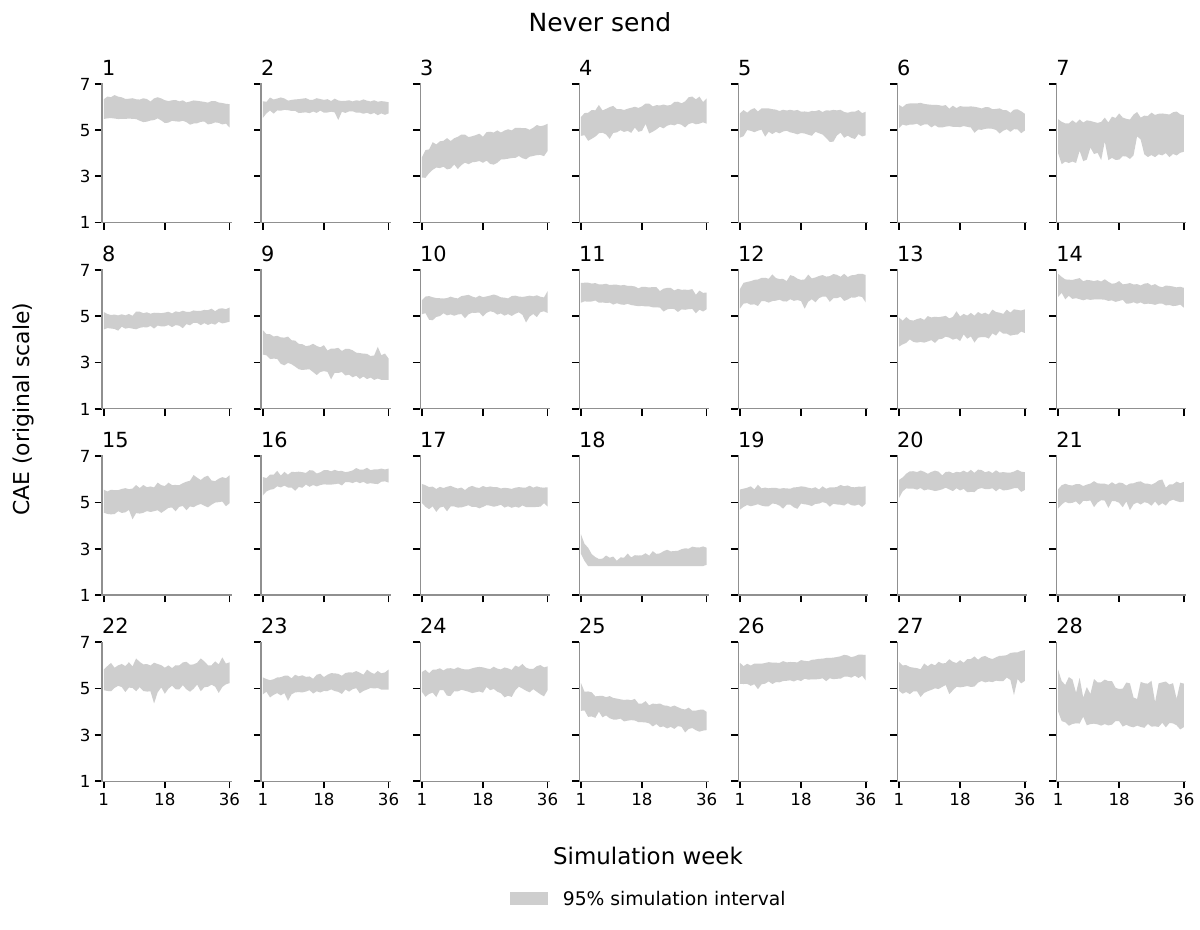}
\caption{Pointwise 95\% simulation intervals for CAE across 36 weeks under the never-send policy. Each panel represents one participant and uses 100 trajectories. Sunday suggestions are fixed at zero. No observed MRT trajectory is superimposed on this target-schedule assessment.}
\label{fig:stability-cae-never}
\end{figure}

\begin{figure}[p]
\centering
\includegraphics[width=\textwidth]{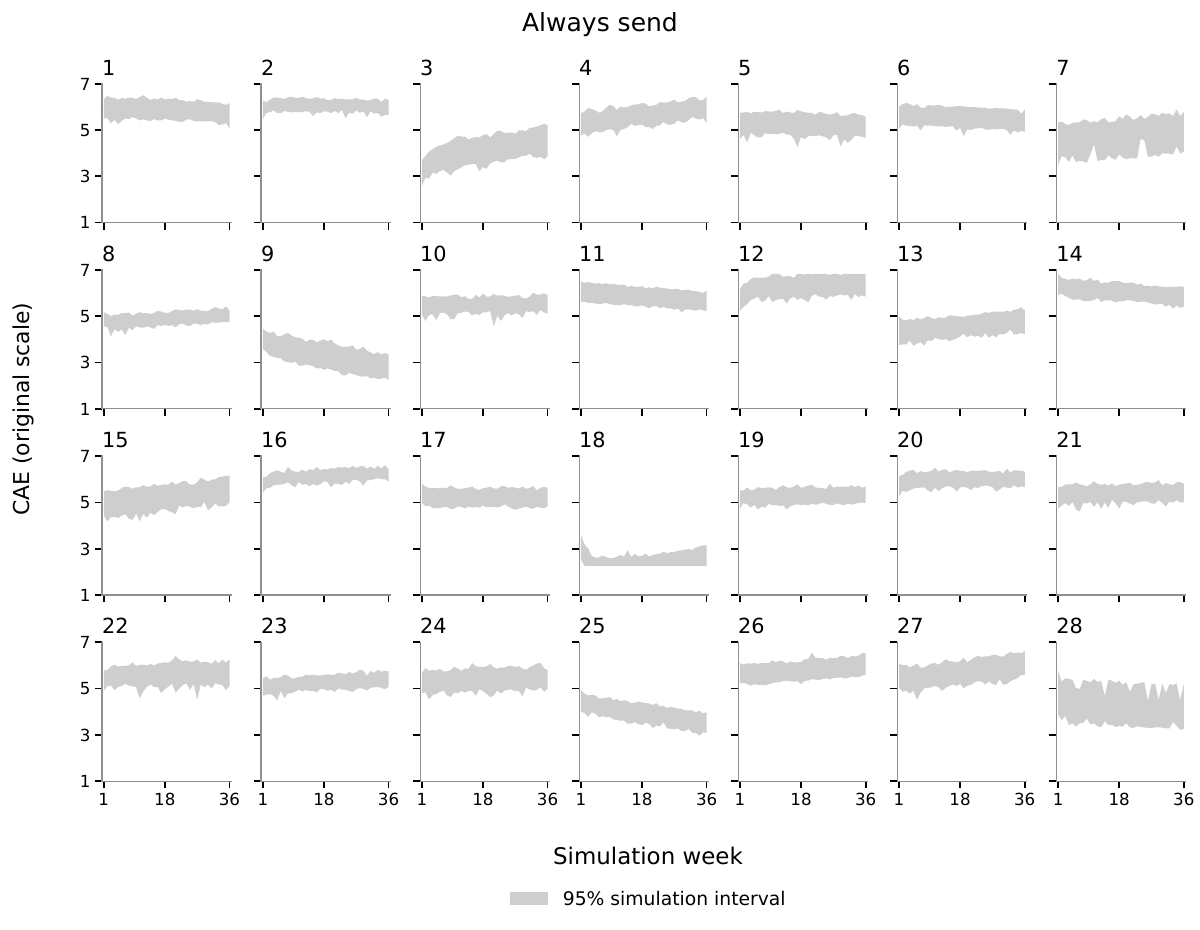}
\caption{Pointwise 95\% simulation intervals for CAE across 36 weeks under the always-send policy. Each panel represents one participant and uses 100 trajectories. Sunday suggestions are fixed at zero. No observed MRT trajectory is superimposed on this target-schedule assessment.}
\label{fig:stability-cae-always}
\end{figure}

\begin{figure}[p]
\centering
\includegraphics[width=\textwidth]{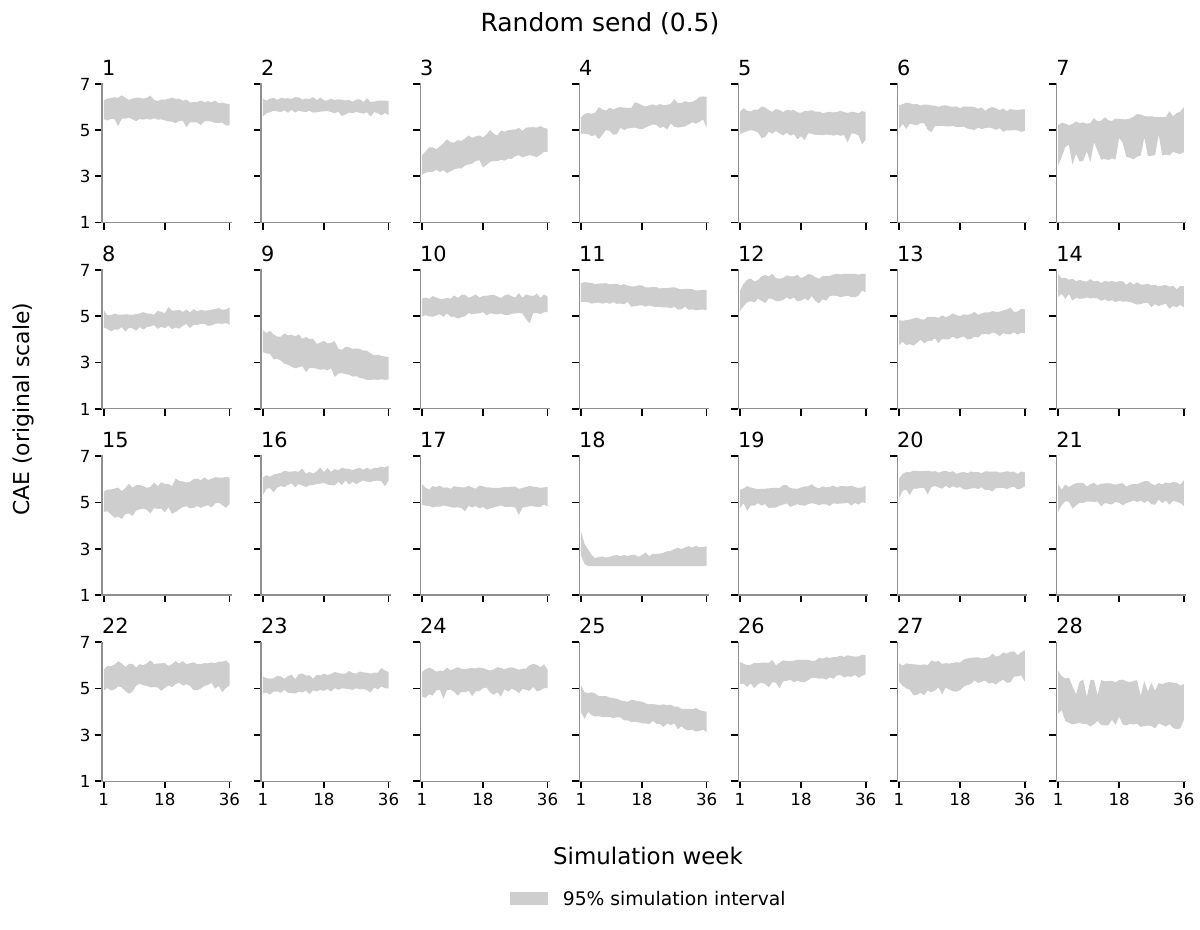}
\caption{Pointwise 95\% simulation intervals for CAE across 36 weeks under the random-send policy. Each panel represents one participant and uses 100 trajectories. Sunday suggestions are fixed at zero. No observed MRT trajectory is superimposed on this target-schedule assessment.}
\label{fig:stability-cae-random}
\end{figure}

\paragraph{Long-horizon stability.}
Because the candidate RL algorithms are evaluated over the planned 36-week study horizon, we also examine simulator behavior beyond the MRT observation period. We generate 100 trajectories for each of the 28 participant models under never-send, always-send, and Bernoulli$(0.5)$ suggestion policies, yielding 8,400 trajectories in total. Figures~\ref{fig:stability-cae-never}--\ref{fig:stability-cae-random} show the corresponding 36-week CAE trajectories under the three policies. All 8,400 trajectories complete with finite generated outcomes. Some participant-specific boundary concentration remains, including latent perceived utility for participant 18 and low FourSC values for participant 22. Thus, the simulator remains numerically stable over the planned study horizon, while retaining policy- and participant-specific differences in long-run behavior.

\clearpage

\clearpage

\clearpage

\clearpage

\begin{table}[p]
\centering
\caption{Participant-specific Bernoulli model-fit diagnostics.}
\label{tab:bernoulli-fit}
\fontsize{8}{9.5}\selectfont
\setlength{\tabcolsep}{2pt}
\renewcommand{\arraystretch}{0.98}

\begin{tabular*}{\textwidth}{@{\extracolsep{\fill}}c*{12}{r}@{}}
\toprule
Participant
& \multicolumn{4}{c}{Page-view occurrence $O$}
& \multicolumn{4}{c}{Fitbit wear $M^{E,FW}$}
& \multicolumn{4}{c}{Daily check-in $M^{E,PJ}$} \\
\cmidrule(lr){2-5}
\cmidrule(lr){6-9}
\cmidrule(lr){10-13}
& $n$ & $\bar z$ & $\bar p$ & BS
& $n$ & $\bar z$ & $\bar p$ & BS
& $n$ & $\bar z$ & $\bar p$ & BS \\
\midrule
1  & 154 & 0.416 & 0.418 & 0.234 & 76 & 0.842 & 0.847 & 0.129 & 77 & 0.714 & 0.717 & 0.168 \\
2  & 154 & 0.377 & 0.375 & 0.224 & 76 & 0.895 & 0.896 & 0.089 & 77 & 0.662 & 0.662 & 0.131 \\
3  & 154 & 0.026 & 0.029 & 0.023 & 77 & 0.883 & 0.879 & 0.086 & 77 & 0.221 & 0.214 & 0.146 \\
4  & 154 & 0.292 & 0.288 & 0.136 & 76 & 0.803 & 0.804 & 0.151 & 77 & 0.169 & 0.187 & 0.096 \\
5  & 154 & 0.130 & 0.130 & 0.111 & 76 & 0.868 & 0.866 & 0.083 & 77 & 0.156 & 0.157 & 0.087 \\
6  & 154 & 0.091 & 0.094 & 0.068 & 76 & 0.816 & 0.817 & 0.137 & 77 & 0.818 & 0.821 & 0.123 \\
7  & 154 & 0.117 & 0.112 & 0.097 & 76 & 0.684 & 0.682 & 0.178 & 77 & 0.130 & 0.124 & 0.102 \\
8  & 154 & 0.468 & 0.468 & 0.180 & 77 & 0.974 & 0.976 & 0.025 & 77 & 0.610 & 0.615 & 0.194 \\
9  & 154 & 0.429 & 0.427 & 0.184 & 77 & 0.740 & 0.744 & 0.122 & 77 & 0.714 & 0.716 & 0.154 \\
10 & 154 & 0.532 & 0.539 & 0.204 & 76 & 0.961 & 0.961 & 0.034 & 77 & 0.532 & 0.520 & 0.221 \\
11 & 154 & 0.468 & 0.469 & 0.135 & 76 & 0.921 & 0.923 & 0.068 & 77 & 0.597 & 0.597 & 0.211 \\
12 & 154 & 0.325 & 0.314 & 0.177 & 77 & 0.299 & 0.296 & 0.091 & 77 & 0.182 & 0.167 & 0.133 \\
13 & 154 & 0.318 & 0.317 & 0.171 & 77 & 0.792 & 0.792 & 0.154 & 77 & 0.753 & 0.762 & 0.165 \\
14 & 154 & 0.338 & 0.337 & 0.212 & 77 & 0.740 & 0.747 & 0.183 & 77 & 0.468 & 0.465 & 0.209 \\
15 & 154 & 0.214 & 0.208 & 0.157 & 76 & 0.947 & 0.952 & 0.037 & 77 & 0.078 & 0.090 & 0.071 \\
16 & 154 & 0.208 & 0.209 & 0.149 & 77 & 0.688 & 0.697 & 0.189 & 77 & 0.403 & 0.406 & 0.216 \\
17 & 154 & 0.636 & 0.631 & 0.082 & 76 & 0.987 & 0.987 & 0.012 & 77 & 0.987 & 0.980 & 0.009 \\
18 & 154 & 0.188 & 0.186 & 0.123 & 76 & 0.974 & 0.973 & 0.024 & 77 & 0.208 & 0.209 & 0.140 \\
19 & 154 & 0.084 & 0.075 & 0.071 & 77 & 0.532 & 0.537 & 0.197 & 77 & 0.078 & 0.083 & 0.063 \\
20 & 154 & 0.273 & 0.264 & 0.183 & 76 & 0.816 & 0.824 & 0.113 & 77 & 0.221 & 0.212 & 0.151 \\
21 & 154 & 0.013 & 0.011 & 0.012 & 77 & 0.961 & 0.958 & 0.019 & 77 & 0.649 & 0.646 & 0.159 \\
22 & 154 & 0.032 & 0.032 & 0.028 & 76 & 0.908 & 0.908 & 0.071 & 77 & 0.078 & 0.078 & 0.064 \\
23 & 154 & 0.149 & 0.150 & 0.121 & 76 & 0.750 & 0.749 & 0.166 & 77 & 0.286 & 0.288 & 0.186 \\
24 & 154 & 0.188 & 0.193 & 0.146 & 77 & 0.948 & 0.947 & 0.038 & 77 & 0.117 & 0.112 & 0.082 \\
25 & 154 & 0.571 & 0.571 & 0.164 & 76 & 0.987 & 0.990 & 0.012 & 77 & 0.701 & 0.701 & 0.200 \\
26 & 154 & 0.500 & 0.499 & 0.166 & 76 & 0.934 & 0.935 & 0.054 & 77 & 0.883 & 0.882 & 0.094 \\
27 & 154 & 0.299 & 0.306 & 0.154 & 76 & 0.316 & 0.306 & 0.144 & 77 & 0.260 & 0.273 & 0.169 \\
28 & 154 & 0.286 & 0.286 & 0.199 & 76 & 0.829 & 0.834 & 0.114 & 77 & 0.558 & 0.560 & 0.244 \\
\bottomrule
\end{tabular*}

\par\vspace{1.5ex}

\begin{tabular*}{\textwidth}{@{\extracolsep{\fill}}c*{12}{r}@{}}
\toprule
Participant
& \multicolumn{4}{c}{Weekly check-in $J^{\mathrm{week}}$}
& \multicolumn{4}{c}{Active status $C^{AD}$}
& \multicolumn{4}{c}{Suggestion interaction $C^{AS}$} \\
\cmidrule(lr){2-5}
\cmidrule(lr){6-9}
\cmidrule(lr){10-13}
& $n$ & $\bar z$ & $\bar p$ & BS
& $n$ & $\bar z$ & $\bar p$ & BS
& $n$ & $\bar z$ & $\bar p$ & BS \\
\midrule
1  & 11 & 1.000 & 0.995 & 0.000 & 12 & 0.583 & 0.555 & 0.232 & 75  & 0.200 & 0.221 & 0.147 \\
2  & 11 & 1.000 & 0.903 & 0.010 & 13 & 0.923 & 0.812 & 0.081 & 87  & 0.276 & 0.297 & 0.122 \\
3  & 11 & 0.545 & 0.561 & 0.229 & 13 & 0.538 & 0.512 & 0.246 & 70  & 0.029 & 0.062 & 0.029 \\
4  & 11 & 0.455 & 0.492 & 0.247 & 7  & 1.000 & 1.000 & 0.000 & 82  & 0.220 & 0.238 & 0.149 \\
5  & 11 & 0.273 & 0.254 & 0.194 & 10 & 0.800 & 0.734 & 0.161 & 67  & 0.119 & 0.152 & 0.097 \\
6  & 11 & 1.000 & 0.998 & 0.000 & 6  & 1.000 & 1.000 & 0.000 & 81  & 0.296 & 0.315 & 0.184 \\
7  & 11 & 0.636 & 0.605 & 0.203 & 2  & 1.000 & 1.000 & 0.000 & 77  & 0.143 & 0.179 & 0.124 \\
8  & 11 & 1.000 & 0.995 & 0.000 & 15 & 0.800 & 0.737 & 0.140 & 75  & 0.813 & 0.809 & 0.126 \\
9  & 11 & 0.909 & 0.920 & 0.079 & 14 & 0.429 & 0.470 & 0.182 & 69  & 0.681 & 0.687 & 0.172 \\
10 & 11 & 0.818 & 0.844 & 0.113 & 18 & 0.667 & 0.593 & 0.229 & 91  & 0.615 & 0.616 & 0.193 \\
11 & 11 & 1.000 & 0.985 & 0.000 & 36 & 0.639 & 0.607 & 0.229 & 73  & 0.890 & 0.876 & 0.089 \\
12 & 11 & 0.545 & 0.586 & 0.248 & 14 & 0.357 & 0.390 & 0.221 & 83  & 0.434 & 0.446 & 0.211 \\
13 & 11 & 0.818 & 0.831 & 0.147 & 34 & 0.529 & 0.520 & 0.231 & 73  & 0.658 & 0.663 & 0.150 \\
14 & 11 & 0.818 & 0.756 & 0.152 & 24 & 0.917 & 0.871 & 0.079 & 66  & 0.500 & 0.510 & 0.215 \\
15 & 11 & 0.273 & 0.238 & 0.174 & 2  & 0.500 & 0.478 & 0.223 & 72  & 0.264 & 0.307 & 0.192 \\
16 & 11 & 1.000 & 0.989 & 0.000 & 23 & 0.348 & 0.374 & 0.210 & 78  & 0.449 & 0.465 & 0.241 \\
17 & 11 & 1.000 & 0.994 & 0.000 & 74 & 0.649 & 0.641 & 0.222 & 81  & 1.000 & 1.000 & 0.000 \\
18 & 11 & 0.818 & 0.762 & 0.129 & 6  & 0.333 & 0.385 & 0.217 & 70  & 0.300 & 0.311 & 0.194 \\
19 & 11 & 0.455 & 0.469 & 0.137 & 6  & 0.833 & 0.692 & 0.156 & 74  & 0.095 & 0.122 & 0.086 \\
20 & 11 & 0.545 & 0.493 & 0.252 & 13 & 0.615 & 0.562 & 0.239 & 67  & 0.343 & 0.365 & 0.218 \\
21 & 11 & 0.545 & 0.512 & 0.175 & 50 & 0.660 & 0.644 & 0.222 & 85  & 0.518 & 0.525 & 0.139 \\
22 & 11 & 0.091 & 0.082 & 0.082 & 3  & 0.667 & 0.542 & 0.236 & 98  & 0.051 & 0.075 & 0.049 \\
23 & 11 & 0.818 & 0.833 & 0.138 & 12 & 1.000 & 1.000 & 0.000 & 110 & 0.218 & 0.230 & 0.156 \\
24 & 11 & 0.091 & 0.088 & 0.081 & 6  & 0.667 & 0.603 & 0.202 & 72  & 0.236 & 0.257 & 0.161 \\
25 & 11 & 1.000 & 0.991 & 0.000 & 25 & 0.280 & 0.304 & 0.193 & 78  & 0.897 & 0.881 & 0.093 \\
26 & 11 & 1.000 & 0.992 & 0.000 & 20 & 0.650 & 0.623 & 0.220 & 50  & 0.980 & 0.934 & 0.021 \\
27 & 11 & 0.364 & 0.340 & 0.228 & 13 & 0.769 & 0.722 & 0.172 & 78  & 0.474 & 0.485 & 0.234 \\
28 & 11 & 1.000 & 0.904 & 0.009 & 21 & 0.714 & 0.675 & 0.204 & 108 & 0.509 & 0.520 & 0.209 \\
\bottomrule
\end{tabular*}
\end{table}

\clearpage

\begin{table}[p]
\centering
\caption{Participant-specific fit diagnostics for prior two-hour
steps and positive page-view intensity.}
\label{tab:auxiliary-decision-fit}
\small
\setlength{\tabcolsep}{4pt}
\renewcommand{\arraystretch}{1.12}
\begin{tabular*}{\textwidth}{@{\extracolsep{\fill}}c*{6}{r}@{}}
\toprule
Participant
& \multicolumn{3}{c}{Prior two-hour steps}
& \multicolumn{3}{c}{Positive page-view intensity} \\
\cmidrule(lr){2-4}\cmidrule(lr){5-7}
& $n$ & Mean residual & RMSE
& $n$ & Mean residual & RMSE \\
\midrule
1  & 134 &  0.005 & 0.808 & 64 &  0.002 & 0.814 \\
2  & 141 &  0.019 & 1.266 & 58 &  0.008 & 0.848 \\
3  & 139 & -0.012 & 0.813 &  4 &  0.000 & 0.002 \\
4  & 132 &  0.008 & 0.670 & 45 &  0.017 & 0.574 \\
5  & 132 & -0.004 & 0.927 & 20 & -0.020 & 0.694 \\
6  & 125 & -0.004 & 1.079 & 14 &  0.002 & 0.332 \\
7  & 113 & -0.002 & 0.821 & 18 &  0.000 & 0.519 \\
8  & 153 &  0.007 & 0.746 & 72 &  0.001 & 0.616 \\
9  & 118 &  0.000 & 0.871 & 66 &  0.003 & 0.934 \\
10 & 150 &  0.007 & 0.778 & 82 & -0.006 & 0.884 \\
11 & 106 &  0.012 & 0.702 & 72 & -0.004 & 0.694 \\
12 &  51 & -0.067 & 1.131 & 50 & -0.007 & 0.555 \\
13 & 136 & -0.002 & 0.858 & 49 &  0.003 & 0.628 \\
14 & 122 &  0.018 & 0.688 & 52 &  0.007 & 0.822 \\
15 & 144 & -0.007 & 1.302 & 33 &  0.032 & 0.589 \\
16 & 114 & -0.004 & 0.618 & 32 & -0.014 & 0.708 \\
17 & 143 & -0.006 & 0.759 & 98 &  0.007 & 0.627 \\
18 & 152 &  0.008 & 0.894 & 29 &  0.001 & 1.024 \\
19 &  87 &  0.010 & 1.069 & 13 &  0.000 & 0.826 \\
20 & 127 & -0.009 & 0.970 & 42 & -0.026 & 0.793 \\
21 & 149 & -0.011 & 1.038 &  2 &  0.000 & 0.000 \\
22 &  70 & -0.009 & 0.528 &  5 &  0.000 & 0.565 \\
23 & 119 &  0.015 & 0.869 & 23 &  0.001 & 0.561 \\
24 & 145 &  0.009 & 0.921 & 29 &  0.017 & 0.811 \\
25 & 152 &  0.007 & 0.864 & 88 &  0.004 & 0.873 \\
26 & 145 &  0.012 & 0.597 & 77 & -0.007 & 0.825 \\
27 &  89 &  0.038 & 0.890 & 46 &  0.034 & 1.208 \\
28 & 132 &  0.004 & 0.617 & 44 &  0.001 & 0.566 \\
\bottomrule
\end{tabular*}
\end{table}

\clearpage

\begin{table}[p]
\centering
\caption{Participant-specific model-fit diagnostics for the two
perceived-utility items.}
\label{tab:utility-items-fit}
\small
\setlength{\tabcolsep}{4pt}
\renewcommand{\arraystretch}{1.12}
\begin{tabular*}{\textwidth}{@{\extracolsep{\fill}}c*{6}{r}@{}}
\toprule
Participant
& \multicolumn{3}{c}{Helpfulness $U_{w,1}$}
& \multicolumn{3}{c}{Pleasantness $U_{w,2}$} \\
\cmidrule(lr){2-4}\cmidrule(lr){5-7}
& $n$ & Mean residual & RMSE
& $n$ & Mean residual & RMSE \\
\midrule
1  & 11 & -0.009 & 0.078 & 11 &  0.001 & 0.007 \\
2  & 10 & -0.006 & 0.167 & 10 &  0.032 & 0.216 \\
3  &  6 &  0.000 & 0.009 &  6 &  0.008 & 0.135 \\
4  &  5 & -0.004 & 0.216 &  5 &  0.019 & 0.208 \\
5  &  3 &  0.000 & 0.000 &  3 &  0.000 & 0.002 \\
6  & 11 &  0.000 & 0.010 & 11 & -0.022 & 0.285 \\
7  &  7 &  0.000 & 0.426 &  7 &  0.000 & 0.211 \\
8  & 11 & -0.019 & 0.159 & 11 &  0.000 & 0.007 \\
9  & 10 &  0.000 & 0.004 & 10 & -0.003 & 0.124 \\
10 &  9 &  0.000 & 0.002 &  9 &  0.000 & 0.002 \\
11 & 11 &  0.001 & 0.095 & 11 &  0.009 & 0.215 \\
12 &  6 &  0.020 & 0.199 &  6 &  0.041 & 0.232 \\
13 &  9 &  0.007 & 0.089 &  9 & -0.012 & 0.155 \\
14 &  8 &  0.005 & 0.420 &  8 & -0.012 & 0.458 \\
15 &  3 &  0.000 & 0.002 &  3 &  0.000 & 0.137 \\
16 & 11 &  0.001 & 0.012 & 11 & -0.005 & 0.054 \\
17 & 11 & -0.002 & 0.091 & 11 & -0.001 & 0.004 \\
18 &  8 & -0.013 & 0.218 &  8 & -0.015 & 0.191 \\
19 &  6 &  0.010 & 0.099 &  6 &  0.004 & 0.090 \\
20 &  6 & -0.004 & 0.357 &  6 &  0.000 & 0.005 \\
21 &  5 &  0.002 & 0.007 &  5 &  0.000 & 0.000 \\
22 &  1 &  0.000 & 0.000 &  1 &  0.000 & 0.000 \\
23 &  9 &  0.000 & 0.094 &  9 &  0.000 & 0.013 \\
24 &  1 &  0.001 & 0.001 &  1 &  0.000 & 0.000 \\
25 & 11 & -0.016 & 0.361 & 11 &  0.011 & 0.356 \\
26 & 11 &  0.000 & 0.095 & 11 &  0.001 & 0.155 \\
27 &  3 &  0.059 & 0.162 &  3 & -0.006 & 0.011 \\
28 & 10 &  0.003 & 0.137 & 10 &  0.022 & 0.297 \\
\bottomrule
\end{tabular*}
\end{table}

\clearpage

\begin{table}[p]
\centering
\caption{Participant-specific one-week lagged residual correlations.}
\label{tab:weekly-residual-correlation}
\fontsize{9}{11}\selectfont
\setlength{\tabcolsep}{3pt}
\renewcommand{\arraystretch}{1.12}
\begin{tabular*}{\textwidth}{@{\extracolsep{\fill}}c*{10}{r}@{}}
\toprule
Participant
& \multicolumn{2}{c}{CAE}
& \multicolumn{2}{c}{Short CAE}
& \multicolumn{2}{c}{$U_{w,1}$}
& \multicolumn{2}{c}{$U_{w,2}$}
& \multicolumn{2}{c}{$J_w$} \\
\cmidrule(lr){2-3}
\cmidrule(lr){4-5}
\cmidrule(lr){6-7}
\cmidrule(lr){8-9}
\cmidrule(lr){10-11}
& $N$ & $\widehat\rho_1$
& $N$ & $\widehat\rho_1$
& $N$ & $\widehat\rho_1$
& $N$ & $\widehat\rho_1$
& $N$ & $\widehat\rho_1$ \\
\midrule
1  & 10 & -0.111 & 10 &  0.113 & 10 & -0.549 & 10 & -0.009 & 10 &  0.167 \\
2  &  9 & -0.723 &  9 &    --- &  9 & -0.305 &  9 &  0.195 & 10 &  0.250 \\
3  &  4 & -0.152 &  4 & -0.789 &  4 & -0.284 &  4 &  0.197 & 10 &  0.370 \\
4  &  1 &    --- &  1 &    --- &  1 &    --- &  1 &    --- & 10 & -0.306 \\
5  &  1 &    --- &  1 &    --- &  1 &    --- &  1 &    --- & 10 &  0.002 \\
6  & 10 & -0.261 & 10 &  0.462 & 10 & -0.263 & 10 &  0.528 & 10 &    --- \\
7  &  5 & -0.372 &  5 & -0.319 &  5 & -0.181 &  5 &  0.055 & 10 &  0.045 \\
8  & 10 & -0.655 & 10 &  0.041 & 10 &  0.113 & 10 &  0.161 & 10 &    --- \\
9  &  8 & -0.414 &  8 & -0.210 &  8 & -0.241 &  8 & -0.118 & 10 & -0.143 \\
10 &  7 & -0.175 &  7 &  0.036 &  7 & -0.177 &  7 & -0.177 & 10 &  0.127 \\
11 & 10 & -0.350 & 10 & -0.320 & 10 & -0.084 & 10 &  0.097 & 10 & -0.449 \\
12 &  2 &    --- &  2 &    --- &  2 &    --- &  2 &    --- & 10 & -0.202 \\
13 &  7 & -0.670 &  7 &  0.371 &  7 & -0.257 &  7 & -0.002 & 10 &  0.379 \\
14 &  5 & -0.728 &  5 & -0.006 &  5 & -0.517 &  5 & -0.464 & 10 & -0.320 \\
15 &  0 &    --- &  0 &    --- &  0 &    --- &  0 &    --- & 10 & -0.525 \\
16 & 10 & -0.594 & 10 & -0.798 & 10 &  0.475 & 10 &  0.577 & 10 & -0.542 \\
17 & 10 & -0.017 & 10 & -0.544 & 10 & -0.105 & 10 & -0.429 & 10 & -0.249 \\
18 &  6 & -0.568 &  6 &  0.372 &  6 & -0.666 &  6 & -0.040 & 10 &  0.055 \\
19 &  5 & -0.687 &  5 & -0.237 &  5 & -0.131 &  5 &  0.180 & 10 &  0.641 \\
20 &  3 & -0.606 &  3 & -0.998 &  3 &  0.870 &  3 & -0.550 & 10 &  0.013 \\
21 &  3 & -0.317 &  3 & -0.915 &  3 & -0.419 &  3 &    --- & 10 &  0.629 \\
22 &  0 &    --- &  0 &    --- &  0 &    --- &  0 &    --- & 10 & -0.113 \\
23 &  6 & -0.390 &  6 & -0.147 &  6 &  0.515 &  6 & -0.428 & 10 & -0.241 \\
24 &  0 &    --- &  0 &    --- &  0 &    --- &  0 &    --- & 10 & -0.060 \\
25 & 10 & -0.662 & 10 &  0.001 & 10 & -0.220 & 10 &  0.181 & 10 &  0.826 \\
26 & 10 & -0.457 & 10 & -0.134 & 10 & -0.133 & 10 &  0.293 & 10 &  0.816 \\
27 &  0 &    --- &  0 &    --- &  0 &    --- &  0 &    --- & 10 & -0.335 \\
28 &  9 & -0.229 &  9 &  0.221 &  9 & -0.221 &  9 &  0.173 & 10 & -0.211 \\
\bottomrule
\end{tabular*}
\end{table}

\clearpage

\begin{table}[p]
\centering
\caption{Participant-specific lagged residual correlations for daily components.}
\label{tab:daily-residual-correlation}
\fontsize{8}{9.5}\selectfont
\setlength{\tabcolsep}{3pt}
\renewcommand{\arraystretch}{0.98}

\begin{tabular*}{\textwidth}{@{\extracolsep{\fill}}c*{8}{r}@{}}
\toprule
Participant
& \multicolumn{4}{c}{Anticipated affect}
& \multicolumn{4}{c}{Fitbit wear} \\
\cmidrule(lr){2-5}\cmidrule(lr){6-9}
& $N_1$ & $\widehat\rho_1$ & $N_7$ & $\widehat\rho_7$
& $N_1$ & $\widehat\rho_1$ & $N_7$ & $\widehat\rho_7$ \\
\midrule
1  & 39 & -0.111 & 39 &  0.026 & 75 &  0.059 & 69 &  0.020 \\
2  & 40 &  0.533 & 37 &  0.173 & 75 &  0.041 & 69 & -0.096 \\
3  &  4 &  0.995 &  5 &  0.612 & 76 &  0.025 & 70 &  0.040 \\
4  &  4 &  0.941 &  1 &    --- & 75 & -0.017 & 69 & -0.159 \\
5  &  4 & -0.997 &  3 &  0.243 & 75 &  0.312 & 69 & -0.021 \\
6  & 51 & -0.241 & 49 &  0.242 & 75 & -0.128 & 69 & -0.140 \\
7  &  2 &    --- &  1 &    --- & 75 & -0.014 & 69 &  0.153 \\
8  & 30 &  0.185 & 29 &  0.048 & 76 &  0.000 & 70 &  0.022 \\
9  & 39 & -0.011 & 39 & -0.061 & 76 &  0.048 & 70 &  0.074 \\
10 & 22 & -0.760 & 19 & -0.007 & 75 &  0.003 & 69 & -0.081 \\
11 & 27 & -0.012 & 33 & -0.158 & 75 & -0.035 & 69 & -0.082 \\
12 &  4 &  0.728 &  3 & -0.725 & 76 & -0.281 & 70 &  0.037 \\
13 & 44 &  0.207 & 39 &  0.171 & 76 &  0.068 & 70 &  0.020 \\
14 & 16 &  0.677 & 20 & -0.348 & 76 &  0.031 & 70 &  0.005 \\
15 &  0 &    --- &  0 &    --- & 75 & -0.074 & 69 &  0.168 \\
16 & 14 &  0.120 & 15 &  0.166 & 76 &  0.018 & 70 & -0.081 \\
17 & 75 & -0.040 & 69 &  0.087 & 75 &  0.088 & 69 &  0.001 \\
18 &  6 &  0.183 &  2 &    --- & 75 & -0.015 & 69 & -0.020 \\
19 &  0 &    --- &  1 &    --- & 76 & -0.084 & 70 &  0.016 \\
20 &  4 &  0.488 &  8 & -0.386 & 75 &  0.100 & 69 & -0.101 \\
21 & 37 &  0.065 & 31 & -0.259 & 76 &  0.554 & 70 &  0.035 \\
22 &  0 &    --- &  0 &    --- & 75 &  0.026 & 69 & -0.138 \\
23 &  6 &  0.186 &  5 &  0.693 & 75 &  0.079 & 69 &  0.065 \\
24 &  3 & -0.997 &  2 &    --- & 76 & -0.184 & 70 & -0.046 \\
25 & 37 & -0.220 & 35 & -0.090 & 75 & -0.026 & 69 & -0.026 \\
26 & 60 & -0.210 & 55 &  0.034 & 75 & -0.127 & 69 &  0.175 \\
27 &  5 & -0.183 &  4 &  0.654 & 75 &  0.025 & 69 &  0.099 \\
28 & 23 & -0.195 & 21 & -0.175 & 75 &  0.108 & 69 &  0.031 \\
\bottomrule
\end{tabular*}

\par\vspace{1.5ex}

\begin{tabular*}{\textwidth}{@{\extracolsep{\fill}}c*{8}{r}@{}}
\toprule
Participant
& \multicolumn{4}{c}{Daily check-in}
& \multicolumn{4}{c}{Active status} \\
\cmidrule(lr){2-5}\cmidrule(lr){6-9}
& $N_1$ & $\widehat\rho_1$ & $N_7$ & $\widehat\rho_7$
& $N_1$ & $\widehat\rho_1$ & $N_7$ & $\widehat\rho_7$ \\
\midrule
1  & 76 & -0.058 & 70 &  0.092 &  1 &    --- &  3 &  0.345 \\
2  & 76 & -0.120 & 70 & -0.118 &  2 &    --- &  1 &    --- \\
3  & 76 & -0.016 & 70 &  0.054 &  2 &    --- &  2 &    --- \\
4  & 76 & -0.064 & 70 & -0.070 &  1 &    --- &  0 &    --- \\
5  & 76 & -0.154 & 70 & -0.135 &  3 & -0.703 &  2 &    --- \\
6  & 76 & -0.142 & 70 &  0.246 &  0 &    --- &  1 &    --- \\
7  & 76 & -0.010 & 70 & -0.042 &  0 &    --- &  0 &    --- \\
8  & 76 &  0.008 & 70 & -0.031 &  3 & -1.000 &  3 &    --- \\
9  & 76 & -0.090 & 70 &  0.159 &  4 &  0.707 &  4 &  0.095 \\
10 & 76 & -0.036 & 70 &  0.006 &  5 & -0.389 &  3 &  0.916 \\
11 & 76 &  0.039 & 70 &  0.370 & 16 &  0.190 & 19 & -0.081 \\
12 & 76 & -0.015 & 70 &  0.052 &  4 &  0.535 &  3 & -0.999 \\
13 & 76 & -0.004 & 70 &  0.066 & 17 &  0.332 & 15 & -0.343 \\
14 & 76 & -0.086 & 70 &  0.219 &  5 & -0.361 &  8 & -0.294 \\
15 & 76 & -0.018 & 70 & -0.088 &  0 &    --- &  0 &    --- \\
16 & 76 & -0.035 & 70 &  0.101 &  8 & -0.209 &  7 & -0.386 \\
17 & 76 &  0.056 & 70 &  0.050 & 71 & -0.051 & 65 &  0.004 \\
18 & 76 &  0.041 & 70 & -0.164 &  1 &    --- &  0 &    --- \\
19 & 76 & -0.275 & 70 &  0.057 &  0 &    --- &  1 &    --- \\
20 & 76 & -0.022 & 70 &  0.228 &  2 &    --- &  3 & -0.493 \\
21 & 76 & -0.107 & 70 &  0.049 & 37 &  0.156 & 31 & -0.327 \\
22 & 76 & -0.110 & 70 & -0.069 &  0 &    --- &  0 &    --- \\
23 & 76 & -0.048 & 70 & -0.052 &  1 &    --- &  2 &    --- \\
24 & 76 &  0.075 & 70 &  0.139 &  2 &    --- &  0 &    --- \\
25 & 76 & -0.042 & 70 & -0.086 &  8 & -0.343 &  5 & -0.309 \\
26 & 76 & -0.027 & 70 &  0.097 &  5 & -0.612 & 10 & -0.093 \\
27 & 76 &  0.026 & 70 & -0.025 &  2 &    --- &  2 &    --- \\
28 & 76 & -0.013 & 70 & -0.033 &  7 & -0.371 &  6 &  0.639 \\
\bottomrule
\end{tabular*}
\end{table}

\clearpage

\begin{table}[p]
\centering
\caption{Participant-specific residual correlations between adjacent AM/PM occasions.}
\label{tab:ampm-residual-correlation}
\fontsize{8}{9.5}\selectfont
\setlength{\tabcolsep}{2.5pt}
\renewcommand{\arraystretch}{0.96}

\begin{tabular*}{\textwidth}{@{\extracolsep{\fill}}c*{10}{r}@{}}
\toprule
\multicolumn{11}{c}{AM $\rightarrow$ same-day PM} \\
\midrule
Participant
& \multicolumn{2}{c}{FourSC}
& \multicolumn{2}{c}{Prior 2h steps}
& \multicolumn{2}{c}{PV occurrence}
& \multicolumn{2}{c}{PV intensity}
& \multicolumn{2}{c}{Interaction} \\
\cmidrule(lr){2-3}\cmidrule(lr){4-5}
\cmidrule(lr){6-7}\cmidrule(lr){8-9}\cmidrule(lr){10-11}
& $N$ & $\widehat\rho$
& $N$ & $\widehat\rho$
& $N$ & $\widehat\rho$
& $N$ & $\widehat\rho$
& $N$ & $\widehat\rho$ \\
\midrule
1  & 61 &  0.110 & 60 & -0.014 & 77 & -0.052 & 15 &  0.085 & 20 &  0.241 \\
2  & 34 & -0.237 & 67 & -0.059 & 77 &  0.020 & 13 &  0.440 & 25 &  0.092 \\
3  &  0 &    --- & 67 & -0.050 & 77 & -0.331 &  0 &    --- & 16 &  1.000 \\
4  & 64 &  0.107 & 59 &  0.127 & 77 & -0.098 &  9 & -0.335 & 26 & -0.109 \\
5  & 58 &  0.103 & 63 &  0.091 & 77 &  0.018 &  2 &    --- & 15 & -0.153 \\
6  & 63 &  0.114 & 54 & -0.039 & 77 & -0.002 &  2 &    --- & 25 &  0.021 \\
7  & 50 &  0.013 & 52 & -0.133 & 77 & -0.051 &  1 &    --- & 20 & -0.175 \\
8  & 76 &  0.019 & 76 &  0.176 & 77 & -0.138 & 13 &  0.121 & 16 &  0.226 \\
9  & 59 & -0.036 & 55 & -0.273 & 77 & -0.096 & 15 &  0.340 & 16 &  0.325 \\
10 & 73 & -0.143 & 73 & -0.051 & 77 & -0.098 & 20 & -0.184 & 25 &  0.138 \\
11 & 19 & -0.185 & 34 &  0.201 & 77 &  0.001 & 18 &  0.438 & 20 & -0.190 \\
12 & 27 & -0.300 & 21 &  0.263 & 77 & -0.171 &  8 &  0.261 & 22 & -0.002 \\
13 & 65 & -0.031 & 61 &  0.297 & 77 & -0.033 & 12 & -0.456 & 21 & -0.332 \\
14 & 56 & -0.011 & 55 &  0.102 & 77 & -0.053 &  7 &  0.199 & 13 &  0.036 \\
15 & 67 &  0.111 & 69 &  0.065 & 77 & -0.192 &  2 &    --- & 17 & -0.075 \\
16 & 47 &  0.020 & 51 & -0.109 & 77 &  0.162 &  5 &  0.664 & 19 &  0.217 \\
17 & 71 & -0.039 & 67 &  0.075 & 77 & -0.253 & 29 &  0.076 & 23 &    --- \\
18 & 75 & -0.043 & 75 &  0.316 & 77 &  0.087 &  5 &  0.408 & 18 &  0.005 \\
19 & 25 & -0.069 & 36 &  0.560 & 77 &  0.060 &  1 &    --- & 20 & -0.046 \\
20 & 59 & -0.004 & 59 &  0.163 & 77 &  0.018 &  7 &  0.260 & 14 &  0.147 \\
21 &  0 &    --- & 74 & -0.110 & 77 & -0.101 &  0 &    --- & 25 & -0.099 \\
22 &  0 &    --- &  0 &    --- & 77 &  0.075 &  0 &    --- & 38 &  0.706 \\
23 & 48 & -0.022 & 52 &  0.111 & 77 & -0.042 &  1 &    --- & 55 &  0.194 \\
24 & 67 & -0.036 & 70 &  0.182 & 77 &  0.041 &  3 &  0.300 & 21 &  0.372 \\
25 & 74 & -0.036 & 75 & -0.088 & 77 & -0.192 & 24 & -0.235 & 20 & -0.067 \\
26 & 71 &  0.042 & 70 &  0.188 & 77 &  0.017 & 21 & -0.015 & 12 & -0.457 \\
27 & 34 &  0.308 & 24 &  0.564 & 77 &  0.013 &  7 &  0.290 & 18 & -0.069 \\
28 & 67 & -0.141 & 63 & -0.211 & 77 & -0.049 &  5 &  0.627 & 53 &  0.508 \\
\bottomrule
\end{tabular*}

\par\vspace{1.5ex}

\begin{tabular*}{\textwidth}{@{\extracolsep{\fill}}c*{10}{r}@{}}
\toprule
\multicolumn{11}{c}{PM $\rightarrow$ next-day AM} \\
\midrule
Participant
& \multicolumn{2}{c}{FourSC}
& \multicolumn{2}{c}{Prior 2h steps}
& \multicolumn{2}{c}{PV occurrence}
& \multicolumn{2}{c}{PV intensity}
& \multicolumn{2}{c}{Interaction} \\
\cmidrule(lr){2-3}\cmidrule(lr){4-5}
\cmidrule(lr){6-7}\cmidrule(lr){8-9}\cmidrule(lr){10-11}
& $N$ & $\widehat\rho$
& $N$ & $\widehat\rho$
& $N$ & $\widehat\rho$
& $N$ & $\widehat\rho$
& $N$ & $\widehat\rho$ \\
\midrule
1  & 60 &  0.002 & 59 &  0.005 & 76 &  0.122 & 17 & -0.091 & 21 & -0.210 \\
2  & 33 &  0.082 & 66 & -0.004 & 76 & -0.099 & 10 & -0.002 & 26 &  0.314 \\
3  &  0 &    --- & 64 &  0.023 & 76 & -0.024 &  0 &    --- & 16 & -0.136 \\
4  & 60 & -0.156 & 58 &  0.124 & 76 & -0.006 &  9 & -0.322 & 23 & -0.205 \\
5  & 53 &  0.049 & 60 &  0.001 & 76 & -0.085 &  1 &    --- & 15 & -0.545 \\
6  & 59 & -0.176 & 50 &  0.150 & 76 &  0.160 &  3 &  0.812 & 25 &  0.099 \\
7  & 45 & -0.028 & 47 &  0.308 & 76 &  0.100 &  2 &    --- & 18 & -0.132 \\
8  & 75 & -0.031 & 75 & -0.079 & 76 &  0.108 & 14 &  0.348 & 20 &  0.115 \\
9  & 56 & -0.053 & 53 &  0.199 & 76 &  0.032 & 19 &  0.248 & 17 &  0.452 \\
10 & 71 &  0.071 & 72 &  0.022 & 76 & -0.042 & 21 &  0.339 & 30 & -0.172 \\
11 & 16 &  0.054 & 31 & -0.198 & 76 & -0.127 & 15 & -0.626 & 17 &  0.228 \\
12 & 23 & -0.032 & 17 &  0.244 & 76 &  0.144 & 10 & -0.536 & 25 &  0.134 \\
13 & 64 & -0.036 & 59 & -0.178 & 76 & -0.058 &  9 &  0.384 & 17 & -0.217 \\
14 & 53 &  0.026 & 50 &  0.058 & 76 & -0.073 &  6 &  0.817 & 16 & -0.176 \\
15 & 64 & -0.059 & 66 & -0.075 & 76 &  0.166 &  6 &  0.271 & 18 &  0.385 \\
16 & 45 & -0.066 & 46 & -0.134 & 76 & -0.039 &  3 & -0.446 & 17 &  0.107 \\
17 & 69 & -0.018 & 66 &  0.224 & 76 &  0.111 & 30 & -0.115 & 19 &    --- \\
18 & 74 & -0.161 & 74 & -0.038 & 76 &  0.032 &  5 & -0.340 & 16 & -0.237 \\
19 & 18 & -0.038 & 31 &  0.358 & 76 & -0.162 &  0 &    --- & 20 & -0.082 \\
20 & 58 & -0.087 & 58 & -0.299 & 76 &  0.006 &  7 &  0.795 & 19 &  0.162 \\
21 &  0 &    --- & 73 & -0.112 & 76 & -0.199 &  0 &    --- & 20 &  0.185 \\
22 &  0 &    --- &  0 &    --- & 76 & -0.105 &  0 &    --- & 30 &  0.357 \\
23 & 44 &  0.043 & 51 &  0.169 & 76 &  0.052 &  2 &    --- & 42 &  0.113 \\
24 & 66 & -0.025 & 67 & -0.053 & 76 & -0.027 &  2 &    --- & 16 & -0.079 \\
25 & 73 &  0.029 & 74 &  0.101 & 76 & -0.119 & 26 &  0.059 & 23 & -0.110 \\
26 & 70 & -0.205 & 69 &  0.060 & 76 &  0.093 & 25 & -0.297 & 16 & -0.103 \\
27 & 33 &  0.220 & 24 & -0.028 & 76 &  0.040 &  6 & -0.278 & 21 & -0.077 \\
28 & 66 &  0.090 & 61 &  0.047 & 76 &  0.153 &  8 & -0.346 & 39 &  0.155 \\
\bottomrule
\end{tabular*}
\end{table}

\clearpage

\begin{table}[p]
\centering
\caption{Participant-specific residual correlations among continuous end-of-week assessments.}
\label{tab:weekly-joint-continuous}
\fontsize{9}{11}\selectfont
\setlength{\tabcolsep}{2pt}
\renewcommand{\arraystretch}{1.15}

\begin{tabular*}{\textwidth}{@{\extracolsep{\fill}}c*{12}{r}@{}}
\toprule
Participant
& \multicolumn{2}{c}{$Y,\widetilde Y$}
& \multicolumn{2}{c}{$Y,U_1$}
& \multicolumn{2}{c}{$Y,U_2$}
& \multicolumn{2}{c}{$\widetilde Y,U_1$}
& \multicolumn{2}{c}{$\widetilde Y,U_2$}
& \multicolumn{2}{c}{$U_1,U_2$} \\
\cmidrule(lr){2-3}\cmidrule(lr){4-5}\cmidrule(lr){6-7}
\cmidrule(lr){8-9}\cmidrule(lr){10-11}\cmidrule(lr){12-13}
& $N$ & $\widehat\rho$
& $N$ & $\widehat\rho$
& $N$ & $\widehat\rho$
& $N$ & $\widehat\rho$
& $N$ & $\widehat\rho$
& $N$ & $\widehat\rho$ \\
\midrule
1  & 11 &  0.392 & 11 &  0.599 & 11 & -0.144 & 11 &  0.334 & 11 &  0.107 & 11 & -0.717 \\
2  & 10 &    --- & 10 & -0.600 & 10 &  0.035 & 10 &    --- & 10 &    --- & 10 &  0.343 \\
3  &  6 & -0.633 &  6 &  0.193 &  6 & -0.493 &  6 &  0.435 &  6 & -0.022 &  6 & -0.794 \\
4  &  5 &  0.566 &  5 &  0.103 &  5 & -0.781 &  5 &  0.609 &  5 & -0.613 &  5 &  0.206 \\
5  &  3 & -0.395 &  3 &    --- &  3 & -0.974 &  3 &    --- &  3 &  0.174 &  3 &    --- \\
6  & 11 & -0.023 & 11 &  0.497 & 11 & -0.291 & 11 & -0.208 & 11 &  0.463 & 11 & -0.096 \\
7  &  7 &  0.368 &  7 & -0.345 &  7 & -0.227 &  7 & -0.258 &  7 & -0.136 &  7 &  0.082 \\
8  & 11 & -0.218 & 11 & -0.361 & 11 &  0.470 & 11 &  0.192 & 11 &  0.179 & 11 & -0.039 \\
9  & 10 &  0.409 & 10 &  0.308 & 10 & -0.127 & 10 & -0.143 & 10 & -0.381 & 10 & -0.139 \\
10 &  9 & -0.733 &  9 & -0.514 &  9 & -0.514 &  9 &  0.321 &  9 &  0.321 &  9 &  1.000 \\
11 & 11 &  0.454 & 11 & -0.024 & 11 & -0.108 & 11 & -0.187 & 11 & -0.117 & 11 & -0.322 \\
12 &  6 &  0.193 &  6 &  0.205 &  6 &  0.301 &  6 &  0.497 &  6 &  0.065 &  6 &  0.781 \\
13 &  9 & -0.020 &  9 &  0.839 &  9 &  0.234 &  9 &  0.239 &  9 & -0.050 &  9 & -0.231 \\
14 &  8 & -0.186 &  8 &  0.550 &  8 &  0.859 &  8 & -0.221 &  8 & -0.252 &  8 &  0.740 \\
15 &  3 & -0.148 &  3 &  0.936 &  3 &  0.989 &  3 &  0.209 &  3 & -0.293 &  3 &  0.874 \\
16 & 11 &  0.438 & 11 & -0.148 & 11 &  0.024 & 11 &  0.143 & 11 & -0.240 & 11 & -0.904 \\
17 & 11 & -0.156 & 11 &  0.165 & 11 &  0.211 & 11 & -0.353 & 11 & -0.361 & 11 &  0.164 \\
18 &  8 & -0.025 &  8 &  0.311 &  8 & -0.237 &  8 & -0.009 &  8 & -0.355 &  8 &  0.533 \\
19 &  6 &  0.309 &  6 &  0.128 &  6 & -0.403 &  6 & -0.868 &  6 &  0.032 &  6 & -0.245 \\
20 &  6 & -0.089 &  6 & -0.259 &  6 &  0.286 &  6 &  0.401 &  6 &  0.308 &  6 & -0.051 \\
21 &  5 & -0.673 &  5 & -0.022 &  5 &    --- &  5 &  0.679 &  5 &    --- &  5 &    --- \\
22 &  1 &    --- &  1 &    --- &  1 &    --- &  1 &    --- &  1 &    --- &  1 &    --- \\
23 &  9 &  0.481 &  9 &  0.045 &  9 & -0.050 &  9 &  0.392 &  9 &  0.246 &  9 & -0.480 \\
24 &  1 &    --- &  1 &    --- &  1 &    --- &  1 &    --- &  1 &    --- &  1 &    --- \\
25 & 11 &  0.094 & 11 & -0.160 & 11 & -0.067 & 11 &  0.072 & 11 & -0.020 & 11 &  0.925 \\
26 & 11 & -0.038 & 11 &  0.448 & 11 &  0.139 & 11 & -0.222 & 11 & -0.101 & 11 & -0.416 \\
27 &  3 & -0.069 &  3 &  0.648 &  3 & -0.519 &  3 &  0.715 &  3 & -0.817 &  3 & -0.987 \\
28 & 10 & -0.075 & 10 &  0.471 & 10 & -0.037 & 10 &  0.190 & 10 & -0.274 & 10 & -0.127 \\
\bottomrule
\end{tabular*}
\end{table}

\begin{table}[p]
\centering
\caption{Participant-specific residual associations between check-in completion and continuous assessments.}
\label{tab:weekly-joint-completion}
\small
\setlength{\tabcolsep}{2pt}
\renewcommand{\arraystretch}{1.15}

\begin{tabular*}{\textwidth}{@{\extracolsep{\fill}}c*{8}{r}@{}}
\toprule
Participant
& \multicolumn{2}{c}{$Y,J^{\mathrm{week}}$}
& \multicolumn{2}{c}{$\widetilde Y,J^{\mathrm{week}}$}
& \multicolumn{2}{c}{$U_1,J^{\mathrm{week}}$}
& \multicolumn{2}{c}{$U_2,J^{\mathrm{week}}$} \\
\cmidrule(lr){2-3}\cmidrule(lr){4-5}
\cmidrule(lr){6-7}\cmidrule(lr){8-9}
& $N$ & $\widehat\rho$
& $N$ & $\widehat\rho$
& $N$ & $\widehat\rho$
& $N$ & $\widehat\rho$ \\
\midrule
1  & 11 & -0.627 & 11 & -0.475 & 11 & -0.498 & 11 & -0.116 \\
2  & 10 &  0.631 & 10 &    --- & 10 & -0.130 & 10 &  0.401 \\
3  &  6 &  0.224 &  6 & -0.388 &  6 & -0.571 &  6 &  0.160 \\
4  &  5 & -0.654 &  5 & -0.365 &  5 &  0.327 &  5 &  0.882 \\
5  &  3 &  0.450 &  3 &  0.643 &  3 &    --- &  3 & -0.642 \\
6  & 11 &    --- & 11 &    --- & 11 &    --- & 11 &    --- \\
7  &  7 &  0.475 &  7 & -0.468 &  7 &  0.058 &  7 &  0.086 \\
8  & 11 &    --- & 11 &    --- & 11 &    --- & 11 &    --- \\
9  & 10 & -0.488 & 10 &  0.109 & 10 & -0.790 & 10 &  0.132 \\
10 &  9 &  0.487 &  9 & -0.165 &  9 & -0.919 &  9 & -0.919 \\
11 & 11 & -0.635 & 11 & -0.066 & 11 &  0.113 & 11 & -0.132 \\
12 &  6 &  0.027 &  6 & -0.386 &  6 &  0.076 &  6 & -0.135 \\
13 &  9 & -0.182 &  9 & -0.381 &  9 & -0.265 &  9 &  0.217 \\
14 &  8 &  0.044 &  8 & -0.835 &  8 & -0.065 &  8 &  0.018 \\
15 &  3 & -0.156 &  3 & -0.954 &  3 & -0.493 &  3 & -0.007 \\
16 & 11 &  0.675 & 11 &  0.188 & 11 & -0.513 & 11 &  0.213 \\
17 & 11 & -0.322 & 11 &  0.255 & 11 &  0.029 & 11 & -0.868 \\
18 &  8 &  0.142 &  8 &  0.737 &  8 &  0.128 &  8 &  0.140 \\
19 &  6 &  0.172 &  6 &  0.807 &  6 & -0.816 &  6 & -0.220 \\
20 &  6 &  0.067 &  6 & -0.179 &  6 & -0.106 &  6 &  0.489 \\
21 &  5 &  0.221 &  5 & -0.008 &  5 &  0.581 &  5 &    --- \\
22 &  1 &    --- &  1 &    --- &  1 &    --- &  1 &    --- \\
23 &  9 & -0.031 &  9 & -0.219 &  9 & -0.087 &  9 & -0.393 \\
24 &  1 &    --- &  1 &    --- &  1 &    --- &  1 &    --- \\
25 & 11 &  0.319 & 11 &  0.221 & 11 & -0.024 & 11 &  0.109 \\
26 & 11 & -0.325 & 11 &  0.296 & 11 & -0.254 & 11 &  0.232 \\
27 &  3 & -0.352 &  3 &  0.958 &  3 &  0.485 &  3 & -0.618 \\
28 & 10 & -0.339 & 10 &  0.297 & 10 & -0.230 & 10 & -0.147 \\
\bottomrule
\end{tabular*}
\end{table}

\clearpage

\begin{table}[p]
\centering
\caption{Participant-specific joint residual correlations for daily components.}
\label{tab:daily-joint-residual-correlation}
\fontsize{9}{11}\selectfont
\setlength{\tabcolsep}{2pt}
\renewcommand{\arraystretch}{1.15}
\begin{tabular*}{\textwidth}{@{\extracolsep{\fill}}c*{12}{r}@{}}
\toprule
Participant
& \multicolumn{2}{c}{AA--FW}
& \multicolumn{2}{c}{AA--PJ}
& \multicolumn{2}{c}{AA--AD}
& \multicolumn{2}{c}{FW--PJ}
& \multicolumn{2}{c}{FW--AD}
& \multicolumn{2}{c}{PJ--AD} \\
\cmidrule(lr){2-3}\cmidrule(lr){4-5}
\cmidrule(lr){6-7}\cmidrule(lr){8-9}
\cmidrule(lr){10-11}\cmidrule(lr){12-13}
& $N$ & $\widehat\rho$
& $N$ & $\widehat\rho$
& $N$ & $\widehat\rho$
& $N$ & $\widehat\rho$
& $N$ & $\widehat\rho$
& $N$ & $\widehat\rho$ \\
\midrule
1 & 55 & -0.085 & 55 & -0.038 & 12 & 0.143 & 77 & 0.017 & 12 & -0.275 & 12 & -0.370 \\
2 & 51 & -0.042 & 51 & -0.444 & 13 & 0.646 & 77 & 0.054 & 13 & -0.219 & 13 & -0.224 \\
3 & 17 & -0.309 & 17 & -0.023 & 13 & 0.184 & 77 & -0.024 & 13 & 0.013 & 13 & 0.178 \\
4 & 13 & 0.060 & 13 & -0.200 & 7 & --- & 77 & -0.100 & 7 & --- & 7 & --- \\
5 & 12 & -0.039 & 12 & 0.108 & 10 & -0.358 & 77 & 0.112 & 10 & 0.541 & 10 & 0.489 \\
6 & 63 & -0.088 & 63 & 0.012 & 6 & --- & 77 & 0.111 & 6 & --- & 6 & --- \\
7 & 10 & 0.575 & 10 & 0.449 & 2 & --- & 77 & 0.293 & 2 & --- & 2 & --- \\
8 & 47 & 0.084 & 47 & 0.179 & 15 & -0.048 & 77 & 0.070 & 15 & -0.638 & 15 & -0.565 \\
9 & 55 & 0.136 & 55 & 0.015 & 14 & 0.713 & 77 & 0.167 & 14 & 0.217 & 14 & -0.370 \\
10 & 41 & -0.004 & 41 & 0.128 & 18 & 0.318 & 77 & 0.075 & 18 & 0.053 & 18 & -0.419 \\
11 & 46 & -0.159 & 46 & -0.025 & 36 & 0.443 & 77 & 0.146 & 36 & -0.002 & 36 & -0.122 \\
12 & 14 & 0.248 & 14 & -0.489 & 14 & 0.059 & 77 & 0.100 & 14 & 0.140 & 14 & 0.200 \\
13 & 58 & -0.004 & 58 & 0.160 & 34 & 0.195 & 77 & 0.165 & 34 & 0.052 & 34 & 0.283 \\
14 & 36 & -0.150 & 36 & -0.018 & 24 & -0.003 & 77 & 0.136 & 24 & -0.189 & 24 & -0.019 \\
15 & 6 & -0.691 & 6 & 0.141 & 2 & --- & 77 & 0.054 & 2 & --- & 2 & --- \\
16 & 31 & 0.031 & 31 & 0.090 & 23 & 0.292 & 77 & 0.019 & 23 & 0.108 & 23 & -0.077 \\
17 & 76 & -0.130 & 76 & -0.042 & 74 & 0.102 & 77 & 0.956 & 74 & -0.013 & 74 & 0.168 \\
18 & 16 & -0.360 & 16 & 0.219 & 6 & -0.643 & 77 & -0.207 & 6 & 0.324 & 6 & -0.339 \\
19 & 6 & -0.125 & 6 & 0.051 & 6 & 0.614 & 77 & 0.049 & 6 & 0.343 & 6 & 0.119 \\
20 & 17 & -0.026 & 17 & 0.083 & 13 & 0.125 & 77 & -0.097 & 13 & 0.272 & 13 & 0.176 \\
21 & 50 & 0.023 & 50 & -0.084 & 50 & 0.196 & 77 & 0.191 & 50 & -0.276 & 50 & -0.054 \\
22 & 6 & -0.230 & 6 & -0.559 & 3 & 0.387 & 77 & -0.090 & 3 & -0.927 & 3 & -0.783 \\
23 & 22 & -0.181 & 22 & -0.105 & 12 & --- & 77 & 0.084 & 12 & --- & 12 & --- \\
24 & 9 & -0.281 & 9 & 0.271 & 6 & 0.320 & 77 & 0.003 & 6 & 0.181 & 6 & -0.259 \\
25 & 54 & -0.040 & 54 & 0.085 & 25 & -0.069 & 77 & -0.119 & 25 & -0.152 & 25 & -0.165 \\
26 & 68 & 0.022 & 68 & -0.030 & 20 & 0.372 & 77 & 0.085 & 20 & -0.056 & 20 & -0.228 \\
27 & 20 & -0.170 & 20 & -0.219 & 13 & 0.310 & 77 & -0.088 & 13 & -0.327 & 13 & -0.039 \\
28 & 43 & 0.144 & 43 & 0.013 & 21 & -0.328 & 77 & -0.025 & 21 & 0.240 & 21 & -0.340 \\
\bottomrule
\end{tabular*}
\end{table}

\clearpage

\begin{table}[p]
\centering
\caption{Participant-specific joint residual correlations between step-count components.}
\label{tab:ampm-joint-steps}
\fontsize{9}{11}\selectfont
\setlength{\tabcolsep}{2pt}
\renewcommand{\arraystretch}{1.15}
\begin{tabular*}{\textwidth}{@{\extracolsep{\fill}}c*{6}{r}@{}}
\toprule
Participant
& \multicolumn{2}{c}{AM FourSC--AM P2h}
& \multicolumn{2}{c}{PM FourSC--PM P2h}
& \multicolumn{2}{c}{AM FourSC--PM P2h} \\
\cmidrule(lr){2-3}\cmidrule(lr){4-5}\cmidrule(lr){6-7}
& $N$ & $\widehat\rho$
& $N$ & $\widehat\rho$
& $N$ & $\widehat\rho$ \\
\midrule
1 & 62 & -0.039 & 63 & -0.004 & 66 & 0.530 \\
2 & 67 & 0.155 & 34 & -0.321 & 72 & 0.293 \\
3 & 67 & 0.053 & 0 & --- & 69 & 0.212 \\
4 & 61 & -0.108 & 64 & 0.046 & 70 & 0.488 \\
5 & 64 & -0.032 & 59 & 0.114 & 63 & 0.300 \\
6 & 59 & -0.014 & 59 & -0.059 & 59 & 0.255 \\
7 & 53 & 0.067 & 51 & -0.056 & 58 & 0.305 \\
8 & 75 & -0.055 & 77 & 0.045 & 76 & 0.269 \\
9 & 56 & 0.270 & 60 & -0.290 & 60 & -0.039 \\
10 & 73 & -0.179 & 73 & 0.058 & 76 & -0.049 \\
11 & 58 & 0.014 & 17 & 0.040 & 35 & 0.235 \\
12 & 22 & 0.317 & 28 & -0.042 & 26 & 0.290 \\
13 & 60 & -0.043 & 67 & 0.005 & 71 & 0.361 \\
14 & 55 & -0.086 & 58 & 0.138 & 63 & 0.572 \\
15 & 71 & 0.020 & 67 & -0.041 & 71 & 0.455 \\
16 & 51 & -0.006 & 49 & 0.042 & 59 & 0.476 \\
17 & 75 & -0.051 & 63 & -0.084 & 67 & 0.541 \\
18 & 75 & 0.141 & 75 & -0.084 & 77 & 0.398 \\
19 & 38 & 0.310 & 30 & -0.115 & 37 & 0.572 \\
20 & 61 & -0.127 & 60 & 0.173 & 62 & 0.383 \\
21 & 75 & 0.042 & 0 & --- & 74 & 0.109 \\
22 & 70 & -0.138 & 0 & --- & 0 & --- \\
23 & 54 & 0.041 & 48 & -0.051 & 60 & 0.513 \\
24 & 73 & 0.218 & 67 & -0.122 & 70 & 0.256 \\
25 & 76 & 0.011 & 73 & -0.082 & 76 & 0.432 \\
26 & 72 & 0.113 & 73 & -0.057 & 70 & 0.441 \\
27 & 24 & -0.150 & 39 & 0.315 & 55 & 0.312 \\
28 & 64 & -0.110 & 67 & 0.050 & 67 & 0.410 \\
\bottomrule
\end{tabular*}
\end{table}

\begin{table}[p]
\centering
\caption{Participant-specific residual correlations between step-count and engagement components at AM occasions.}
\label{tab:ampm-joint-physical-engagement-am}
\fontsize{9}{11}\selectfont
\setlength{\tabcolsep}{2pt}
\renewcommand{\arraystretch}{1.15}
\begin{tabular*}{\textwidth}{@{\extracolsep{\fill}}c*{12}{r}@{}}
\toprule
Participant
& \multicolumn{2}{c}{FourSC--$O$}
& \multicolumn{2}{c}{FourSC--PV$^+$}
& \multicolumn{2}{c}{FourSC--AS}
& \multicolumn{2}{c}{P2h--$O$}
& \multicolumn{2}{c}{P2h--PV$^+$}
& \multicolumn{2}{c}{P2h--AS} \\
\cmidrule(lr){2-3}\cmidrule(lr){4-5}
\cmidrule(lr){6-7}\cmidrule(lr){8-9}
\cmidrule(lr){10-11}\cmidrule(lr){12-13}
& $N$ & $\widehat\rho$
& $N$ & $\widehat\rho$
& $N$ & $\widehat\rho$
& $N$ & $\widehat\rho$
& $N$ & $\widehat\rho$
& $N$ & $\widehat\rho$ \\
\midrule
1 & 68 & 0.007 & 34 & -0.108 & 32 & 0.104 & 65 & 0.343 & 32 & 0.407 & 31 & 0.236 \\
2 & 72 & -0.006 & 28 & 0.337 & 44 & 0.090 & 69 & 0.245 & 28 & 0.114 & 41 & 0.038 \\
3 & 70 & -0.186 & 4 & -0.264 & 32 & -0.759 & 69 & 0.090 & 4 & -0.293 & 33 & -0.035 \\
4 & 72 & -0.088 & 21 & 0.130 & 38 & 0.026 & 62 & -0.009 & 19 & 0.040 & 34 & 0.047 \\
5 & 65 & -0.204 & 9 & -0.065 & 30 & -0.251 & 67 & 0.043 & 8 & -0.708 & 31 & -0.008 \\
6 & 65 & 0.241 & 7 & -0.259 & 38 & 0.152 & 63 & 0.035 & 7 & 0.041 & 38 & -0.031 \\
7 & 59 & -0.126 & 10 & 0.466 & 30 & -0.269 & 53 & 0.251 & 10 & 0.298 & 26 & 0.092 \\
8 & 76 & 0.016 & 31 & -0.115 & 36 & 0.083 & 76 & 0.107 & 32 & 0.028 & 37 & 0.165 \\
9 & 61 & -0.190 & 24 & -0.108 & 21 & -0.254 & 57 & -0.369 & 24 & -0.385 & 21 & -0.344 \\
10 & 76 & 0.047 & 33 & 0.504 & 46 & 0.237 & 74 & 0.004 & 34 & -0.019 & 46 & 0.003 \\
11 & 61 & -0.312 & 32 & 0.292 & 35 & -0.151 & 71 & -0.151 & 39 & 0.020 & 39 & -0.080 \\
12 & 27 & -0.153 & 10 & -0.035 & 16 & 0.026 & 23 & -0.080 & 9 & 0.005 & 13 & 0.160 \\
13 & 72 & -0.009 & 22 & -0.184 & 34 & 0.127 & 62 & -0.059 & 19 & 0.308 & 30 & -0.147 \\
14 & 63 & -0.051 & 21 & 0.095 & 30 & -0.098 & 57 & -0.150 & 19 & 0.310 & 28 & -0.393 \\
15 & 73 & -0.138 & 15 & -0.176 & 36 & -0.422 & 73 & 0.008 & 15 & 0.375 & 34 & -0.002 \\
16 & 59 & -0.080 & 14 & 0.075 & 35 & -0.030 & 53 & -0.266 & 13 & -0.286 & 30 & -0.223 \\
17 & 75 & -0.024 & 56 & -0.232 & 41 & --- & 76 & 0.178 & 56 & -0.154 & 41 & --- \\
18 & 77 & 0.063 & 17 & -0.063 & 35 & 0.163 & 75 & 0.048 & 17 & -0.149 & 35 & -0.025 \\
19 & 40 & 0.025 & 2 & --- & 18 & -0.196 & 42 & 0.157 & 2 & --- & 18 & 0.259 \\
20 & 65 & -0.047 & 21 & -0.170 & 30 & -0.109 & 63 & 0.165 & 20 & -0.133 & 28 & 0.415 \\
21 & 75 & -0.066 & 2 & --- & 42 & -0.251 & 75 & 0.028 & 2 & --- & 42 & 0.144 \\
22 & 76 & -0.090 & 5 & -0.200 & 59 & -0.202 & 70 & -0.137 & 4 & 0.043 & 54 & -0.209 \\
23 & 63 & -0.254 & 13 & -0.103 & 46 & -0.050 & 58 & 0.025 & 14 & 0.095 & 44 & -0.217 \\
24 & 73 & 0.071 & 15 & -0.332 & 33 & 0.073 & 73 & -0.097 & 15 & 0.197 & 33 & 0.286 \\
25 & 77 & 0.017 & 50 & -0.111 & 38 & 0.273 & 76 & -0.080 & 49 & -0.160 & 37 & -0.002 \\
26 & 72 & -0.034 & 42 & -0.167 & 21 & -0.106 & 72 & 0.031 & 42 & -0.124 & 21 & -0.378 \\
27 & 55 & 0.218 & 25 & 0.080 & 23 & 0.124 & 24 & 0.222 & 10 & 0.379 & 13 & 0.044 \\
28 & 68 & -0.035 & 22 & -0.148 & 46 & -0.182 & 64 & 0.147 & 22 & 0.139 & 43 & 0.225 \\
\bottomrule
\end{tabular*}
\end{table}

\begin{table}[p]
\centering
\caption{Participant-specific residual correlations between step-count and engagement components at PM occasions.}
\label{tab:ampm-joint-physical-engagement-pm}
\fontsize{9}{11}\selectfont
\setlength{\tabcolsep}{2pt}
\renewcommand{\arraystretch}{1.15}
\begin{tabular*}{\textwidth}{@{\extracolsep{\fill}}c*{12}{r}@{}}
\toprule
Participant
& \multicolumn{2}{c}{FourSC--$O$}
& \multicolumn{2}{c}{FourSC--PV$^+$}
& \multicolumn{2}{c}{FourSC--AS}
& \multicolumn{2}{c}{P2h--$O$}
& \multicolumn{2}{c}{P2h--PV$^+$}
& \multicolumn{2}{c}{P2h--AS} \\
\cmidrule(lr){2-3}\cmidrule(lr){4-5}
\cmidrule(lr){6-7}\cmidrule(lr){8-9}
\cmidrule(lr){10-11}\cmidrule(lr){12-13}
& $N$ & $\widehat\rho$
& $N$ & $\widehat\rho$
& $N$ & $\widehat\rho$
& $N$ & $\widehat\rho$
& $N$ & $\widehat\rho$
& $N$ & $\widehat\rho$ \\
\midrule
1 & 65 & -0.018 & 26 & -0.110 & 35 & -0.240 & 69 & -0.158 & 27 & -0.202 & 36 & 0.020 \\
2 & 35 & -0.076 & 11 & 0.585 & 19 & 0.144 & 72 & 0.156 & 25 & -0.229 & 37 & -0.022 \\
3 & 0 & --- & 0 & --- & 0 & --- & 70 & 0.081 & 0 & --- & 32 & 0.344 \\
4 & 64 & 0.041 & 20 & -0.432 & 32 & -0.172 & 70 & -0.175 & 21 & 0.112 & 36 & -0.017 \\
5 & 60 & -0.010 & 7 & -0.434 & 25 & -0.114 & 65 & -0.088 & 8 & -0.191 & 28 & -0.462 \\
6 & 71 & 0.068 & 4 & 0.184 & 33 & 0.192 & 62 & -0.141 & 4 & -0.013 & 29 & 0.249 \\
7 & 52 & -0.060 & 6 & 0.120 & 22 & -0.366 & 60 & 0.006 & 6 & -0.761 & 25 & 0.380 \\
8 & 77 & 0.105 & 40 & -0.101 & 38 & 0.099 & 77 & -0.124 & 40 & 0.037 & 38 & -0.029 \\
9 & 61 & 0.094 & 35 & -0.133 & 32 & 0.303 & 61 & -0.028 & 35 & 0.118 & 32 & -0.109 \\
10 & 73 & -0.241 & 44 & 0.192 & 41 & -0.094 & 76 & -0.023 & 47 & 0.031 & 44 & -0.274 \\
11 & 24 & -0.039 & 13 & 0.042 & 10 & -0.054 & 35 & 0.114 & 12 & 0.040 & 16 & 0.168 \\
12 & 32 & -0.104 & 18 & 0.024 & 20 & 0.118 & 28 & 0.262 & 16 & 0.637 & 16 & 0.591 \\
13 & 68 & -0.065 & 23 & 0.237 & 31 & -0.106 & 74 & 0.068 & 25 & -0.542 & 32 & 0.242 \\
14 & 60 & 0.032 & 21 & -0.105 & 23 & 0.051 & 65 & 0.194 & 21 & 0.303 & 25 & 0.220 \\
15 & 68 & -0.023 & 15 & 0.039 & 31 & 0.033 & 71 & 0.111 & 17 & 0.427 & 31 & 0.150 \\
16 & 53 & -0.165 & 7 & -0.243 & 26 & 0.061 & 61 & 0.099 & 8 & 0.220 & 30 & 0.254 \\
17 & 71 & -0.059 & 39 & -0.026 & 37 & --- & 67 & 0.066 & 36 & 0.049 & 34 & --- \\
18 & 75 & 0.114 & 12 & -0.459 & 34 & 0.323 & 77 & -0.080 & 12 & 0.044 & 35 & -0.108 \\
19 & 30 & -0.102 & 1 & --- & 11 & -0.183 & 45 & 0.070 & 6 & -0.208 & 22 & 0.077 \\
20 & 62 & 0.089 & 12 & 0.113 & 22 & 0.053 & 64 & 0.257 & 13 & 0.014 & 21 & 0.190 \\
21 & 0 & --- & 0 & --- & 0 & --- & 74 & 0.148 & 0 & --- & 39 & -0.023 \\
22 & 0 & --- & 0 & --- & 0 & --- & 0 & --- & 0 & --- & 0 & --- \\
23 & 50 & 0.152 & 5 & -0.676 & 37 & 0.132 & 61 & -0.083 & 9 & 0.763 & 46 & 0.079 \\
24 & 70 & 0.132 & 12 & -0.781 & 34 & -0.002 & 72 & 0.081 & 13 & -0.260 & 33 & -0.120 \\
25 & 74 & -0.120 & 37 & -0.079 & 39 & -0.216 & 76 & 0.069 & 38 & 0.019 & 40 & -0.158 \\
26 & 75 & -0.094 & 31 & 0.012 & 27 & 0.048 & 73 & 0.071 & 31 & -0.174 & 26 & 0.141 \\
27 & 39 & -0.272 & 9 & 0.221 & 22 & -0.396 & 65 & -0.114 & 10 & 0.313 & 34 & -0.255 \\
28 & 70 & 0.113 & 19 & -0.077 & 50 & 0.071 & 68 & 0.079 & 18 & 0.378 & 48 & -0.150 \\
\bottomrule
\end{tabular*}
\end{table}

\begin{table}[p]
\centering
\caption{Participant-specific joint residual correlations among page-view and interaction components.}
\label{tab:ampm-joint-engagement}
\fontsize{8}{9.5}\selectfont
\setlength{\tabcolsep}{2pt}
\renewcommand{\arraystretch}{0.96}

\begin{tabular*}{\textwidth}{@{\extracolsep{\fill}}c*{6}{r}@{}}
\toprule
\multicolumn{7}{c}{AM} \\
\midrule
Participant
& \multicolumn{2}{c}{$O$--PV$^+$}
& \multicolumn{2}{c}{$O$--AS}
& \multicolumn{2}{c}{PV$^+$--AS} \\
\cmidrule(lr){2-3}\cmidrule(lr){4-5}\cmidrule(lr){6-7}
& $N$ & $\widehat\rho$
& $N$ & $\widehat\rho$
& $N$ & $\widehat\rho$ \\
\midrule
1 & 34 & 0.084 & 35 & 0.478 & 15 & 0.497 \\
2 & 30 & -0.011 & 48 & 0.447 & 16 & 0.116 \\
3 & 4 & 0.154 & 35 & 0.532 & 3 & 0.277 \\
4 & 21 & 0.141 & 41 & 0.448 & 14 & -0.167 \\
5 & 10 & -0.200 & 34 & 0.436 & 5 & -0.382 \\
6 & 9 & -0.004 & 45 & 0.813 & 9 & 0.700 \\
7 & 12 & 0.301 & 42 & 0.610 & 7 & -0.031 \\
8 & 32 & -0.062 & 37 & 0.690 & 22 & -0.043 \\
9 & 27 & -0.171 & 29 & 0.674 & 16 & -0.013 \\
10 & 34 & 0.098 & 46 & 0.514 & 26 & 0.652 \\
11 & 41 & -0.059 & 41 & 0.746 & 33 & 0.230 \\
12 & 22 & -0.211 & 42 & 0.550 & 16 & 0.685 \\
13 & 24 & 0.267 & 38 & 0.801 & 15 & -0.106 \\
14 & 27 & 0.005 & 38 & 0.655 & 18 & 0.407 \\
15 & 16 & -0.034 & 37 & 0.865 & 10 & 0.561 \\
16 & 20 & 0.370 & 41 & 0.883 & 15 & -0.175 \\
17 & 56 & -0.058 & 41 & --- & 41 & --- \\
18 & 17 & 0.023 & 35 & 0.879 & 10 & -0.272 \\
19 & 4 & 0.299 & 38 & 0.940 & 2 & --- \\
20 & 25 & -0.197 & 37 & 0.877 & 11 & 0.118 \\
21 & 2 & --- & 43 & 0.332 & 1 & --- \\
22 & 5 & -0.085 & 60 & 0.755 & 5 & 0.116 \\
23 & 14 & 0.029 & 55 & 0.346 & 11 & 0.030 \\
24 & 16 & -0.098 & 34 & 0.339 & 6 & 0.353 \\
25 & 50 & -0.005 & 38 & 0.440 & 34 & 0.210 \\
26 & 44 & -0.171 & 22 & 0.995 & 21 & 0.145 \\
27 & 33 & -0.134 & 36 & 0.717 & 15 & -0.143 \\
28 & 25 & -0.108 & 53 & 0.526 & 16 & -0.079 \\
\bottomrule
\end{tabular*}

\par\vspace{1.5ex}

\begin{tabular*}{\textwidth}{@{\extracolsep{\fill}}c*{6}{r}@{}}
\toprule
\multicolumn{7}{c}{PM} \\
\midrule
Participant
& \multicolumn{2}{c}{$O$--PV$^+$}
& \multicolumn{2}{c}{$O$--AS}
& \multicolumn{2}{c}{PV$^+$--AS} \\
\cmidrule(lr){2-3}\cmidrule(lr){4-5}\cmidrule(lr){6-7}
& $N$ & $\widehat\rho$
& $N$ & $\widehat\rho$
& $N$ & $\widehat\rho$ \\
\midrule
1 & 30 & -0.187 & 40 & 0.358 & 17 & 0.495 \\
2 & 28 & -0.036 & 39 & -0.057 & 18 & 0.176 \\
3 & 0 & --- & 35 & 0.205 & 0 & --- \\
4 & 24 & -0.426 & 41 & 0.537 & 15 & 0.117 \\
5 & 10 & 0.206 & 33 & -0.009 & 5 & -0.142 \\
6 & 5 & -0.478 & 36 & 0.052 & 3 & 0.983 \\
7 & 6 & -0.032 & 35 & 0.613 & 3 & -0.632 \\
8 & 40 & 0.042 & 38 & 0.536 & 30 & -0.042 \\
9 & 39 & 0.081 & 40 & 0.224 & 25 & 0.309 \\
10 & 48 & -0.047 & 45 & 0.532 & 33 & 0.145 \\
11 & 31 & 0.076 & 32 & -0.703 & 23 & 0.232 \\
12 & 28 & 0.115 & 41 & 0.399 & 17 & 0.366 \\
13 & 25 & -0.253 & 35 & 0.278 & 15 & -0.135 \\
14 & 25 & 0.020 & 28 & 0.542 & 9 & -0.393 \\
15 & 17 & 0.008 & 35 & 0.985 & 7 & -0.353 \\
16 & 12 & -0.363 & 37 & 0.382 & 9 & 0.479 \\
17 & 42 & 0.266 & 40 & --- & 39 & --- \\
18 & 12 & -0.173 & 35 & 0.467 & 6 & 0.834 \\
19 & 9 & -0.043 & 36 & 0.808 & 7 & -0.526 \\
20 & 17 & 0.308 & 30 & 0.673 & 8 & -0.047 \\
21 & 0 & --- & 42 & -0.292 & 0 & --- \\
22 & 0 & --- & 38 & 0.092 & 0 & --- \\
23 & 9 & 0.235 & 55 & 0.532 & 9 & -0.057 \\
24 & 13 & 0.117 & 38 & 0.290 & 6 & 0.839 \\
25 & 38 & 0.057 & 40 & 0.382 & 30 & 0.600 \\
26 & 33 & 0.318 & 28 & -0.029 & 23 & -0.337 \\
27 & 13 & 0.454 & 42 & 0.456 & 10 & 0.168 \\
28 & 19 & 0.055 & 55 & 0.399 & 17 & -0.257 \\
\bottomrule
\end{tabular*}
\end{table}

\clearpage

\clearpage

\clearpage

\clearpage

\clearpage

\clearpage

\section{Experimental Design and Additional Results}
\label{app:experiments}

We evaluate the seven RL configurations in Table~\ref{tab:rl-variants} together with a non-learning reference policies: random send with probability $0.5$. Each policy is evaluated in the vanilla MRT-calibrated testbed and in the alternative simulation environments described in Section~\ref{ssec:ste-tuning}.

Each simulated participant is followed for $36$ weeks. Week 1 is an initial micro-randomization period in which activity suggestions are assigned with probability $0.5$. Beginning in week 2, the RL policies are updated weekly according to Algorithm~\ref{alg:rlsvi1}. The non-learning reference policies use the same simulated observation process and weekly schedule but do not update from accumulating data.

Within each Monte Carlo replication, all candidate policies are evaluated on the same sampled set of participant types. We additionally use common random numbers for exogenous simulator stochasticity whenever possible, so that policy comparisons are paired at the simulated-participant level. We perform 200 Monte Carlo replications for each combination of simulation environment and policy. 

\subsection{Evaluation Metrics}
\label{subsec:evaluation-metrics}

The primary performance criterion is cumulative weekly CAE, matching the scientific objective of the RL system. For simulated participant $i$ under policy $\pi$, define
\[
G_i^\pi(W)
=
\sum_{w=1}^{W}Y_{i,w}^\pi,
\]
where $W=36$ for the final-horizon analysis.

Because policies are evaluated on matched simulated participants, the primary policy contrast is computed within replication relative to the never-send reference policy:
\[
D_r^\pi(W)
=
\frac{1}{N}
\sum_{i=1}^{N}
\left[
G_{r,i}^\pi(W)
-
G_{r,i}^{\mathrm{never}}(W)
\right].
\]
We report the Monte Carlo mean
\[
\widehat\Delta^\pi(W)
=
\frac{1}{R}
\sum_{r=1}^{R}
D_r^\pi(W)
\]
and its Monte Carlo standard error,
\[
\operatorname{SE}_{\mathrm{MC}}
\left(
\widehat\Delta^\pi(W)
\right)
=
\frac{
\operatorname{SD}\{D_1^\pi(W),\ldots,D_R^\pi(W)\}
}{
\sqrt R
}.
\]

\subsection{Additional Results}
\label{subsec:additional-results}

Figure~2 of the main paper summarizes cumulative CAE but does not reveal how the learned policies differ in their action-selection behavior. We therefore examine the week-specific send probabilities for the morning and afternoon decision slots, denoted $\pi_{\mathrm{AM}}$ and $\pi_{\mathrm{PM}}$.

\begin{figure}[htbp]
    \centering
    \includegraphics[width=\linewidth]{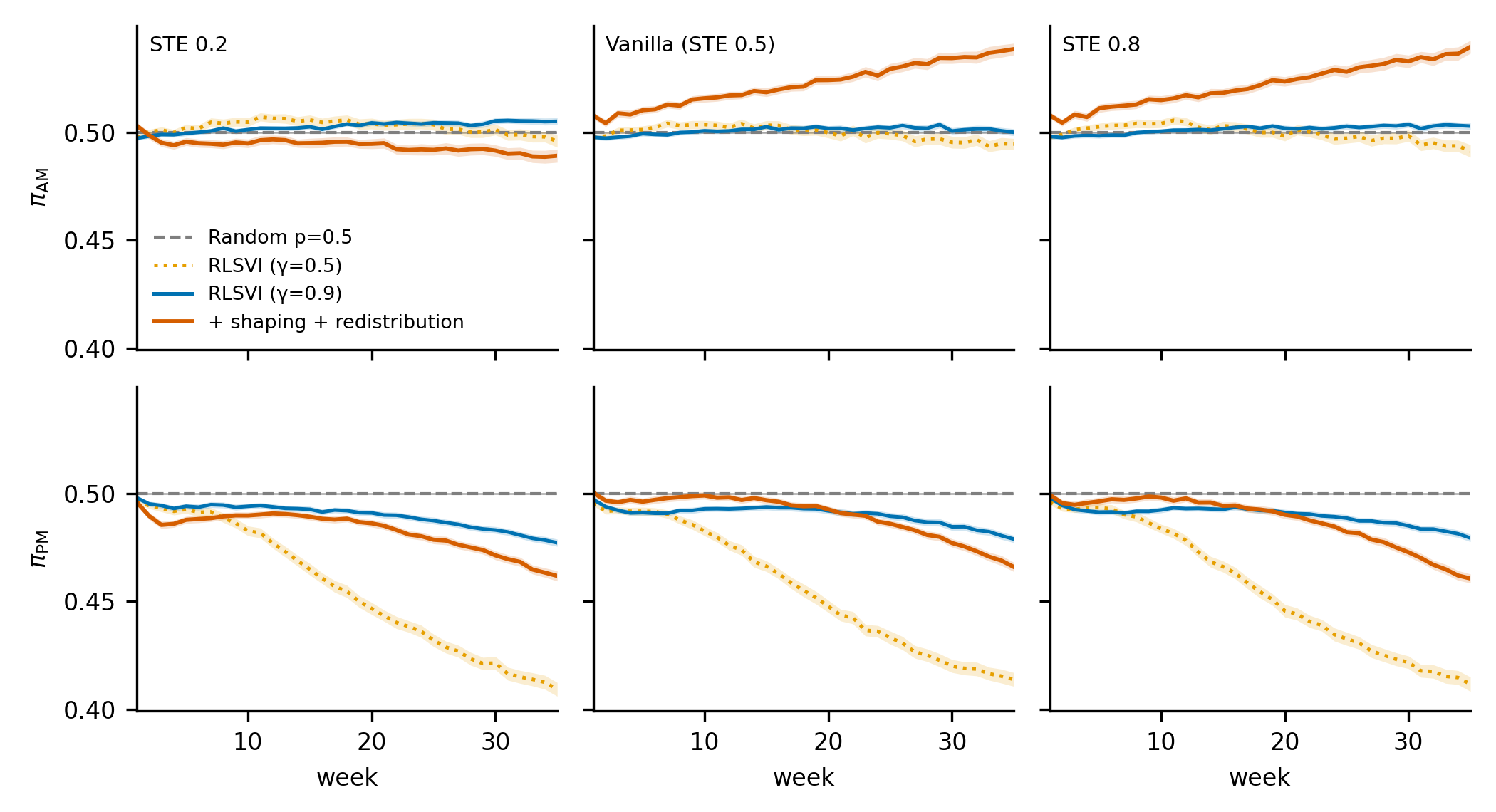}
    \caption{Population-average send probability by decision slot and week, with 95\% confidence intervals across 200 simulation replications. Rows correspond to morning and afternoon decision slots; columns correspond to STE $\approx 0.2$, vanilla (STE $\approx 0.5$), and STE $\approx 0.8$ environments. The gray dashed line denotes random send ($p=0.5$).}
    \label{fig:pi-am-pm}
\end{figure}

Figure~\ref{fig:pi-am-pm} shows distinct policy behaviors across RL variants. Standard RLSVI with $\gamma=0.9$ remains close to $0.5$ in both decision slots. With $\gamma=0.5$, the morning probability remains near $0.5$, whereas the afternoon probability decreases to approximately $0.42$ across all three environments, reducing the overall intervention rate. Because this morning--afternoon asymmetry changes little as the intervention signal strengthens, it does not appear to reflect learning of the amplified proximal treatment pathway.

In contrast, the shaping-plus-redistribution variant primarily reallocates intervention probability across decision slots while leaving the overall intervention rate relatively stable. In the vanilla and STE $\approx 0.8$ environments, the morning probability increases to approximately $0.53$ and the afternoon probability decreases to approximately $0.47$, whereas little separation is observed under STE $\approx 0.2$. This pattern is consistent with the cumulative-CAE results in Figure 2 of the main paper: the advantage of redistribution becomes more pronounced as the underlying intervention signal strengthens.

\begin{figure}[htbp]
    \centering
    \includegraphics[width=\linewidth]{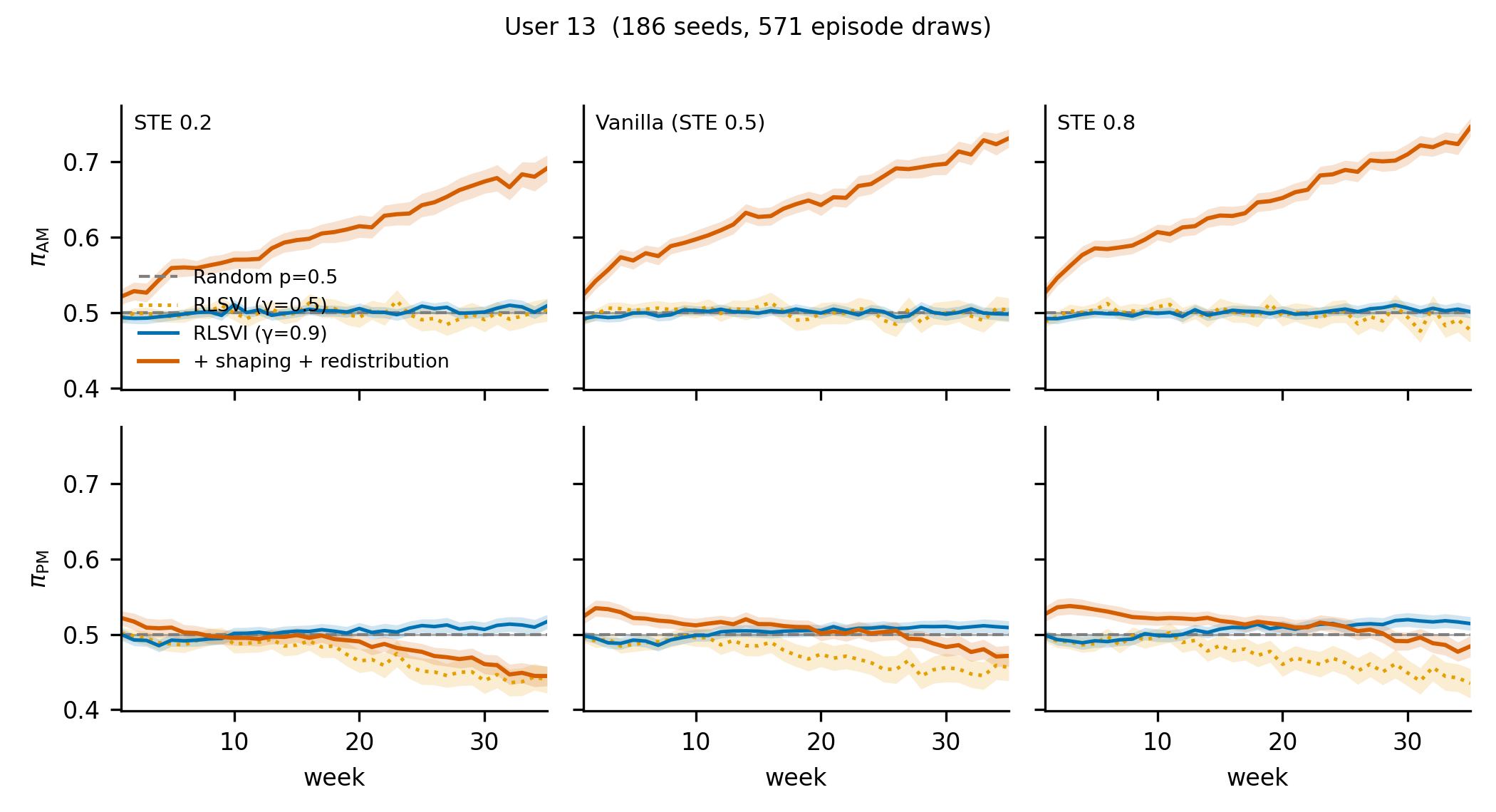}
    \caption{Participant-specific send probability by decision slot and week for an illustrative participant (user 13), based on 186 replications in which this participant model was sampled. The fitted environment for this participant favors morning activity suggestions.}
    \label{fig:user-pi-am-pm}
\end{figure}

Figure~\ref{fig:user-pi-am-pm} illustrates the corresponding participant-specific behavior for a participant whose fitted intervention effect is stronger in the morning than in the afternoon.  In this user's fitted mediator models, a
suggestion raises anticipated affect by $0.047$ in the morning versus $0.007$ in the afternoon; the implied fourSC action effect is
likewise positive in the morning and near zero in the afternoon.
Under shaping with learned redistribution, the morning send probability increases to approximately $0.71$ by the end of the study, while the afternoon probability remains near $0.5$. Standard RLSVI with $\gamma=0.9$ shows little differentiation between slots, whereas $\gamma=0.5$ again primarily reduces the afternoon probability. Although this single participant is illustrative, the result shows how reward redistribution can translate heterogeneous mediator signals into participant-specific temporal patterns of intervention delivery.

\end{document}